\documentclass[twocolumn]{aastex631}

\usepackage[caption=false]{subfig}
\usepackage{array}
\newcolumntype{H}{>{\setbox0=\hbox\bgroup}c<{\egroup}@{}}

\usepackage{color}

\newcommand{\Lya}{Ly$\alpha$}
\newcommand{\OII}{[O\,{\sc ii}]}
\newcommand{\OIII}{[O\,{\sc iii}]}
\newcommand{\OIIIuv}{O\,{\sc iii}]}
\newcommand{\Ha}{H$\alpha$}
\newcommand{\Hb}{H$\beta$}
\newcommand{\Hg}{H$\gamma$}
\newcommand{\Hd}{H$\delta$}
\newcommand{\CIII}{C{\sc iii}]}
\newcommand{\CIV}{C{\sc iv}}
\newcommand{\HeII}{He{\sc ii}}

\newcommand{\NeIII}{[Ne\,{\sc iii}]}
\newcommand{\NII}{[N\,{\sc ii}]}
\newcommand{\SII}{[S\,{\sc ii}]}
\newcommand{\fesc}{$f_{\rm esc}$}

\newcommand{\xiion}{$\xi_{\rm ion}$}

\newcommand{\Oabundance}{$12+\log ({\rm O/H})$}
\newcommand{\Zsun}{$Z_{\odot}$}

\newcommand{\HII}{H{\sc ii}}

\newcommand{\ebv}{E(B$-$V)}

\newcommand{\Msun}{$M_{\odot}$}

\newcommand{\Muv}{$M_{\rm UV}$}
\newcommand{\Mi}{$M_{i}$}

\accepted{in ApJ Supplement Series: Jun 05, 2022}

\shorttitle{Metallicity Diagnostics Established by a Local Galaxy Census}
\shortauthors{Nakajima et al.}

\begin{document}

\title{EMPRESS. V. Metallicity Diagnostics of Galaxies over \Oabundance\ $\simeq 6.9 - 8.9$\\
Established by a Local Galaxy Census: Preparing for JWST Spectroscopy}

\correspondingauthor{Kimihiko Nakajima}
\email{kimihiko.nakajima@nao.ac.jp}

\author[0000-0003-2965-5070]{Kimihiko Nakajima}
\affiliation{National Astronomical Observatory of Japan, 2-21-1 Osawa, Mitaka, Tokyo 181-8588, Japan}

\author[0000-0002-1049-6658]{Masami Ouchi}
\affiliation{National Astronomical Observatory of Japan, 2-21-1 Osawa, Mitaka, Tokyo 181-8588, Japan}
\affiliation{Institute for Cosmic Ray Research, The University of Tokyo, 5-1-5 Kashiwanoha, Kashiwa, Chiba 277-8582, Japan}
\affiliation{Kavli Institute for the Physics and Mathematics of the Universe (WPI), University of Tokyo, Kashiwa, Chiba 277-8583, Japan}

\author[0000-0002-5768-8235]{Yi Xu}
\affiliation{Institute for Cosmic Ray Research, The University of Tokyo, 5-1-5 Kashiwanoha, Kashiwa, Chiba 277-8582, Japan}
\affiliation{Department of Astronomy, Graduate School of Science, The University of Tokyo, 7-3-1 Hongo, Bunkyo, Tokyo 113-0033, Japan}

\author{Michael Rauch}
\affiliation{Observatories of the Carnegie Institution for Science, 813 Santa Barbara St., Pasadena, CA 91101, USA}

\author[0000-0002-6047-430X]{Yuichi Harikane}
\affiliation{Institute for Cosmic Ray Research, The University of Tokyo, 5-1-5 Kashiwanoha, Kashiwa, Chiba 277-8582, Japan}
\affiliation{Department of Physics and Astronomy, University College London, Gower Street, London WC1E 6BT, UK}

\author{Moka Nishigaki}
\affiliation{Department of Astronomical Science, The Graduate University for Advanced Studies, SOKENDAI, 2-21-1 Osawa, Mitaka, Tokyo, 181-8588, Japan}

\author[0000-0001-7730-8634]{Yuki Isobe}
\affiliation{Institute for Cosmic Ray Research, The University of Tokyo, 5-1-5 Kashiwanoha, Kashiwa, Chiba 277-8582, Japan}
\affiliation{Department of Physics, Graduate School of Science, The University of Tokyo, 7-3-1 Hongo, Bunkyo, Tokyo 113-0033, Japan}

\author[0000-0002-3801-434X]{Haruka Kusakabe}
\affiliation{Observatoire de Gen\`{e}ve, Universit\'e de Gen\`{e}ve, 51 Chemin de P\'egase, 1290 Versoix, Switzerland}

\author[0000-0002-7402-5441]{Tohru Nagao}
\affiliation{Research Center for Space and Cosmic Evolution, Ehime University, Matsuyama, Ehime 790-8577, Japan}

\author[0000-0001-9011-7605]{Yoshiaki Ono}
\affiliation{Institute for Cosmic Ray Research, The University of Tokyo, 5-1-5 Kashiwanoha, Kashiwa, Chiba 277-8582, Japan}

\author[0000-0003-3228-7264]{Masato Onodera}
\affiliation{Subaru Telescope, National Astronomical Observatory of Japan, National Institutes of Natural Sciences (NINS), 650 North Aohoku Place, Hilo, HI 96720, USA}
\affiliation{Department of Astronomical Science, The Graduate University for Advanced Studies, SOKENDAI, 2-21-1 Osawa, Mitaka, Tokyo, 181-8588, Japan}

\author[0000-0001-6958-7856]{Yuma Sugahara}
\affiliation{National Astronomical Observatory of Japan, 2-21-1 Osawa, Mitaka, Tokyo 181-8588, Japan}
\affiliation{Waseda Research Institute for Science and Engineering, Faculty of Science and Engineering, Waseda University, 3-4-1, Okubo, Shinjuku, Tokyo 169-8555, Japan}

\author[0000-0002-1418-3309]{Ji Hoon Kim}
\affiliation{Subaru Telescope, National Astronomical Observatory of Japan, National Institutes of Natural Sciences (NINS), 650 North Aohoku Place, Hilo, HI 96720, USA}
\affiliation{Metaspace, 36 Nonhyeon-ro, Gangnam-gu, Seoul 06312, Republic of Korea}
\affiliation{Astronomy Program, Department of Physics and Astronomy, Seoul National University, 1 Gwanak-ro, Gwanak-gu, Seoul 08826, Republic of Korea}
\affiliation{SNU Astronomy Research Center, Seoul National University, 1 Gwanak-ro, Gwanak-gu, Seoul 08826, Republic of Korea}

\author[0000-0002-3852-6329]{Yutaka Komiyama}
\affiliation{National Astronomical Observatory of Japan, 2-21-1 Osawa, Mitaka, Tokyo 181-8588, Japan}
\affiliation{Department of Astronomical Science, The Graduate University for Advanced Studies, SOKENDAI, 2-21-1 Osawa, Mitaka, Tokyo, 181-8588, Japan}

\author[0000-0003-1700-5740]{Chien-Hsiu Lee}
\affiliation{NSF's National Optical-Infrared Astronomy Research Laboratory, 950 N Cherry Ave., Tucson, AZ 86719, USA}

\author[0000-0001-7869-2551]{Fakhri S. Zahedy}
\affiliation{Observatories of the Carnegie Institution for Science, 813 Santa Barbara St., Pasadena, CA 91101, USA}



\begin{abstract}

We present optical-line gas metallicity diagnostics established by 
the combination of local SDSS galaxies
and the largest compilation of extremely metal-poor galaxies (EMPGs)
including new EMPGs identified by the Subaru EMPRESS survey.
A total of 103 EMPGs are included
that cover a large parameter space of magnitude 
(\Mi\ $=-19$ to $-7$)
and \Hb\ equivalent width ($10 - 600$\,\AA), i.e., 
wide ranges of stellar mass and star-formation rate.
Using reliable metallicity measurements of the direct method for these galaxies, 
we derive the relationships between strong optical-line ratios and 
gas-phase metallicity 
over the range of \Oabundance\ $\simeq 6.9 - 8.9$ 
corresponding to $0.02 - 2$ solar metallicity \Zsun.
We confirm that R23-index, (\OIII$+$\OII)/\Hb, is the most accurate 
metallicity indicator
with the metallicity uncertainty of $0.14$\,dex over the range 
among various popular metallicity indicators.
The other metallicity indicators show large scatters in the metal-poor range
($\lesssim 0.1$\,\Zsun).
It is explained by our \verb+CLOUDY+ photoionization modeling that, 
unlike R23-index, 
the other metallicity indicators do not use a sum of singly and doubly 
ionized lines and cannot trace both low and high ionization gas.
We find that the accuracy of the metallicity indicators is significantly
improved, if one uses \Hb\ equivalent width measurements 
that tightly correlate with ionization states.
In this work, 
we also present the relation of physical properties with 
UV-continuum slope $\beta$
and ionization production rate \xiion\ derived with GALEX data for the EMPGs,
and provide local anchors of galaxy properties together 
with the optical-line metallicity indicators
that are available in the form of ASCII table
and useful for forthcoming JWST spectroscopic studies.

\end{abstract}


\section{Introduction} \label{sec:intro}

Before the next generation of powerful telescopes such as 
the \textsl{James Webb Space Telescope} (JWST) and 
the $30$\,m-class extremely large telescopes will come online, 
there is an increasing awareness of the importance of low-redshift, 
young, and low-mass star-forming galaxies 
as probes of systems in the early universe. 
Although such local galaxies would have different characteristics and 
formation processes from high redshift galaxies, they are still useful to discuss
which characteristics we can use to address big questions such as the 
star-formation of galaxies in the early phase of galaxy evolution, 
their subsequent evolution, and 
the role of galaxies during the reionization process.

An important knowledge we can learn from the low-redshift galaxies 
is the emergent emission-line spectrum as a function of galaxies' key properties
such as metallicity and ionization parameter in the inter-stellar medium (ISM)
as well as the shape of ionizing spectrum.
Once the relationships are confirmed in conjunction with theoretical models, 
we can extend the knowledge toward
higher-redshifts to address early galaxies' properties via spectroscopic studies.
Gas-phase metallicity is one of the most crucial quantities, as it reflects 
the star-formation/explosion history (\citealt{MM2019} for a review).
It has thus long been studied how to estimate gas-phase metallicity in distant galaxies.

The most accurate gas metallicity estimate is provided if the electron temperature 
($T_e$) is directly measured via auroral lines such as \OIII$\lambda 4363$.
This is called the direct $T_e$ method, and most reliably used to identify 
the primitive galaxies (e.g., \citealt{kojima2020, izotov2018_lowZ}) and examine the 
scaling relations such as the stellar mass -- metallicity relation (e.g., \citealt{AM2013}).
The direct $T_e$ method is not always available, however, due to the faint nature of
the auroral lines.
Instead, stronger metal lines are focused and substituted to estimate metallicities
for faint galaxies especially at high redshift.
Strong metal lines divided by a hydrogen line, such as 
(\OIII$\lambda\lambda 5007, 4959$$+$\OII$\lambda 3727$)$/$\Hb\ (R23-index)
and \NII$\lambda 6584$$/$\Ha\ (N2-index),
as well as other metal line ratios 
have been historically investigated and proposed as gas-phase metallicity indicators 
both observationally 
(empirically; \citealt{pagel1979, edmunds1984, vanZee1998, PP2004, pilyugin2005, stasinska2006, nagao2006_metallicity, maiolino2008, marino2013, pilyugin2016, curti2017, curti2020})
and theoretically with photoionization models 
(\citealt{mcgaugh1991, KD2002, blanc2015, strom2018}; 
see \citealt{MM2019} for a comprehensive review).
\citet{maiolino2008} make use of individual low-metallicity galaxies as compiled in
\citet{nagao2006_metallicity} and derive the strong lines' diagnostics
in an empirical manner.
However, the authors have to rely on the photoionization model fitting for the 
high metallicity galaxies, because the detection of auroral lines becomes more challenging
from galaxies with a higher metallicity and a lower gas temperature.
This would cause a systematic change in line ratios as a function of metallicity 
between the low and high metallicity regimes (see also \citealt{nagao2006_metallicity}).
Recently, \citet{curti2017} and the subsequent study of \citet{curti2020} obtain
the relations at such a high-metallicity regime (\Oabundance\ $\simeq 8.1-8.9$)
based on the direct $T_e$ method by stacking spectra of Sloan Digital Sky Survey (SDSS). 
Nevertheless, the authors do not fully compile rare, individual metal-poor objects
for developing the diagnostics (\Oabundance\ $\gtrsim 7.7$). 
Accordingly, a comprehensive study of metallicity diagnostics 
using a larger sample at low-metallicity and covering the full metallicity range
is now required.

Moreover, a possible dependence of the empirical metallicity diagnostics on
the ionization state needs to be examined. 
\citet{pilyugin2016} discuss the effect of ionization state on the metallicity calibrators
(see also \citealt{pilyugin2005,KD2002,izotov2019_lowZcandidates}).
It is suggested to use several metal lines probing both high and low ionization gas 
to correct for the effect especially at the low-metallicity regime.
Such a correction, 
which is ideally feasible even when limited sets of emission lines are available, 
is particularly important 
if the diagnostics are applied to high redshift objects, because
a typical ionization parameter is suggested to become higher in galaxies 
at a higher redshift (e.g., \citealt{NO2014}).

From another viewpoint of what can be learned about early galaxies from local observations,
recent work has been particularly focusing on the properties
in the rest-frame ultra-violet (UV) wavelength. 
For example, \citet{izotov2016_nature,izotov2016_4more} observe $z\simeq 0.3$ galaxies 
with an extremely large equivalent width (EW) of \OIII$\lambda 5007$ 
called green pea galaxies in the rest-frame below the Lyman limit, 
and confirm they present Lyman continuum (LyC) leakage with a moderate escape fraction
(\fesc\ $=0.02 - 0.7$, median is \fesc\ $\sim 0.1$; see also \citealt{izotov2018_46per,izotov2018_5more}).
Because such intense emission lines are thought to be more common in galaxies 
at higher redshift (e.g., \citealt{smit2014,khostovan2016,reddy2018_mosdef,tang2019,nakajima2020}),
the correlation between \fesc\ and emission lines' visibility needs to be further 
investigated and clarified (see also \citealt{NO2014,faisst2016_xiion,plat2019,ramambason2020}).
An expanded survey at low-redshifts is thus on-going with Hubble Space Telescope 
(HST; GO 15626, PI: A. Jaskot; see also \citealt{flury2022}).

The non-ionizing UV wavelength spectrum also provides useful indicators and is actively 
studied for local galaxies, 
including \Lya\ emission (e.g., \citealt{henry2015,verhamme2017,jaskot2019,izotov2020}),
additional prominent UV emission lines 
such as \CIII$\lambda 1909$, \CIV$\lambda 1549$, and \HeII$\lambda 1640$
(e.g., \citealt{berg2016,senchyna2017}), 
and also absorption lines from the ISM (e.g., \citealt{chisholm2018}).
Especially, intense high ionization emission lines such as \HeII\ and \CIV\ are 
often identified in young, metal-poor galaxies 
(\citealt{senchyna2019}). 
This also supports the importance of local young, metal-poor galaxies as analogs of high redshift systems
since intense high ionization UV emission lines are more frequently found at high redshift
remarkably in the reionization era
(\citealt{erb2010,stark2014,berg2018,stark2015_c3, stark2015_c4, stark2017, laporte2017_agn, mainali2018, hutchison2019, jiang2021, topping2021}).
Detailed modelings are also emerging to accurately interpret the UV emission lines
(e.g., \citealt{feltre2016,gutkin2016,JR2016,nakajima2018_vuds,byler2018,hirschmann2019}).
The investigation of the UV spectrum in galaxies will be further explored in the local 
universe with HST/COS (GO 15840, PI: D. Berg; see also \citealt{berg2022}).

Despite the importance, many low redshift sources studied in details (i.e., followed-up
with the UV instruments) are relatively massive with a stellar mass 
M$_{\star}=10^7$--$10^9$\,M$_{\odot}$
and bright, evolved galaxies whose metallicity are modest, 
as low as the sub-solar metallicity value (Z $=0.1$--$0.3\,Z_{\odot}$).
The next focus in the community is thus to explore the properties of further young, low-mass, and
metal-poor galaxies, ideally as primordial as first-generation galaxies in the early universe
(e.g., \citealt{wise2012}).
These will also enable us to determine the metallicity diagnostics at the low-metallicity end.
In this paper, we investigate such metal-poor galaxies, especially highlighting those
with a metallicity below $10$\,\%\ of the solar value
(Z$<0.1$\,Z$_{\odot}$, equivalently \Oabundance\ $<7.69$ based on the solar 
chemical composition of \citealt{asplund2009})
which are called extremely metal-poor galaxies, or dubbed ``EMPGs''.
This paper is a part of a program named EMPRESS \citep{kojima2020} to 
systematically sample EMPGs and examine their detailed properties.
In EMPRESS, we use the deep and wide multi-wavelength imaging data of 
Subaru/Hyper Suprime-Cam (HSC) Subaru Strategic Program (SSP; 
\citealt{aihara2018}).
At $z\lesssim 0.03$, low-mass, actively star-forming galaxies present $g$- and $r$-band 
excesses with intense nebular emission lines of \OIII$\lambda\lambda 5007,4959$ and \Ha, 
respectively. In particular, a metal-poorer galaxy tends to present a redder 
($g-r$) color with a weaker contribution of metal line of \OIII\ relative to \Ha.
Although a similar approach using a broadband excess is successfully developed 
to select intense \OIII\ emitting galaxies such as 
green pea galaxies \citep{cardamone2009} at $z\sim 0.3$ 
and blueberry galaxies \citep{yang2017} at $z\sim 0.02$, 
they should not be extremely metal-poor by definition.
We adopt a machine learning technique to reliably search for EMPGs whose strong hydrogen lines 
while moderate-to-weak metal lines are imprinted on the photometric broadband 
SEDs.
A similar idea is also adopted by \citet{SS2019} to search for EMPGs based on
the photometric data. 
A series of spectroscopic follow-up has been conducted for the EMPG candidates
of EMPRESS to confirm their metallicities and spectroscopically characterize
the properties particularly about the presence of massive stars in metal-poor 
galaxies (\citealt{kojima2021,isobe2021_fe}), 
stellar feedbacks \citep{xu2022}, and the shape of ionizing spectrum 
to explain the high ionization emission lines \citep{umeda2022}.
This paper presents another spectroscopic observation recently conducted 
to enlarge the EMPG sample, 
and develop the metallicity diagnostics at the lowest metallicity range.
Furthermore, we present the fundamental UV properties 
such as the UV continuum slope $\beta$ and the ionizing photon production 
efficiency \xiion\ 
to examine the young stellar population at the extremely metal-poor regime.

The paper is organized as follows. 
We describe our sample of EMPGs in \S\ref{sec:samples}. 
This includes newly-identified EMPGs by EMPRESS, and compiles
previously known metal-poor objects from the literature.
In \S\ref{sec:diagnostics}, we present the metallicity diagnostics over the 
metallicity range \Oabundance\ $\simeq 6.9-8.9$ and discuss the scatters 
in the metal-poor regime in detail. Prescriptions are also proposed 
to improve the accuracy of the metallicity indicators 
by correcting for the variation of ionization state.
In \S\ref{sec:UV_properties}, we present the UV properties of the compiled EMPGs 
using the \textsl{GALEX} photometric data, and then examine the dependencies 
of the UV properties on metallicity and so on.
Finally, we summarize our findings in \S\ref{sec:summary}.
Throughout the paper we assume a solar chemical composition 
following \citet{asplund2009}, and adopt a concordance cosmology with 
$\Omega_\Lambda\ =0.7$, $\Omega_M =0.3$ and $H_0=70$\,km\,s$^{-1}$\,Mpc$^{-1}$.
All magnitudes are given in the AB system \citep{OG1983}.

\begin{splitdeluxetable*}{lcccccccccBcccccccccc}
\tablecaption{Extinction-corrected emission line fluxes of MagE sources
\label{tbl:fluxes_mage}}
\tablewidth{0.99\columnwidth}
\tabletypesize{\scriptsize}
\tablehead{
\colhead{ID}
& \colhead{[O\,{\sc ii}]$\lambda3726$}
& \colhead{[O\,{\sc ii}]$\lambda3729$}
& \colhead{[Ne\,{\sc iii}]$\lambda3869$}
& \colhead{He\,{\sc i}$\lambda 3889$}
& \colhead{[Ne\,{\sc iii}]$\lambda3967$}
& \colhead{H$\epsilon$}
& \colhead{H$\delta$}
& \colhead{H$\gamma$}
& \colhead{[O\,{\sc iii}]$\lambda4363$}
& \colhead{He\,{\sc ii}$\lambda4686$}
& \colhead{H$\beta$}
& \colhead{[O\,{\sc iii}]$\lambda4959$}
& \colhead{[O\,{\sc iii}]$\lambda5007$}
& \colhead{He\,{\sc i}$\lambda 5876$}
& \colhead{H$\alpha$}
& \colhead{[N\,{\sc ii}]$\lambda6583$}
& \colhead{[S\,{\sc ii}]$\lambda6716$}
& \colhead{[S\,{\sc ii}]$\lambda6731$}
& \colhead{He\,{\sc i}$\lambda 7065$}
} 
\startdata
HSC J0845+0131 & $15.04 \pm 1.55$ & $27.11 \pm 2.12$ & $19.74 \pm 2.32$ & $21.70 \pm 2.92$ & $9.36 \pm 1.44$ & $<4.96$ & $21.93 \pm 3.18$ & $44.26 \pm 2.89$ & $8.55 \pm 2.05$ & $<1.00$ & $100.00 \pm 2.41$ & $100.16 \pm 3.39$ & $294.30 \pm 3.60$ & $21.56 \pm 3.23$ & $267.19 \pm 11.39$ & $3.57 \pm 0.84$ & $3.99 \pm 0.64$ & $1.99 \pm 0.45$ & $2.80 \pm 0.57$ \\ 
HSC J0912-0104  & -- & -- & --& $<9.48$  & -- & --& $18.45 \pm 5.16$ & $41.25 \pm 7.57$  & -- & --& $100.00 \pm 5.17$ & $69.93 \pm 6.26$ & $195.11 \pm 5.42$ & $14.95 \pm 2.28$ & $266.59 \pm 7.92$  & -- & -- & -- & --\\ 
HSC J0935-0115 & $8.59 \pm 0.52$ & $12.41 \pm 0.64$ & $19.87 \pm 1.19$ & $17.60 \pm 0.77$ & $5.13 \pm 0.55$ & $13.71 \pm 2.15$ & $24.76 \pm 0.85$ & $44.52 \pm 1.60$ & $8.15 \pm 0.51$ & $2.44 \pm 0.35$ & $100.00 \pm 2.56$ & $83.37 \pm 3.09$ & $241.61 \pm 11.68$ & $8.65 \pm 0.52$ & $253.27 \pm 12.98$ & $0.70 \pm 0.12$ & $2.28 \pm 0.21$ & $1.03 \pm 0.14$ & $1.71 \pm 0.10$ \\ 
HSC J1210-0103  & -- & --& $<27.93$  & --& $<20.78$  & -- & --& $<32.78$  & -- & --& $100.00 \pm 15.83$ & $111.69 \pm 11.13$ & $276.62 \pm 17.05$  & --& $314.88 \pm 18.41$ & $<7.48$ & $20.52 \pm 6.31$ & $<4.06$  & --\\ 
HSC J1237-0016 & $17.56 \pm 2.13$ & $25.56 \pm 2.55$ & $51.67 \pm 4.23$ & $23.60 \pm 2.50$ & $19.72 \pm 2.15$ & $<6.72$ & $26.86 \pm 1.82$ & $48.79 \pm 1.45$ & $16.71 \pm 1.44$ & $<1.28$ & $100.00 \pm 2.12$ & $191.76 \pm 4.69$ & $528.68 \pm 15.56$ & $11.31 \pm 1.34$ & $284.04 \pm 10.61$  & --& $3.20 \pm 0.74$ & $3.09 \pm 0.71$ & $<1.11$ \\ 
HSC J1401-0040 & $34.96 \pm 1.35$ & $47.48 \pm 1.69$ & $31.22 \pm 1.24$ & $13.70 \pm 1.39$ & $10.68 \pm 0.68$ & $13.15 \pm 3.85$ & $26.82 \pm 1.60$ & $50.28 \pm 1.37$ & $12.38 \pm 1.17$ & $3.06 \pm 0.57$ & $100.00 \pm 1.77$ & $161.50 \pm 4.17$ & $474.83 \pm 14.80$ & $9.37 \pm 0.75$ & $287.26 \pm 7.34$ & $2.65 \pm 0.29$ & $8.89 \pm 0.39$ & $6.15 \pm 0.31$ & $4.50 \pm 0.87$ \\ 
HSC J1407-0047 & $35.23 \pm 5.16$ & $46.99 \pm 6.10$ & $23.46 \pm 3.53$ & $<6.31$ & $10.54 \pm 1.99$ & $<7.09$ & $29.00 \pm 5.18$ & $47.07 \pm 4.95$ & $<2.63$  & --& $100.00 \pm 3.52$ & $103.18 \pm 5.16$ & $295.95 \pm 9.76$ & $9.23 \pm 0.73$ & $278.69 \pm 7.17$  & --& $9.41 \pm 2.34$ & $10.27 \pm 2.45$  & --\\ 
HSC J1411-0032 & $35.55 \pm 1.65$ & $44.86 \pm 1.88$ & $52.63 \pm 2.36$ & $24.02 \pm 1.51$ & $21.34 \pm 1.22$ & $<9.04$ & $30.26 \pm 1.94$ & $53.09 \pm 1.84$ & $10.35 \pm 1.02$  & --& $100.00 \pm 2.95$ & $254.87 \pm 4.96$ & $747.01 \pm 17.88$ & $10.56 \pm 0.72$ & $300.00 \pm 5.57$ & $3.96 \pm 0.53$  & --& $5.27 \pm 0.45$  & --\\ 
HSC J1452+0241 & $6.46 \pm 2.03$ & $10.54 \pm 2.76$ & $18.32 \pm 3.76$ & $18.22 \pm 2.98$ & $5.22 \pm 1.74$ & $15.66 \pm 3.11$ & $30.22 \pm 2.00$ & $46.33 \pm 2.33$ & $7.82 \pm 1.53$ & $<1.34$ & $100.00 \pm 1.94$ & $80.68 \pm 2.60$ & $252.60 \pm 5.97$ & $13.19 \pm 1.48$ & $282.58 \pm 7.05$ & $<1.17$ & $1.63 \pm 0.46$ & $1.25 \pm 0.38$ & $2.82 \pm 0.78$ \\ 
SDSS J1044+0353 & $9.55 \pm 0.51$ & $12.34 \pm 0.56$ & $31.54 \pm 1.90$ & $17.50 \pm 0.55$ & $10.58 \pm 0.83$ & $13.66 \pm 3.86$ & $24.43 \pm 0.58$ & $47.83 \pm 0.76$ & $13.36 \pm 0.23$ & $1.80 \pm 0.13$ & $100.00 \pm 2.46$ & $145.96 \pm 4.84$ & $431.31 \pm 16.68$ & $8.64 \pm 0.29$ & $265.60 \pm 8.74$ & $0.82 \pm 0.08$ & $2.11 \pm 0.06$ & $1.60 \pm 0.05$ & $2.89 \pm 0.09$ \\ 
SDSS J1253-0312 & $39.46 \pm 0.86$ & $45.43 \pm 0.91$ & $50.52 \pm 1.17$ & $18.96 \pm 0.53$ & $15.26 \pm 0.54$ & $14.08 \pm 0.75$ & $25.67 \pm 0.82$ & $49.34 \pm 1.69$ & $11.15 \pm 0.44$ & $1.72 \pm 0.12$ & $100.00 \pm 3.46$ & $244.90 \pm 10.26$ & $756.67 \pm 44.04$ & $10.97 \pm 0.28$ & $273.73 \pm 18.77$ & $14.60 \pm 0.38$ & $7.23 \pm 0.27$ & $7.66 \pm 0.25$ & $3.65 \pm 0.12$ \\ 
SDSS J1323-0132 & $8.09 \pm 0.47$ & $9.53 \pm 0.52$ & $55.49 \pm 1.90$ & $19.94 \pm 0.56$ & $17.42 \pm 0.85$ & $<5.19$ & $26.65 \pm 0.82$ & $49.49 \pm 1.11$ & $22.11 \pm 0.61$ & $0.80 \pm 0.22$ & $100.00 \pm 2.23$ & $263.32 \pm 9.99$ & $777.85 \pm 38.28$ & $9.67 \pm 0.36$ & $288.28 \pm 11.34$ & $0.84 \pm 0.04$  & -- & --& $2.49 \pm 0.20$ \\ 
SDSS J1418+2102 & $19.72 \pm 0.63$ & $28.26 \pm 0.85$ & $39.40 \pm 1.05$ & $16.13 \pm 0.44$ & $12.27 \pm 0.51$ & $13.34 \pm 4.37$ & $26.91 \pm 0.48$ & $47.91 \pm 0.57$ & $14.91 \pm 0.43$ & $2.22 \pm 0.25$ & $100.00 \pm 1.09$ & $173.77 \pm 5.31$ & $555.68 \pm 9.16$ & $10.07 \pm 0.21$ & $283.65 \pm 6.16$ & $1.69 \pm 0.10$ & $4.60 \pm 0.11$ & $3.41 \pm 0.09$ & $3.21 \pm 0.16$ \\ 
\enddata
\tablecomments{%
Flux ratios relative to \Hb\ corrected for Galactic and the intrinsic dust extinction
and multiplied by $100$.
Upper-limit values at the $1\sigma$ level.
}
\end{splitdeluxetable*}

\section{Galaxy Samples} \label{sec:samples}

\subsection{Individual Metal Poor Galaxies} \label{ssec:samples_empgs}

This section introduces the EMPG sample. 
We first explain EMPGs selected by EMPRESS, and provide some new galaxies
recently identified by our own spectroscopic observation.
In addition, we compile metal-poor galaxies from the literature to build
the largest, up-to-date EMPG sample as detailed below.

\subsubsection{EMPRESS EMPGs: Earlier data} \label{sssec:samples_empgs_empress_earlier}

Early spectroscopic observations undertaken for the EMPRESS sample 
were described in \citet{kojima2020} and \citet{isobe2021_fe}.
As a pilot spectroscopic observation, \citet{kojima2020} presented data for 
$10$ EMPG candidates, three of which were selected from the HSC-SSP catalog 
and $7$ were from the SDSS. 
Using \OIII$\lambda 4363$ as an electron temperature probe, 
the EMPRESS team confirmed $2$ new EMPGs including HSC J1631+4426
showing the lowest metallicity ever reported. Of the other eight candidates, 
seven were also confirmed to be actively star-forming galaxies with emission lines 
as intense as the two EMPGs but slightly metal-enriched ($0.1$--$0.5$\,\Zsun).
Hereafter such moderately metal-poor galaxies are simply called ``MPGs'' for short.
The remaining single candidate was faint and not used in this paper due to a lack
of \OIII$\lambda 4363$ detection, although its strong emission lines indicated 
a very low metallicity. 
In \citet{isobe2021_fe}, another spectroscopic follow-up for $13$ EMPG candidates,
all of which were HSC-selected, was performed with Keck/LRIS. 
Ten were confirmed to be intense emission line galaxies, 
four of which were EMPGs and five were MPGs ($0.1$--$0.2$\,\Zsun).
The other object was suggested to be an EMPG but excluded in this paper's
compilation due to its large uncertainty of metallicity ($\Delta \log$(O/H) $>0.2$). 
Taken together, the $6$ EMPGs as well as the $12$ MPGs are added in the compilation 
from the early EMPRESS work.

\begin{figure*}[t]
    \begin{center}
     \includegraphics[bb=0 0 514 448, width=0.99\textwidth]{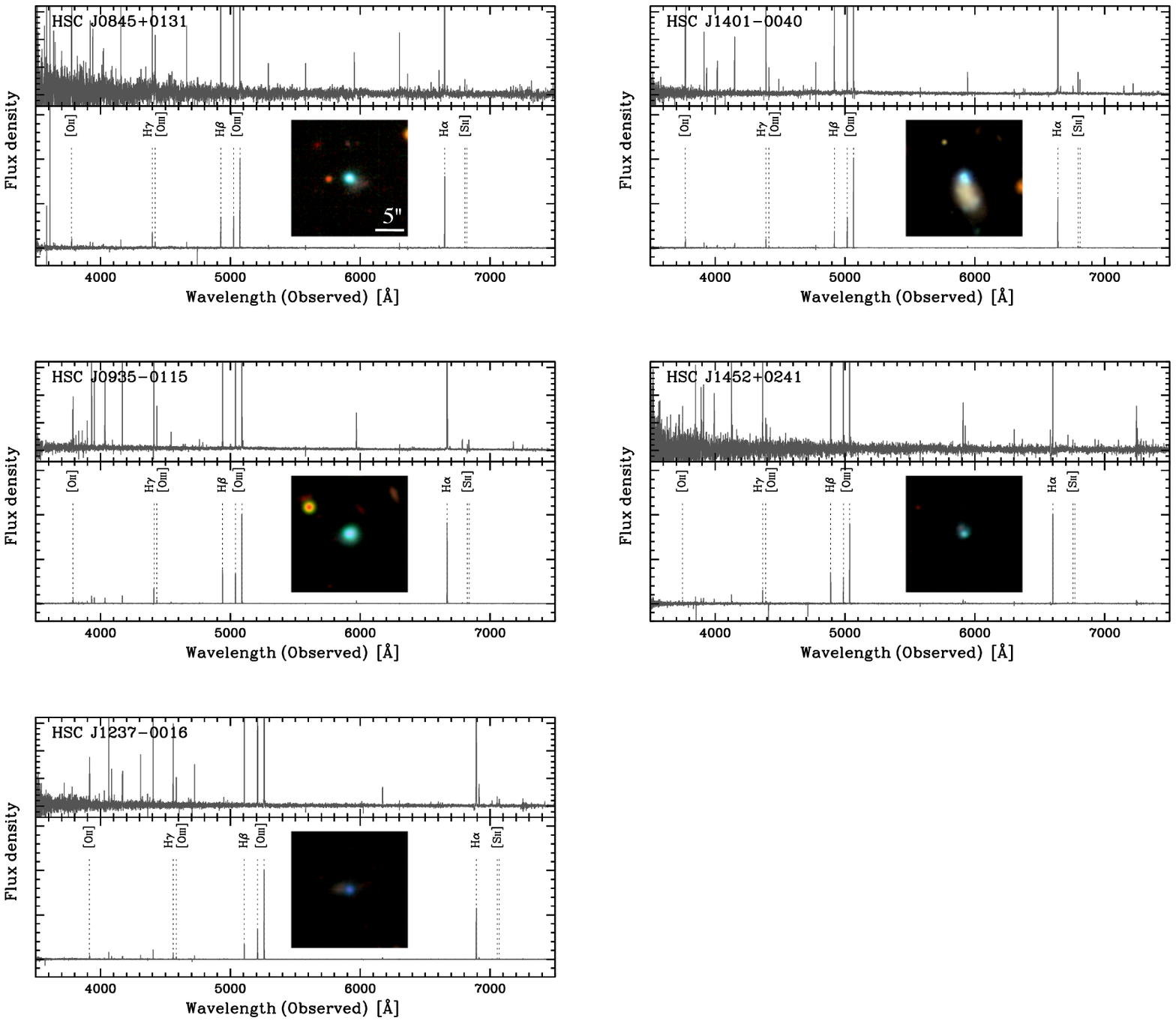}
    \caption{%
    	The newly identified $5$ EMPRESS EMPGs with their optical spectra and 
 		images. 
		For each object the bottom panel shows the MagE spectrum arbitrary normalized 
		based on the strong emission lines, while the top panels decreases the scale of the 
		vertical axis by a factor of $20$ to see the weaker lines.
		The inserted panel presents $20^{\prime\prime}\times20^{\prime\prime}$ 
		$gri$ composite image from HSC centered on the EMPG core. 
		North is up and east is to the left.
		}
    \label{fig:spectra_mage_empgs}
    \end{center} 
\end{figure*}

\subsubsection{EMPRESS EMPGs: New data} \label{sssec:samples_empress_new}

Following the success of the machine-learning selection of EMPGs by 
EMPRESS \citep{kojima2020}, we have expanded spectroscopic follow-up 
observations to increase the numbers of confirmed EMPGs and 
characterize their properties in a statistical manner.

The third spectroscopic observation was conducted with Magellan/MagE
on UT 2021 February 10 in clear conditions with a seeing of $0.5$--$0.9^{\prime\prime}$.
Ten EMPG candidates were selected for the follow-up from the HSC candidate
catalog. We chose 7 from the catalog originally provided by \citet{kojima2020}, 
and the remaining 3 from a newly provided, enlarged EMPRESS sample. 
The new sample was based on an upgraded machine learning classifier with further metal-poor 
EMPG training models (Z $=10^{-3}$--$0.01$\,\Zsun; based on photoionization models 
from \citealt{nakajima2018_vuds}) 
and the latest HSC-SSP DR (S20A; \citealt{aihara2018}). 
In addition to the HSC candidates, we observed 4 previously known (E)MPGs 
(\Oabundance\ $=7.4$--$8.0$; \citealt{kniazev2003,izotov2012_shock})
selected from SDSS, 
to examine the kinematics of low-mass galaxies \citep{xu2022}.
Full details of the observation and data reduction procedures are presented in 
a companion paper of \citep{xu2022}.
Briefly, we found nine out of the ten HSC candidates in which we identified multiple 
emission lines over the wavelength range of $\lambda_{\rm obs}=3500$--$10000$\,\AA\
that are suggestive of metal-poor galaxies at $z\lesssim 0.05$.
In this paper, we report the properties of the nine new galaxies of HSC 
as well as the four SDSS galaxies 
(hereafter, MagE sources).

Table \ref{tbl:fluxes_mage} lists the key line intensities normalized by \Hb\ and their 
$1\sigma$ errors for the $13$ MagE sources.
We measure the flux of each of the lines by fitting a Gaussian profile plus 
a constant continuum. 
The measured fluxes are then corrected for Galactic extinction based on the
\citet{SF2011}'s map as well as the extinction curve of \citet{cardelli1989}.
We use the Balmer lines of \Ha, \Hb, \Hg, and \Hd\ to estimate the dust attenuation
by simultaneously fitting the fluxes assuming the \citet{calzetti2000}' attenuation curve.
We do not correct for the potential stellar absorption around the Balmer lines 
which would be negligible 
($\sim 1$--$2$\,\AA\ in metal-poor galaxies; e.g., \citealt{izotov2012}) 
for our sources with very large EWs of emission lines
(EW(\Hb) $\sim 100$--$550$\,\AA).

\begin{deluxetable}{lcccc}[t]
\tablecaption{Physical properties of MagE sources
\label{tbl:properties_mage}}
\tablewidth{0.99\columnwidth}
\tabletypesize{\scriptsize}
\tablehead{
\colhead{ID}
& \colhead{EW(H$\beta$)}
& \colhead{$T_e$(O\,{\sc iii})}
& \colhead{$N_e$(O\,{\sc ii})}
& \colhead{12+log(O/H)}
\\
& (\AA)
& ($10^3\,$K)
& (cm$^{-3}$)
& 
}
\startdata
HSC J0845+0131 & $220.7 \pm 30.0$ & $18.4 \pm 2.5$ & -- & $7.35 \pm 0.13$ \\ 
HSC J0912-0104 & $142.9 \pm 32.4$ & -- & -- & -- \\ 
HSC J0935-0115 & $266.5 \pm 9.7$ & $20.1 \pm 1.0$ & $30^{+100}_{-30}$ & $7.17 \pm 0.07$ \\ 
HSC J1210-0103 & $180.7 \pm 18.1$$^{(\dag)}$ & -- & -- & -- \\ 
HSC J1237-0016 & $225.2 \pm 19.9$ & $19.3 \pm 1.0$ & $20^{+220}_{-20}$ & $7.55 \pm 0.08$ \\ 
HSC J1401-0040 & $133.5 \pm 4.7$ & $17.3 \pm 0.9$ & $110^{+70}_{-60}$ & $7.64 \pm 0.06$ \\ 
HSC J1407-0047 & $98.9 \pm 9.0$ & -- & $130^{+350}_{-130}$ & -- \\ 
HSC J1411-0032 & $288.4 \pm 24.7$ & $13.0 \pm 0.5$ & $200^{+100}_{-80}$ & $8.12 \pm 0.06$ \\ 
HSC J1452+0241 & $550.8 \pm 6.4$$^{(\dag)}$ & $19.1 \pm 2.2$ & $<710$ & $7.21 \pm 0.12$ \\ 
SDSS J1044+0353 & $307.4 \pm 8.8$ & $19.1 \pm 0.5$ & $170^{+110}_{-90}$ & $7.45 \pm 0.04$ \\ 
SDSS J1253-0312 & $226.8 \pm 8.0$ & $13.3 \pm 0.4$ & $350^{+50}_{-50}$ & $8.09 \pm 0.06$ \\ 
SDSS J1323-0132 & $227.2 \pm 6.2$ & $18.1 \pm 0.6$ & $310^{+150}_{-120}$ & $7.74 \pm 0.05$ \\ 
SDSS J1418+2102 & $269.7 \pm 5.2$ & $17.6 \pm 0.3$ & $40^{+50}_{-40}$ & $7.64 \pm 0.04$ \\ 
\enddata
\tablecomments{%
EW(\Hb) are given at the rest-frame value. 
EWs are determined based on the spectra except for those with ($\dag$).
($\dag$) Because their continuums are not detected in the spectra, EW(\Hb) are 
estimated from the photometrically inferred EW(\Ha) divided by the typical
EW ratio of $5.47$ (see \S\ref{ssec:UV_properties_xiion}).
Metallicities are based on the direct $T_e$ method and given for those with 
a \OIII$\lambda 4363$ detection.
Upper-limit of $N_e$ is given at the $1\sigma$ level. 
For HSC J0845+0131, $N_e$ is constrained to be $<130$\,cm$^{-3}$ 
at the $2\sigma$ level.
}
\end{deluxetable}

One of the key properties to characterize our sources is gas-phase metallicity. 
We identify the temperature-sensitive line of \OIII$\lambda 4363$ in 
$10$ of the $13$ sources.
For the $10$ objects, $6$ of which are HSC-selected, we determine the oxygen abundance
using the direct $T_e$ method and use it as a proxy for gas-phase metallicity.
We first estimate electron number densities and electron temperatures using 
\textsf{getTemDen}, a package of a Python tool PyNeb \citep{luridiana2015}.
We use \OII$\lambda\lambda 3726/3729$ doublet ratio to estimate the density
of O$^{+}$ ions ($N_e$(O\,{\sc ii})), and assume a homogeneous density in the entire gas. 
The density varies from $20$ to $350$\,cm$^{-3}$ for the nine objects whose
\OII\ doublet are significantly detected. 
The \OIII$\lambda\lambda4363/5007$ ratio is then input together
with the measured density into \textsf{getTemDen} to derive the electron temperature
of O$^{2+}$ zone ($T_e$(O\,{\sc iii})). 
For the 2 sources without a direct measurement of density, 
we assume a typical value of $100$\,cm$^{-3}$, 
consistent with the upper-limits of $N_e$ for both of the sources.
We have also confirmed the density assumption has little impact on the derived 
metallicity values ($\Delta log$(O/H) $<0.01$\,dex) 
by changing the assumed density from $1$ to $1000$\,cm${-3}$.
The temperature of O$^{+}$ zone, $T_e$(O\,{\sc ii}), is extrapolated from $T_e$(O\,{\sc iii}) 
employing the prescription of \citet{izotov2006}.
We derive the abundances of O$^{+}$$/$H$^{+}$ with the \OII$\lambda\lambda 3726,3729$ 
(\OII$\lambda 3727$ hereafter as a sum of the doublet) to \Hb\  
and $T_e$(O\,{\sc ii}), and O$^{2+}$$/$H$^{+}$ with the \OIII$\lambda\lambda 4959,5007$ to \Hb\
and $T_e$(O\,{\sc iii}) using the PyNeb package \textsf{getIonAbundance}. 
For a possible higher ionization abundance of O$^{3+}$$/$H$^{+}$, we follow the 
approximation given by \citet{izotov2006}. Specifically, we add the O$^{3+}$ component
in the calculation if the \HeII$\lambda 4686$ emission line is detected. The contribution is
small for the strongest case (O$^{3+}$$/$(O$^{+}$+O$^{2+}$) $<0.012$) 
which is consistent with the previous EMPG studies. 
The measured oxygen abundances as well as $T_e$(O\,{\sc iii}) and $N_e$(O\,{\sc ii})
are summarized in Table \ref{tbl:properties_mage}.
Five of the 6 HSC-selected EMPG candidates have a metallicity below the $10$\,\%\ 
solar value (\Oabundance\ $<7.69$) and are thus confirmed being EMPGs.
The spectra and optical images for the newly identified EMPRESS-EMPGs are 
illustrated in Figure \ref{fig:spectra_mage_empgs}.
One source, J1411-0032, is turned out to be slightly metal-enriched system
as similarly identified in earlier studies of EMPRESS 
(\citealt{kojima2020,isobe2021_fe}), and added to the MPG catalog.
For the other three HSC-selected EMPG candidates, the non-detection of 
\OIII$\lambda 4363$ is still consistent with a metal-poor 
(i.e., high electron temperature) gas due to the rather faint nature. 
Indeed, their metallicities are indirectly inferred to be as low as those of EMPGs 
(\Oabundance\ $=7.1$--$7.4$)
based on the empirical metallicity indicators of 
R23-index 
and the upper-limit of N2-index 
(optimized for sources with EW(\Hb)=$100$--$200$\,\AA; see \S\ref{ssec:diagnostics_scatters}).
Future deep spectroscopy will reveal the properties in more details for the 
unconfirmed candidates.

\begin{figure*}[t]
    \begin{flushleft}
        \subfloat{
            \includegraphics[bb=40 163 567 617, width=0.49\textwidth]{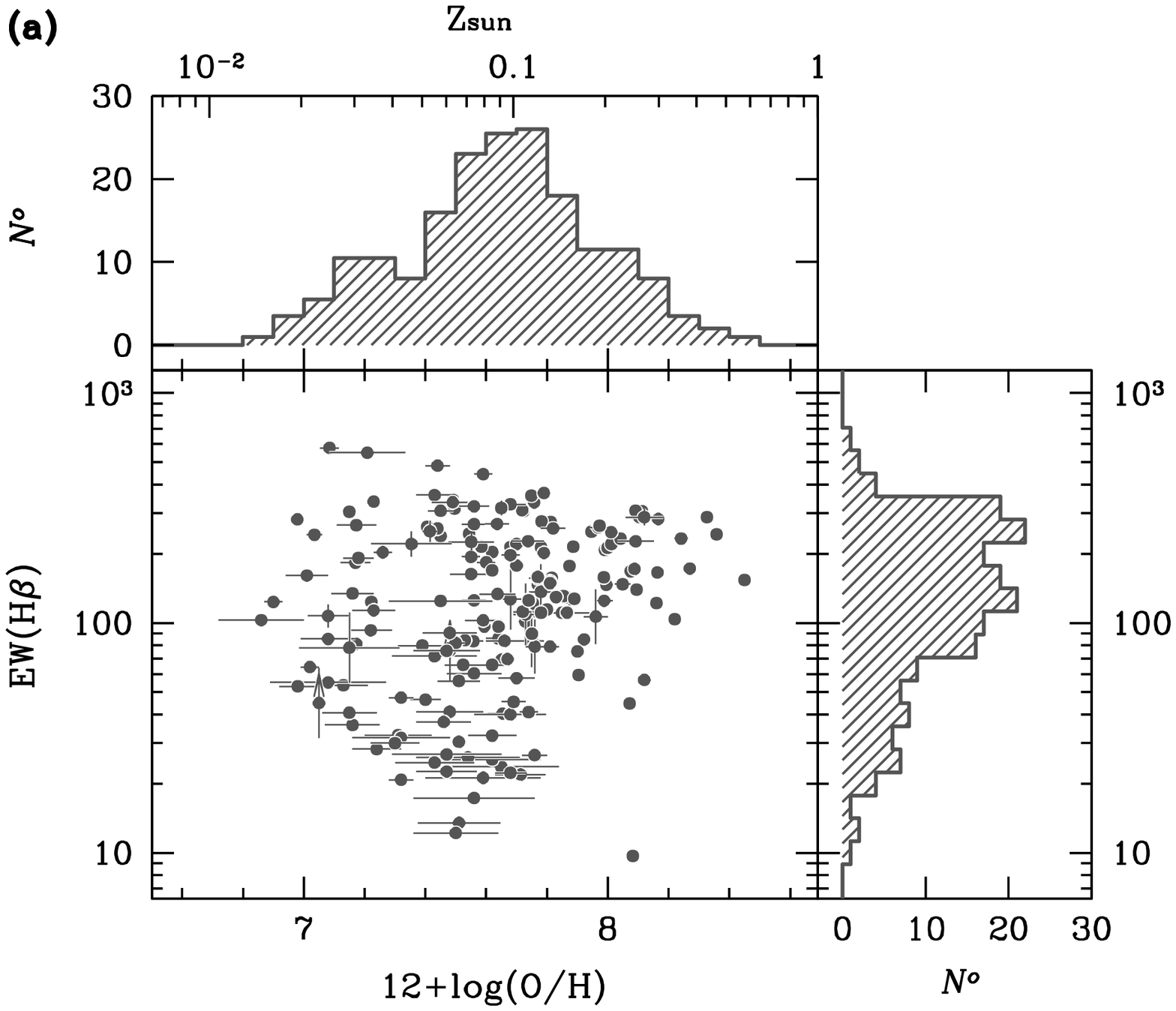}
        }
        \subfloat{
            \includegraphics[bb=-60 7 323 299, width=0.36\textwidth]{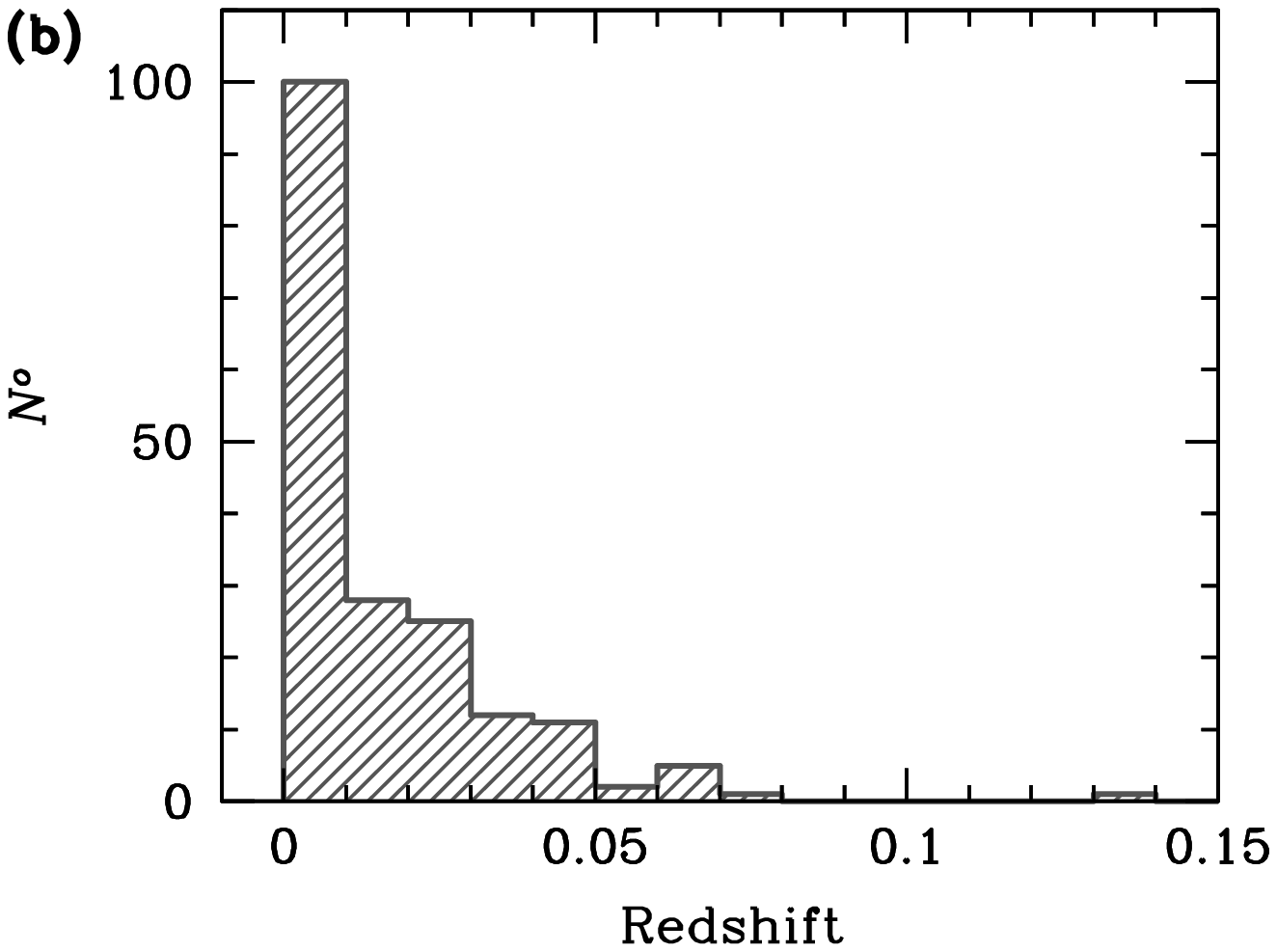}
        }
        
        \subfloat{
            \includegraphics[bb=40 163 567 495, width=0.49\textwidth]{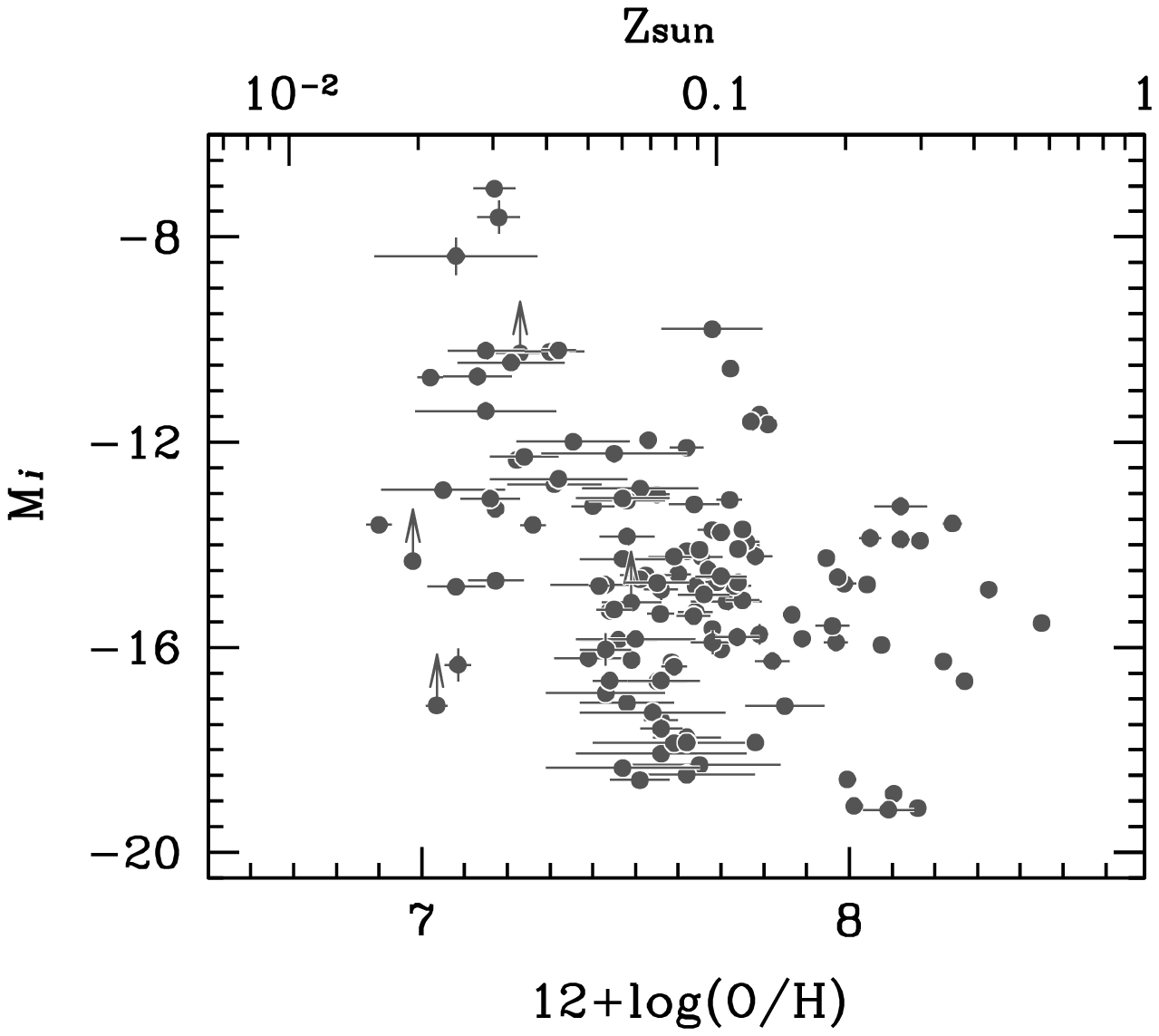}
        }
        \subfloat{
            \includegraphics[bb=40 163 567 495, width=0.49\textwidth]{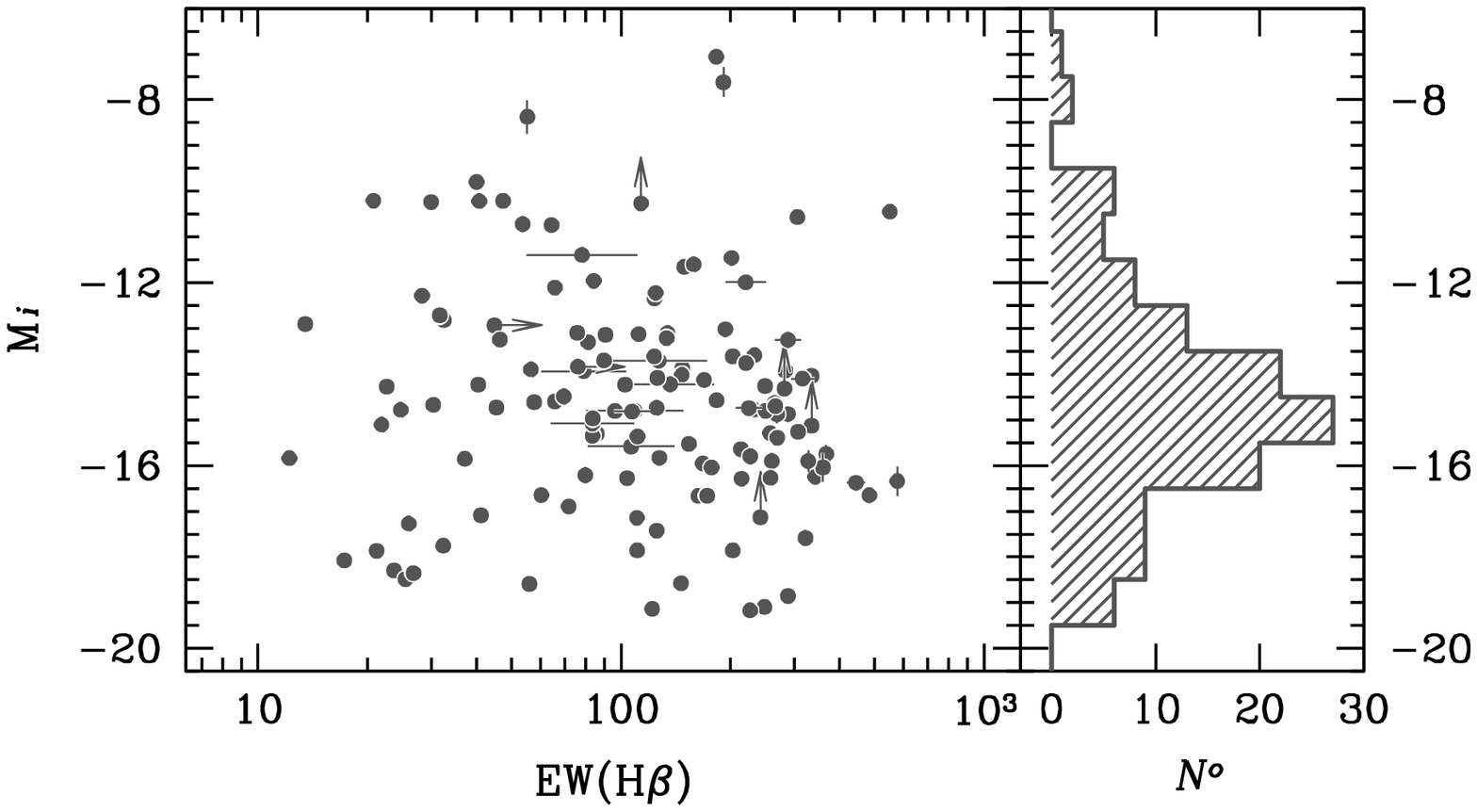}
        }
        \caption{%
        	(a) Scatter plot matrix showing the distributions of \Oabundance, EW(\Hb), and \Mi\
		for the compiled metal-poor galaxies (both EMPGs and MPGs) in this paper.
		(b) Redshift distribution of the compiled (E)MPGs.
		We compile EMPGs (i.e., \Oabundance\ $<7.69$) as complete
		as possible in the local universe ($95$\,\%\ are at $z<0.05$), 
		while include the MPGs only when found in the EMPG literature.
        }
        \label{fig:dist_Mi_ewhb_Z_redshift}
    \end{flushleft} 
\end{figure*}


In addition, a further spectroscopic follow-up has been conducted under EMPRESS
and will be detailed in a series of forthcoming papers (e.g., Nishigaki et al. 2022 in prep.). 
In this paper, we additionally exploit the results of $3$ new EMPGs from the latest 
observations.

In summary, there are 8 EMPGs and one MPG newly identified in the EMPRESS sample. 
These 9 EMPRESS galaxies as well as the 
4 SDSS galaxies (2 EMPGs and 2 MPGs)
whose properties are re-measured with the MagE spectra are added to our sample 
and used for the following analysis.
We do not use the remaining 3 faint HSC galaxies in this paper, 
although their strong emission lines indicate they are as metal-poor as EMPGs. 
This spectroscopic observation thus re-confirms the ability of EMPRESS 
to select EMPGs in the local universe from the photometric catalog.

\subsubsection{Other EMPGs} \label{sssec:samples_empgs_others}

In addition to the (E)MPGs provided by EMPRESS,
we compile observations of comparably low-metallicity objects 
from the literature 
that have a firm determination of gas-phase metallicity based on \OIII$\lambda 4363$.
We find $154$ objects in total, without any duplication, in the literature
\citep{kniazev2003,TI2005,pustilnik2005,izotov2006_two,izotov2009,izotov2012,izotov2012_shock,izotov2018_lowZ,izotov2019_lowZspec,izotov2020,izotov2021,IT2007,guseva2007,pustilnik2010,skillman2013,hirschauer2016,sanchez-almeida2016,hsyu2017}
which include isolated galaxies such as emission line galaxies and blue compact dwarfs
as well as metal-poor clumps and \HII-regions in a nearby galaxy.
Only the sources with a reliable determination of oxygen abundance have been collected,
with the direct $T_e$ method using a suite of the optical oxygen and hydrogen 
emission lines as for the EMPRESS sources (\S \ref{sssec:samples_empress_new}). 
Some nearby objects lack \OII$\lambda 3727$ especially in the SDSS spectra.
For these objects, the O$^{+}$$/$H$^{+}$ abundance has been derived with the 
\OII$\lambda\lambda7319,7330$.
Sources with a large uncertainty of metallicity, $\Delta$ log(O/H) $>0.2$, have been removed
from the compilation.
For those without line flux/EW values reported in the literature 
\citep{kniazev2003,sanchez-almeida2016}, 
we have used the SDSS Data Release 16 SkyServer%
\footnote{
\url{http://skyserver.sdss.org/dr16}
}
to retrieve the spectroscopic data. 
Among the $154$ compiled objects, 
$87$ sources present a metallicity below \Oabundance\ $=7.69$
and are classified as EMPGs.
The remaining, relatively low-metallicity objects (\Oabundance\ $=7.7$--$8.4$) are MPGs
that are supplementarily provided from the same literature as used for 
the EMPG compilation.
The MPG subsample is therefore not complete in terms of metallicity but just served 
as a reference sample.
\\

In summary, our sample contains $103$ EMPGs and $82$ MPGs 
($N^o = 185$ in total),
increasing the sample size of EMPGs by a factor of three as compared 
to that of previous work (e.g., \citealt{nagao2006_metallicity}).
We note that no obvious active galactic nuclei (AGN) activity is identified 
in the compiled (E)MPG sample based on the BPT diagram
\citep{kauffmann2003}. 
Although we may not fully exclude the possibility of metal-poor AGN in the sample 
with the BPT diagram according to the photoionization predictions
\citep{kewley2013_theory},
we assume no AGN in the sample in the following analysis.
Figure \ref{fig:dist_Mi_ewhb_Z_redshift} (a) presents the distributions of metallicity, EW(\Hb), 
and $i$-band absolute magnitude \Mi\ 
for the compiled sources.
We derive \Mi\ based on the $i$- or $z$-band broadband photometric data of HSC or SDSS,
in addition to the luminosity distance. 
We choose the broadband which is free from an intense \Ha\ emission for each of the objects
(see \S\ref{ssec:UV_properties_xiion} for more details).
The EMPGs cover large parameter spaces of \Mi\ from $-19$ to $-7$
and EW(\Hb) from $10$ to $600$\,\AA, 
i.e., wide ranges of stellar mass and specific star-formation rate, respectively.
Assuming a mass to optical luminosity relation typically seen in (E)MPGs 
\citep{kojima2020, xu2022}, 
the \citet{kennicutt1998} relation, and a \citet{chabrier2003} initial mass function,
the ranges correspond to the stellar masses of $10^{8}$\,\Msun\ down to below $10^4$\,\Msun,
and to the specific star-formation rates of a few Gyr$^{-1}$ up to $\sim 200$\,Gyr$^{-1}$.
Notably, the EMPRESS survey enlarges the sample of the largest EW(\Hb) (E)MPGs
(e.g., see EW(\Hb) of the HSC-selected objects in Table \ref{tbl:properties_mage}).
Figure \ref{fig:dist_Mi_ewhb_Z_redshift} (b) shows the redshift distribution of the sample
confirming that most of the (E)MPGs ($95$\.\%) are found at $z<0.05$.
Their UV properties are also derived and discussed later in this paper 
(\S\ref{sec:UV_properties}).

\begin{figure*}[t]
    \begin{center}
        \subfloat{
            \includegraphics[bb=23 161 522 698, width=0.40\textwidth]{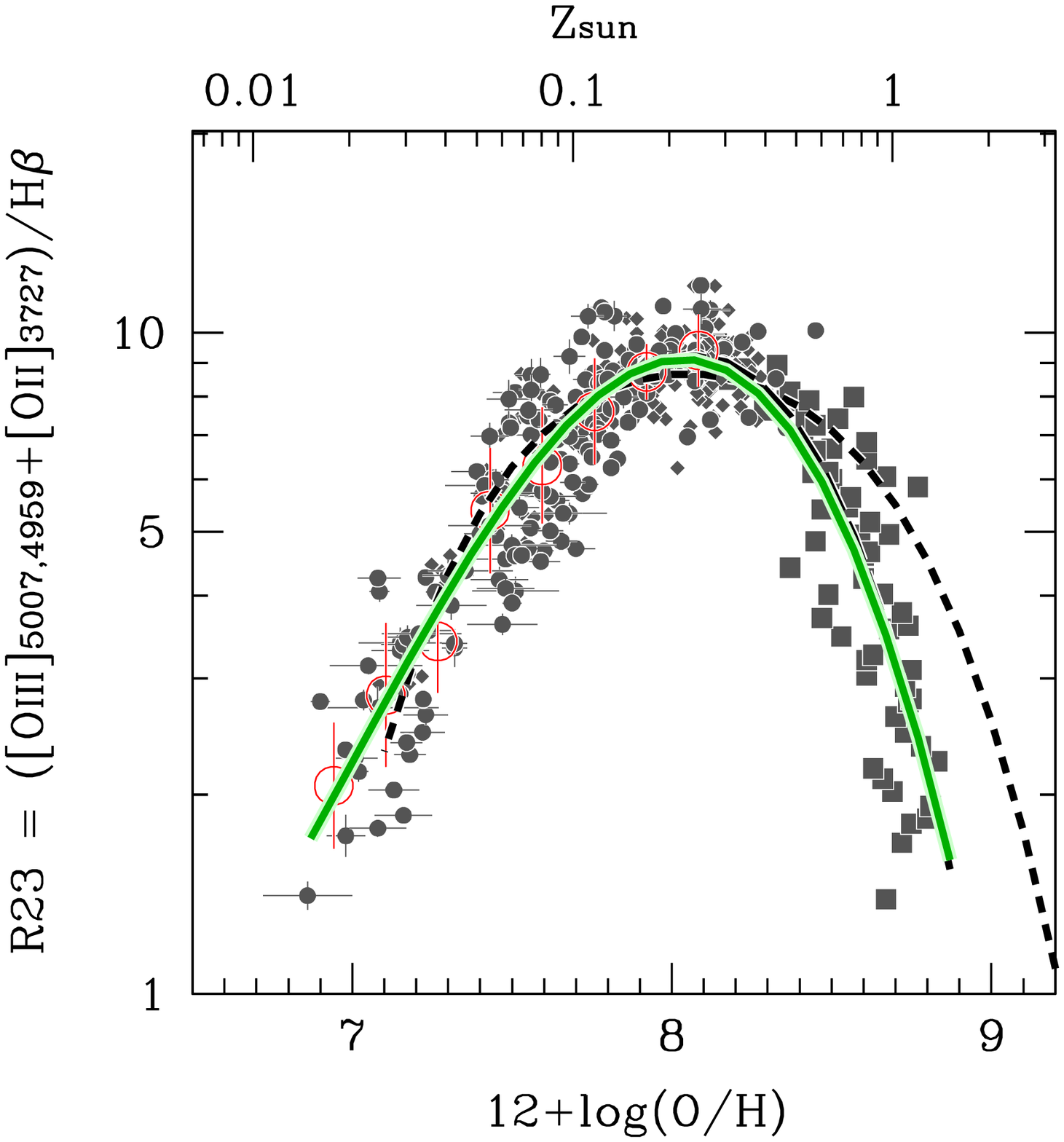}
        }
        \hspace{0.05\textwidth}
        \subfloat{
            \includegraphics[bb=23 161 522 698, width=0.40\textwidth]{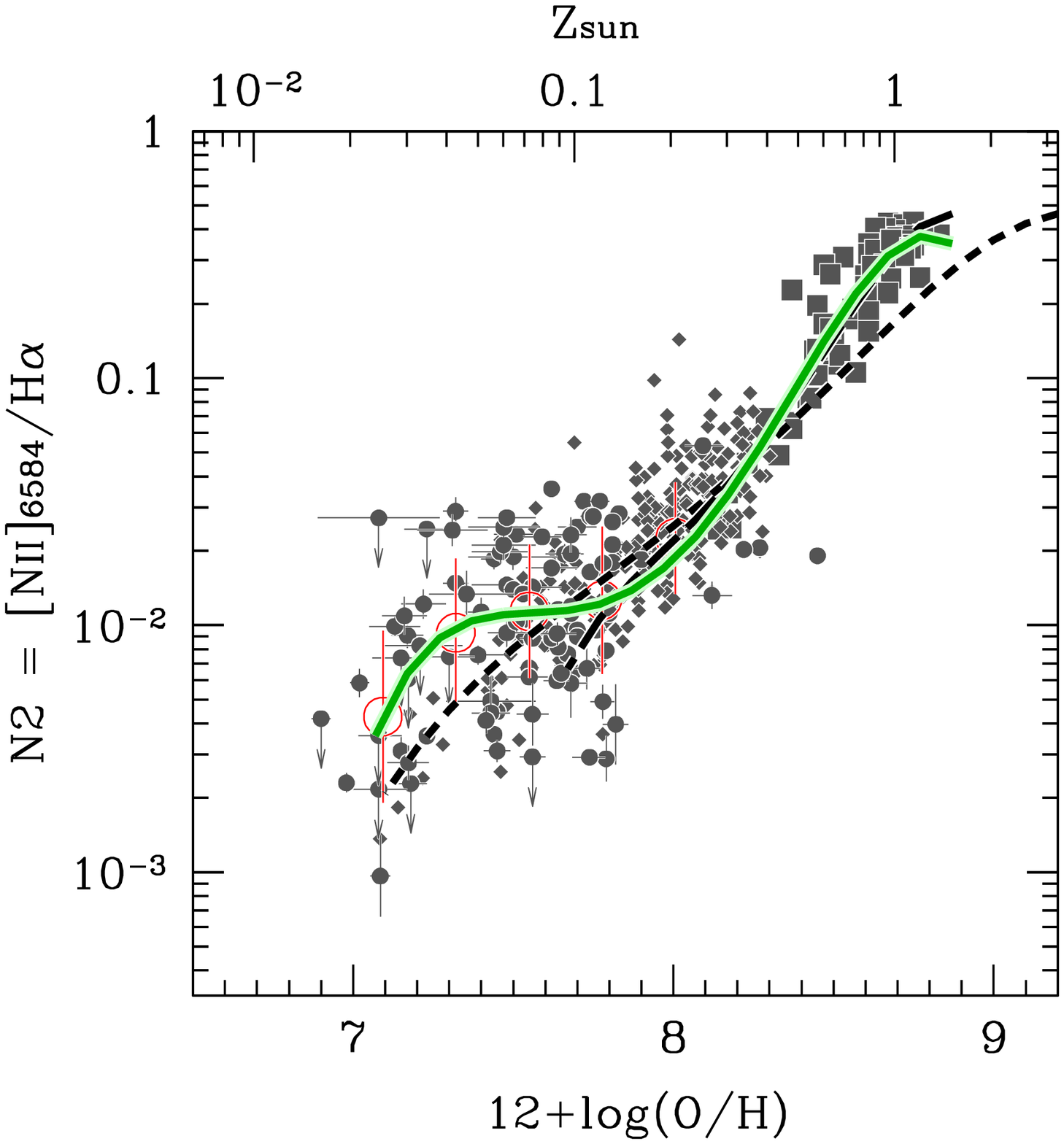}
        }
        
        \caption{%
		Relationship between metallicity and the strong line ratios of 
		R23-index (left) and N2-index (right).
		Our compiled (E)MPGs are shown with gray filled circles (individual) and 
		red open circles (binned average, with the standard deviation 
		of the binned distribution as the error).
		Gray filled squares denote stacked SDSS spectra in the high-metallicity regime
		\citep{curti2017,curti2020}.
		Small gray diamonds are metal-poor galaxies compiled by
		\citet{nagao2006_metallicity}
		just for a reference (i.e., not used for the following best-fit function).
		All the measurements of metallicity are done in a consistent manner 
		based on the direct $T_e$ method.
		Green curve presents our best-fit function, 
		whereas the black short-dashed and long-dashed curve illustrates 
		the function of \citet{maiolino2008} and \citet{curti2017,curti2020},
		respectively.
        }
        \label{fig:Z_empirical_all1}
    \end{center} 
\end{figure*}


\subsection{Stacked SDSS Galaxies} \label{ssec:samples_sdss}

To compensate the high-metallicity range in determining the metallicity diagnostics 
(\S\ref{sec:diagnostics}),
we exploit the analyses of \citet{curti2017,curti2020}
which are based on $120,000$ galaxies' spectra from SDSS.
Sources that are dominated by an AGN activity are excluded from the sample
based on the BPT diagram \citep{kauffmann2003}.
The authors stack SDSS spectra in bins of strong line ratios
to detect the \OIII$\lambda 4363$ in a statistical manner
and determine the metallicity based on the direct $T_{\rm e}$ method.
Each bin contains $10 - \sim 6000$ galaxies for the stacking analysis.
The resulting stacked spectra show metallicities ranging from \Oabundance $=8.1$
to $\sim 8.9$.

\begin{figure*}[t]
    \begin{center}
        \subfloat{
            \includegraphics[bb=23 161 522 698, width=0.3\textwidth]{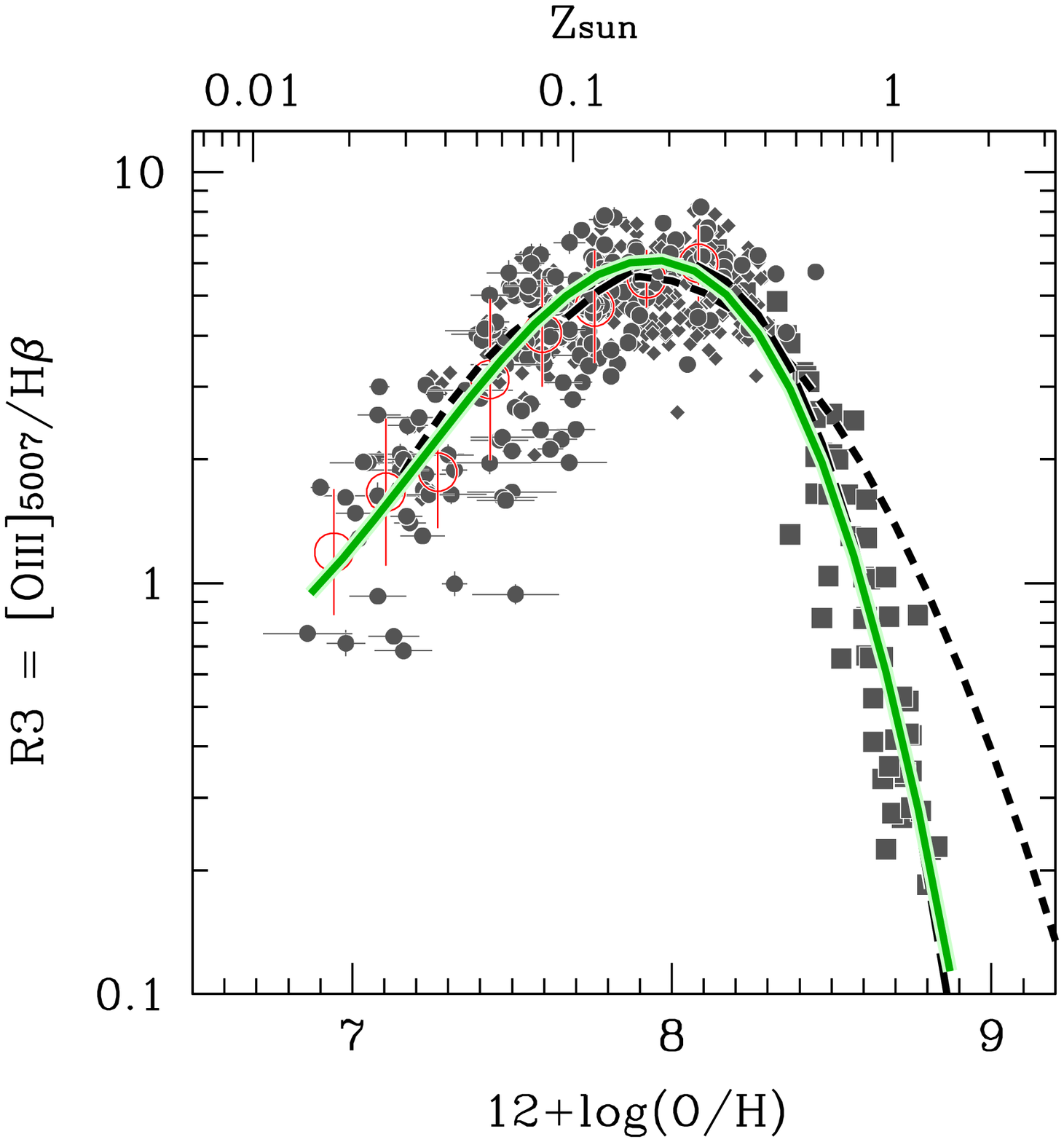}
        }
        \subfloat{
            \includegraphics[bb=23 161 522 698, width=0.3\textwidth]{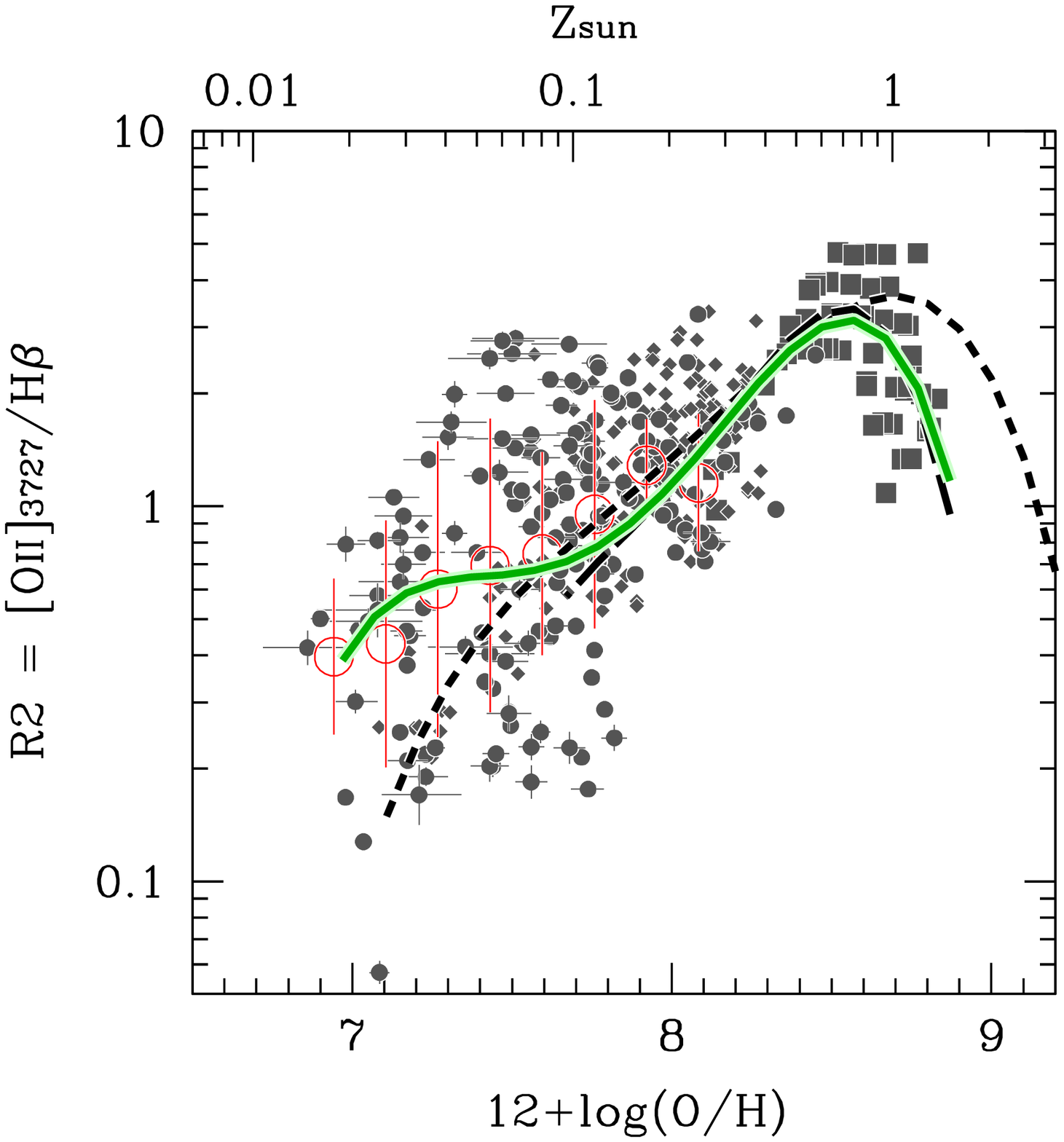}
        }
        \subfloat{
            \includegraphics[bb=23 161 522 698, width=0.3\textwidth]{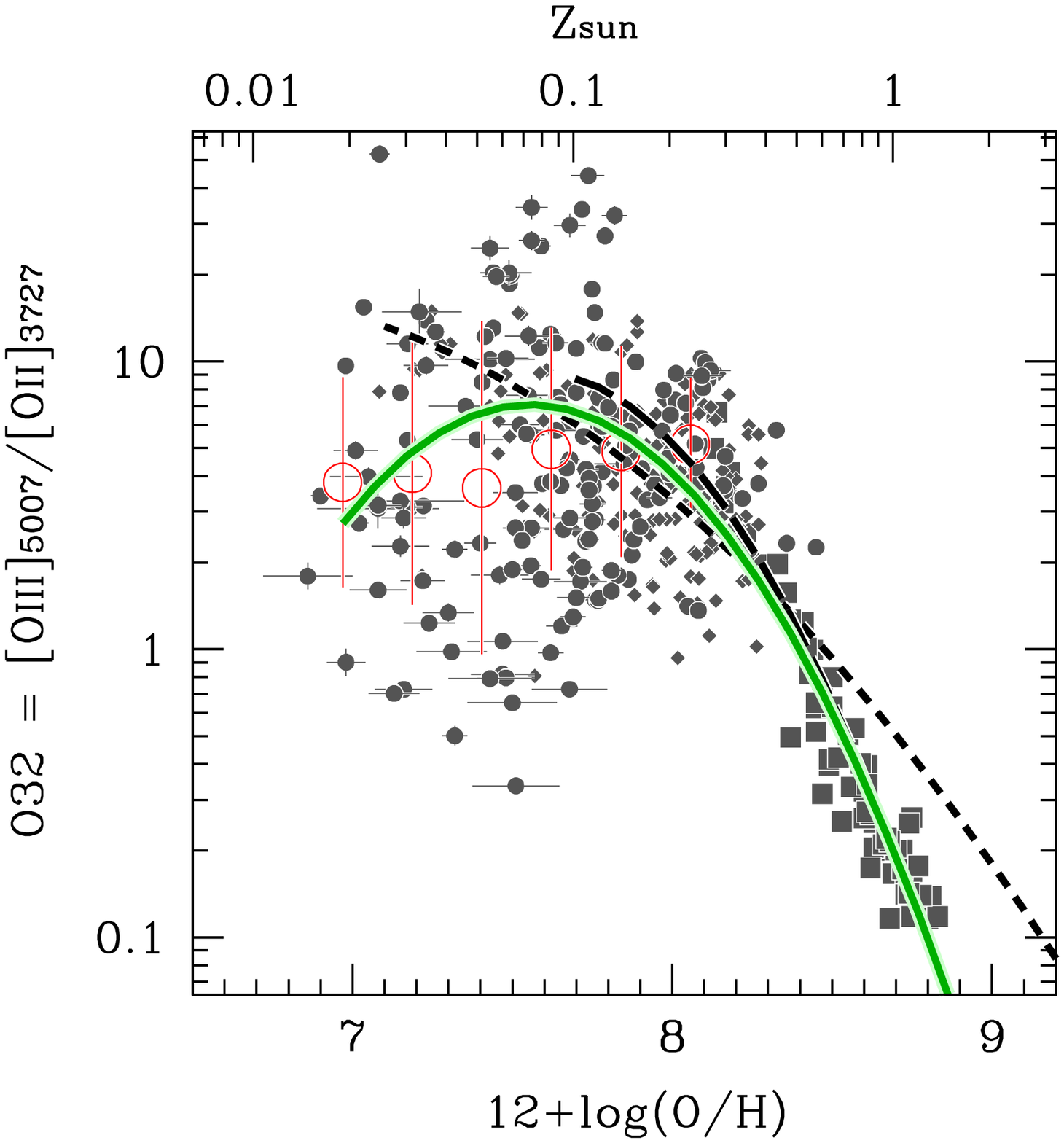}
        }
        
        \subfloat{
            \includegraphics[bb=23 161 522 698, width=0.3\textwidth]{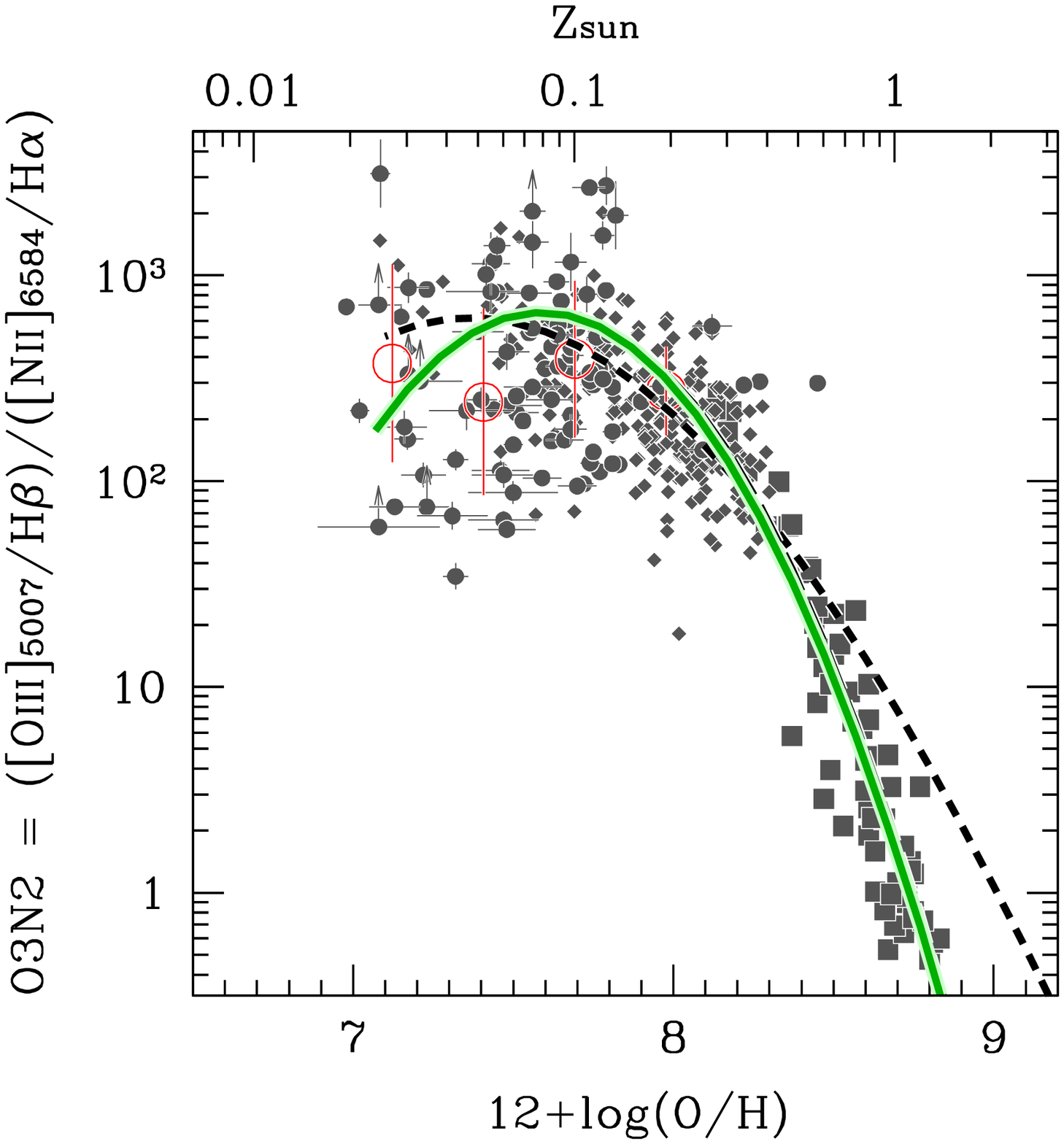}
        }
        \subfloat{
            \includegraphics[bb=23 161 522 698, width=0.3\textwidth]{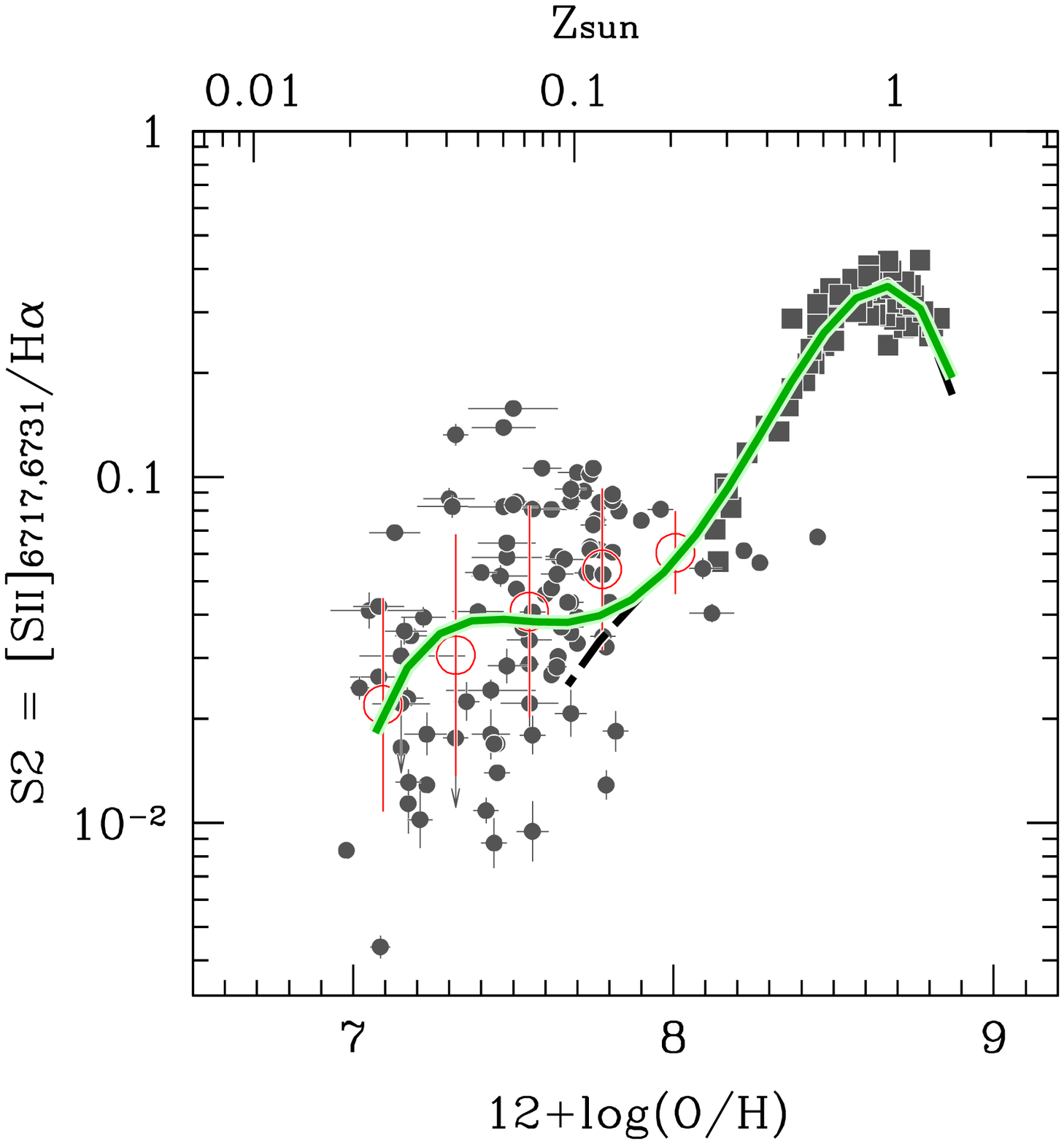}
        }
        \subfloat{
            \includegraphics[bb=23 161 522 698, width=0.3\textwidth]{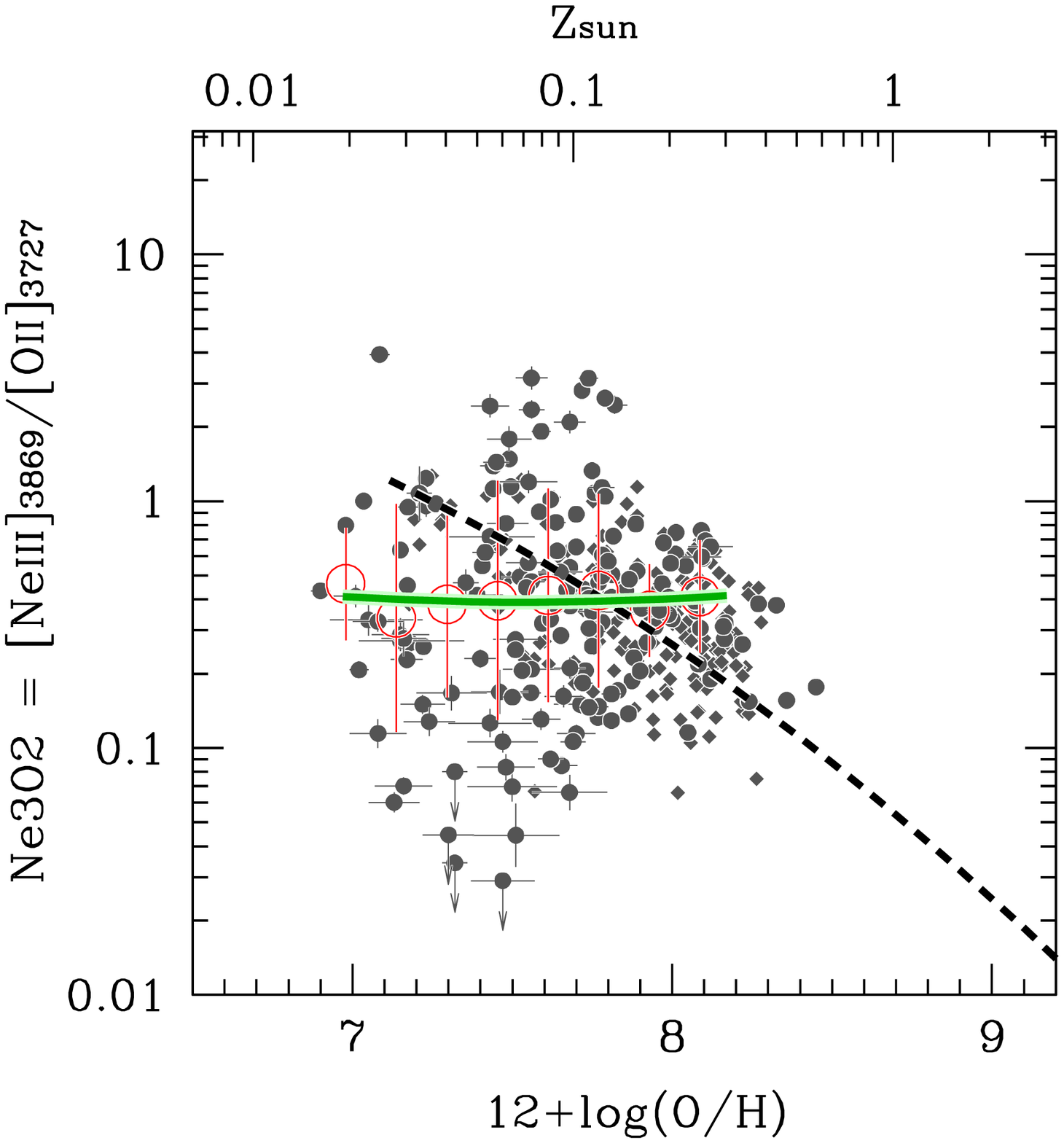}
        }
        \caption{%
        	Same as Figure \ref{fig:Z_empirical_all1} but for the line ratios of 
		R3, R2, O32, O3N2, S2, and Ne3O2 (from top left to bottom right).
        }
        \label{fig:Z_empirical_all2}
    \end{center} 
\end{figure*}


\section{Strong line diagnostics of metallicity} 
\label{sec:diagnostics}

\begin{figure*}[t]
    \begin{center}
        \subfloat{
            \includegraphics[bb=23 161 522 698, width=0.40\textwidth]{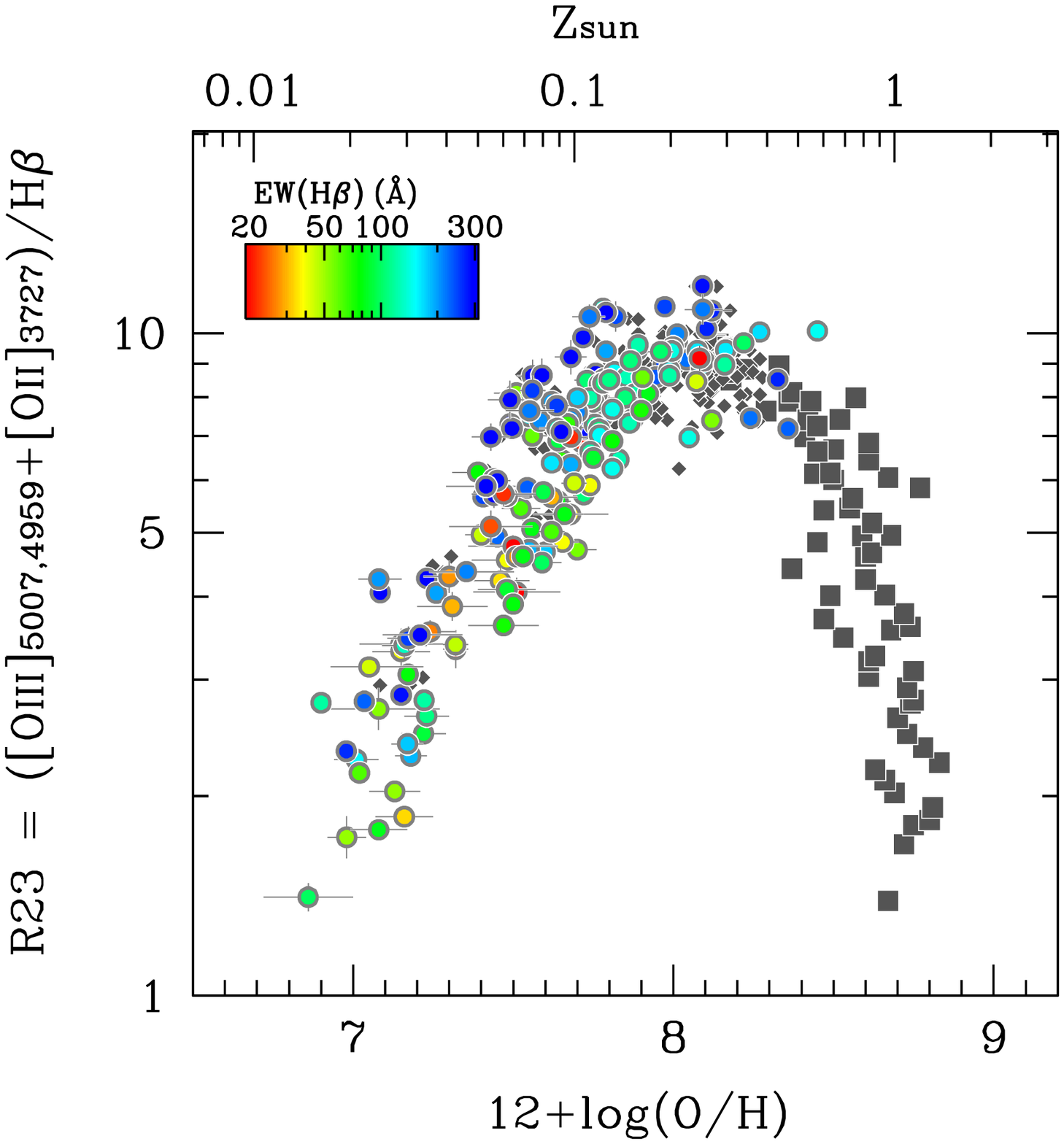}
        }
        \hspace{0.05\textwidth}
        \subfloat{
            \includegraphics[bb=23 161 522 698, width=0.40\textwidth]{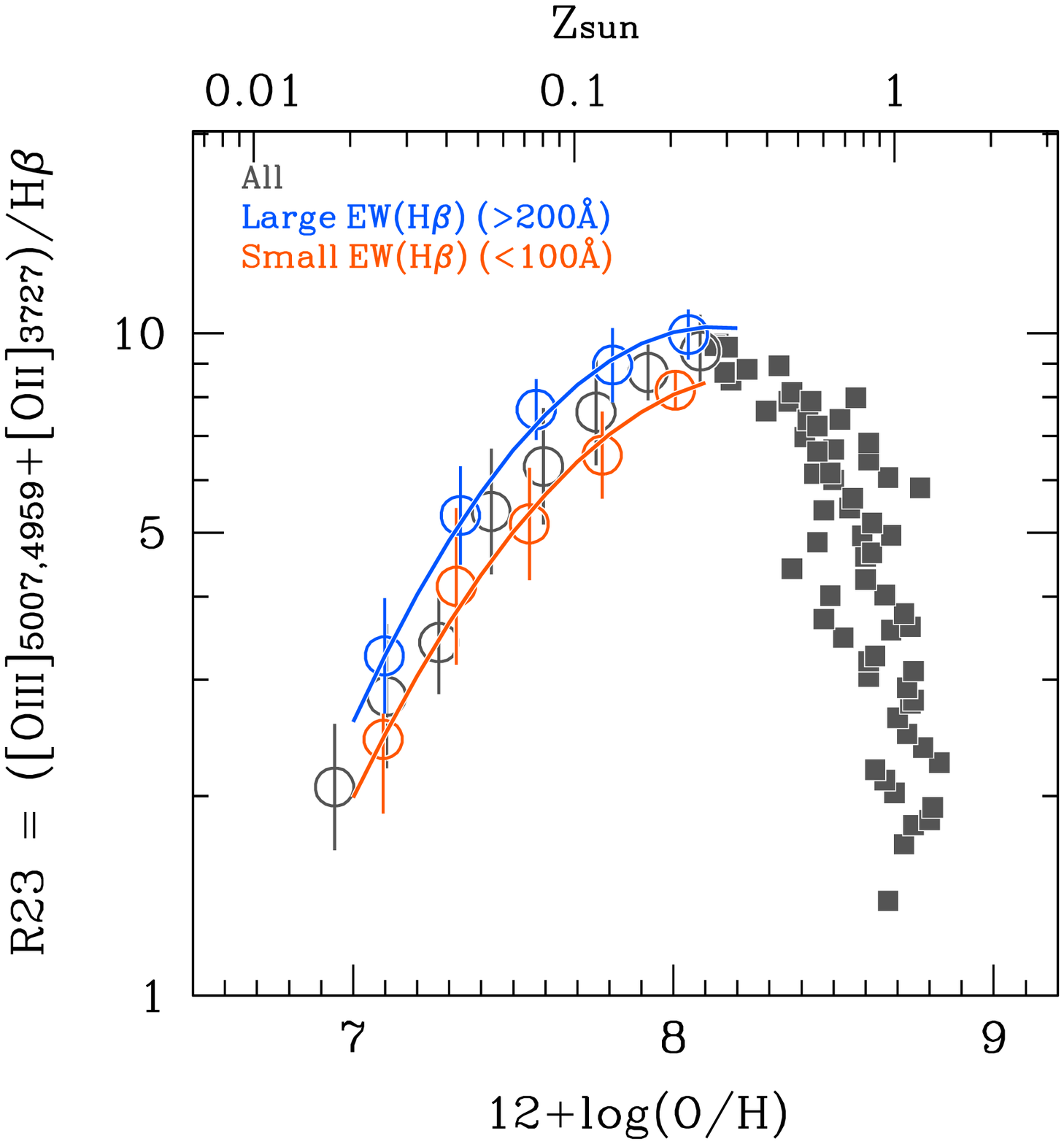}
        }
        
        \subfloat{
            \includegraphics[bb=23 161 522 698, width=0.40\textwidth]{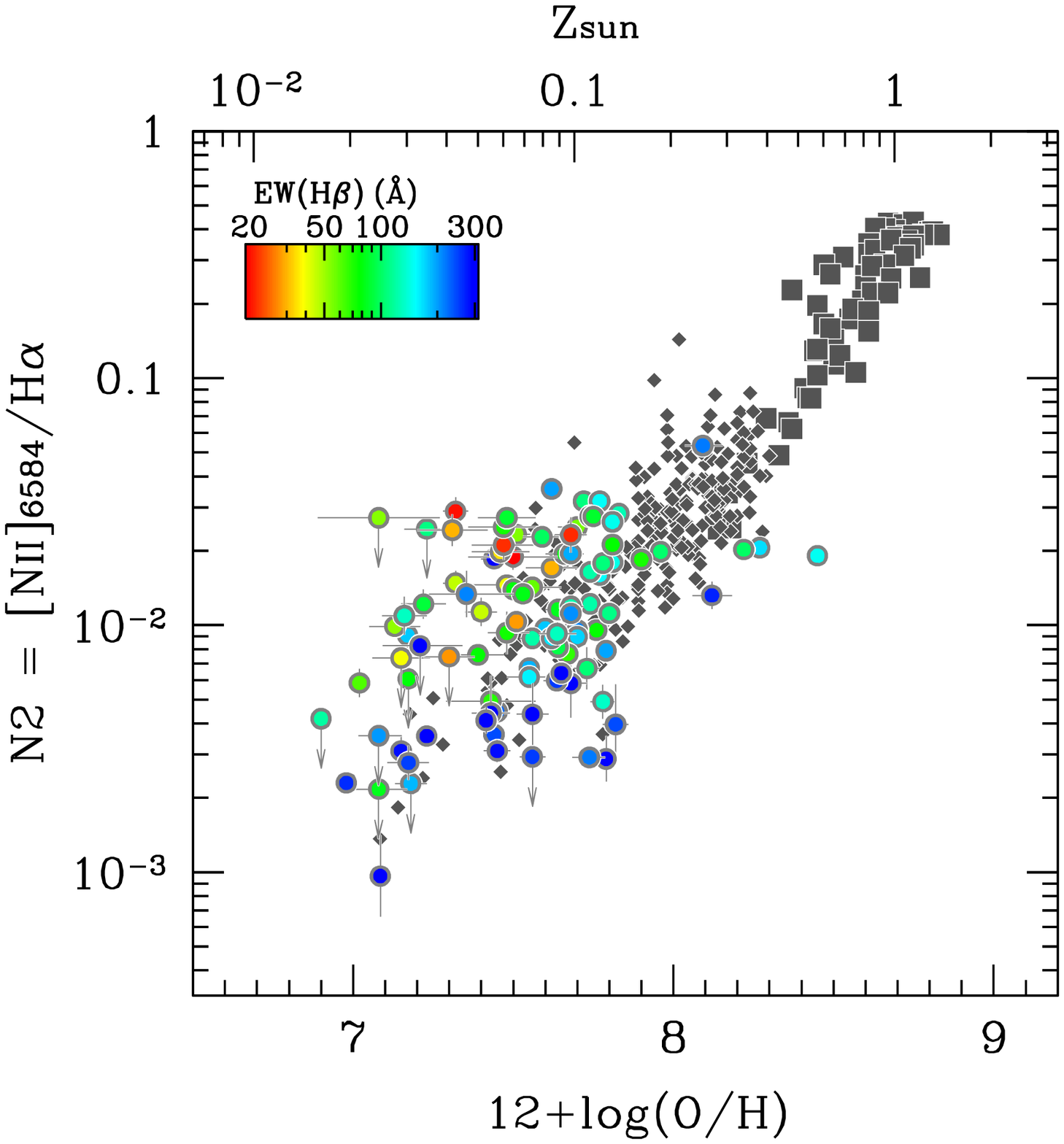}
        }
        \hspace{0.05\textwidth}
        \subfloat{
            \includegraphics[bb=23 161 522 698, width=0.40\textwidth]{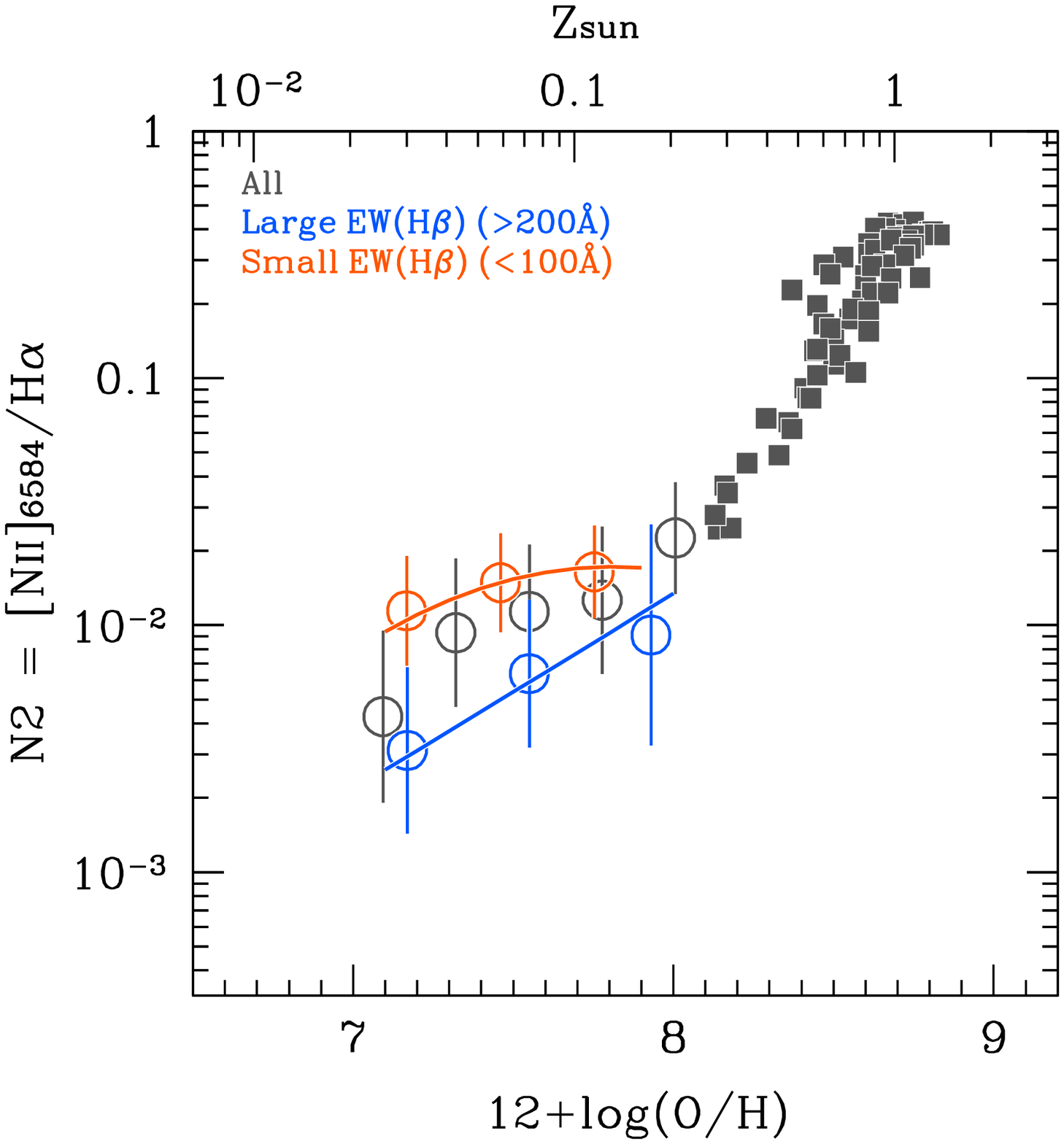}
        }
        \caption{%
        	Same as Figure \ref{fig:Z_empirical_all1} but highlighting that 
		the dispersion in the indicator of R23-index (top) and N2-index (bottom)
		is correlated with EW(\Hb).
		In the left panels, our compile (E)MPGs are color-coded according to EW(\Hb).
		Their binned average relationships are presented in the right panels, where 
		blue and orange open circles are based on the Large and Small EW(\Hb)
		objects (EW $>200$\,\AA\ and $<100$\,\AA), respectively, 
		while gray open circles are based on the full sample of the compiled (E)MPGs.
		The best-fit functions to the Large and the Small EW(\Hb) subsamples 
		are over-plotted with the same color-code. 
        }
        \label{fig:Z_empirical_ewhb1}
    \end{center} 
\end{figure*}


\begin{figure*}[t!]
    \begin{center}
        \subfloat{
            \includegraphics[bb=23 161 522 698, width=0.3\textwidth]{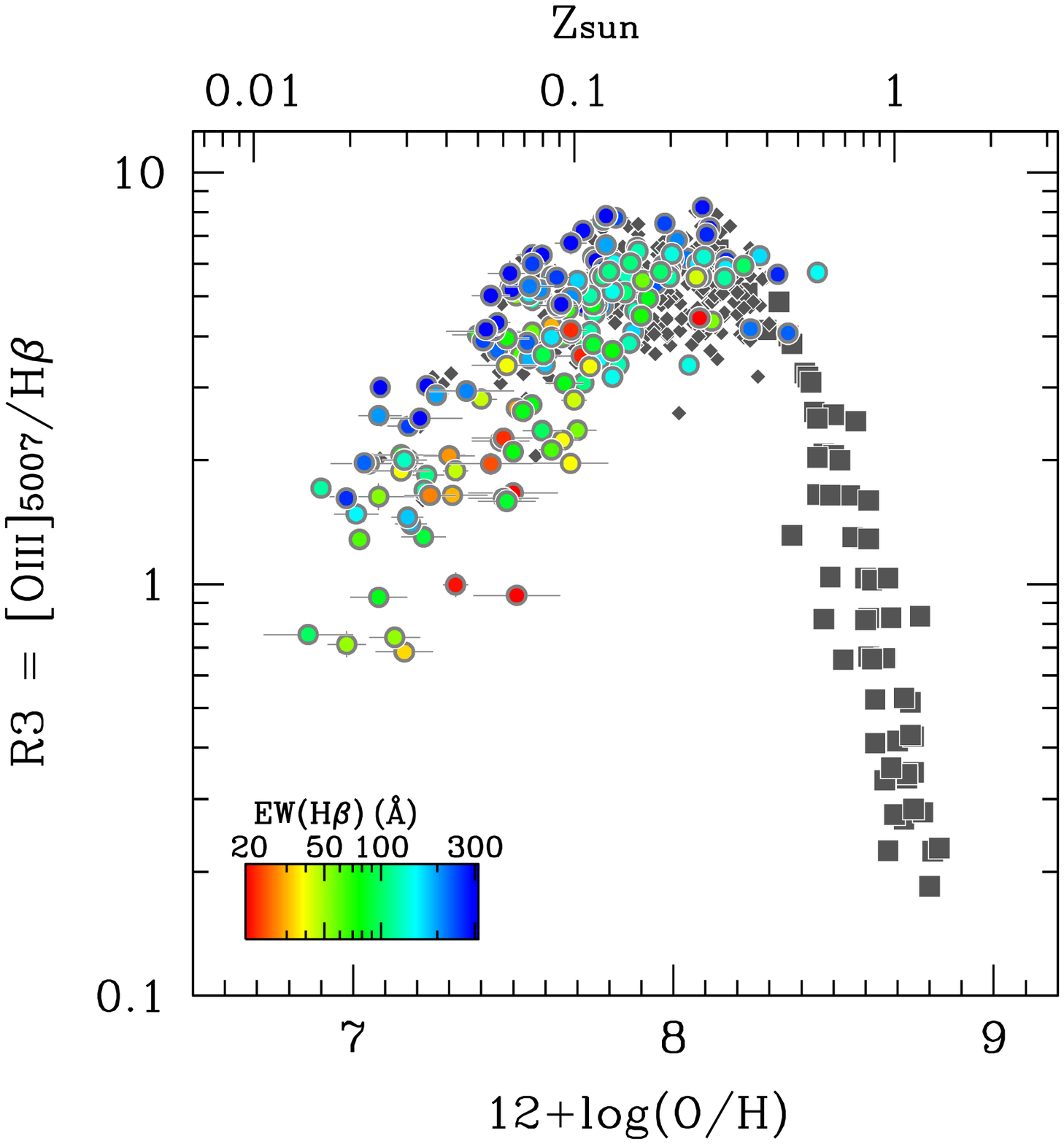}
        }
        \hspace{0.05\textwidth}
        \subfloat{
            \includegraphics[bb=23 161 522 698, width=0.3\textwidth]{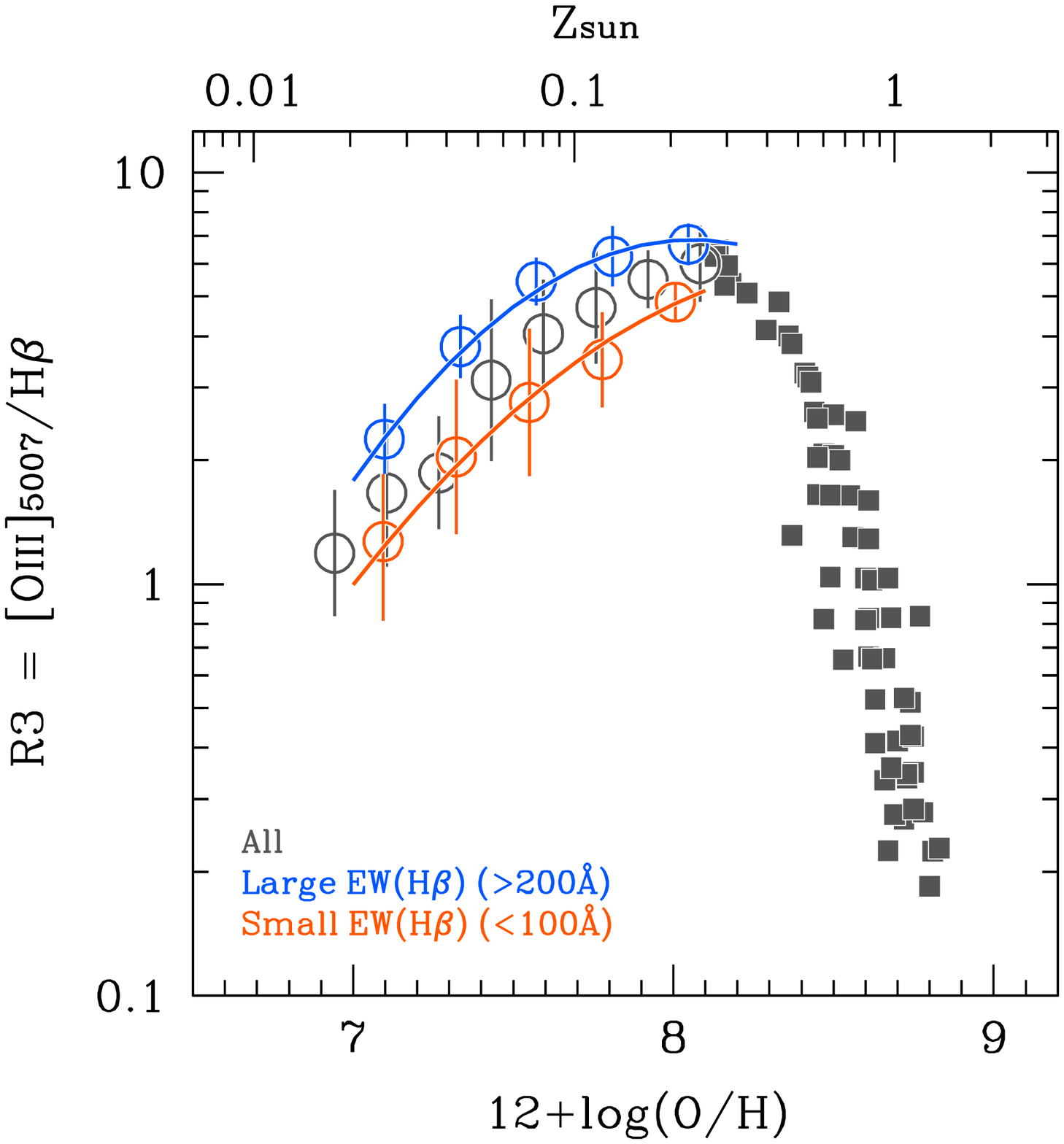}
        }
        
        \subfloat{
            \includegraphics[bb=23 161 522 698, width=0.3\textwidth]{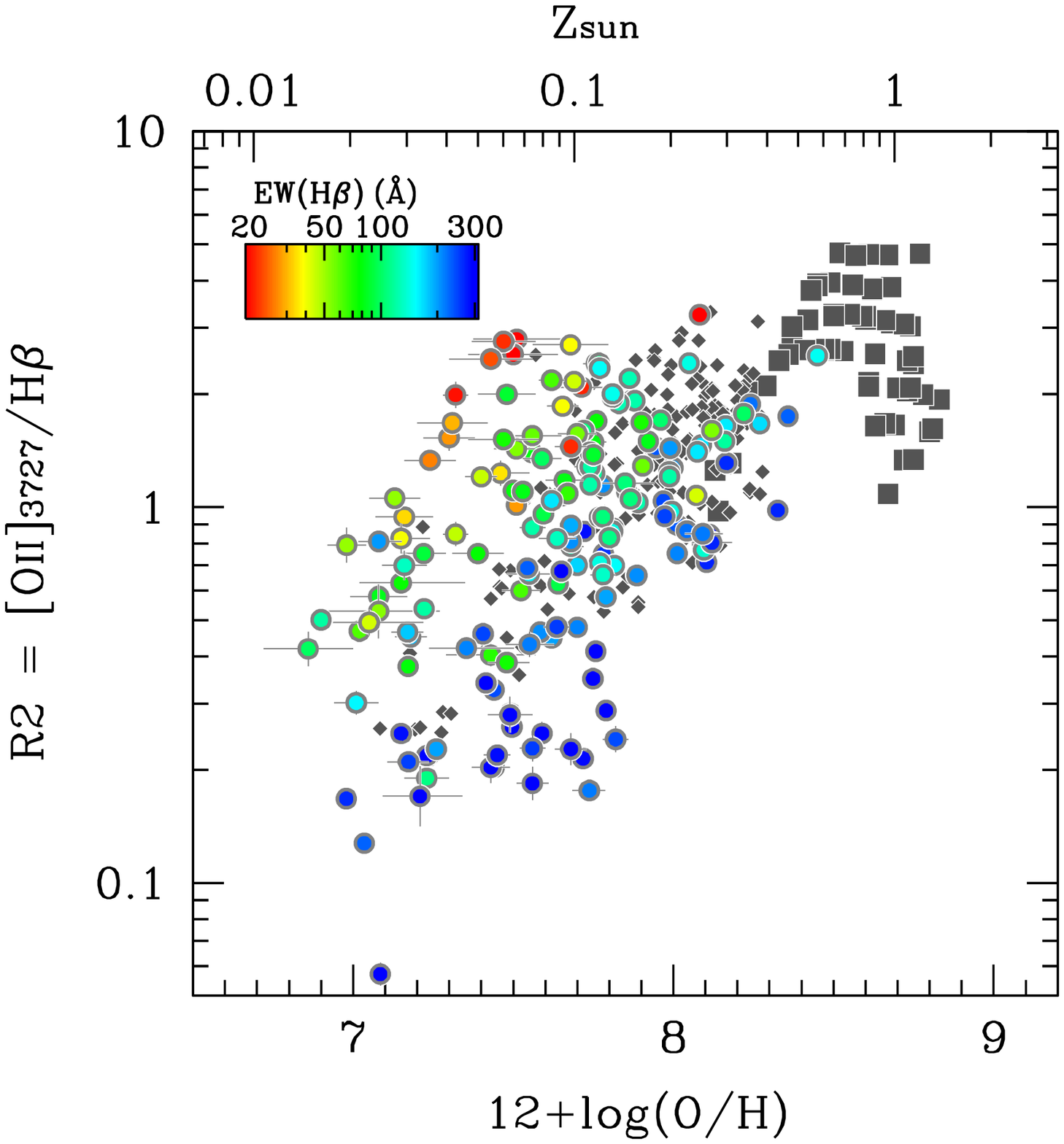}
        }
        \hspace{0.05\textwidth}
        \subfloat{
            \includegraphics[bb=23 161 522 698, width=0.3\textwidth]{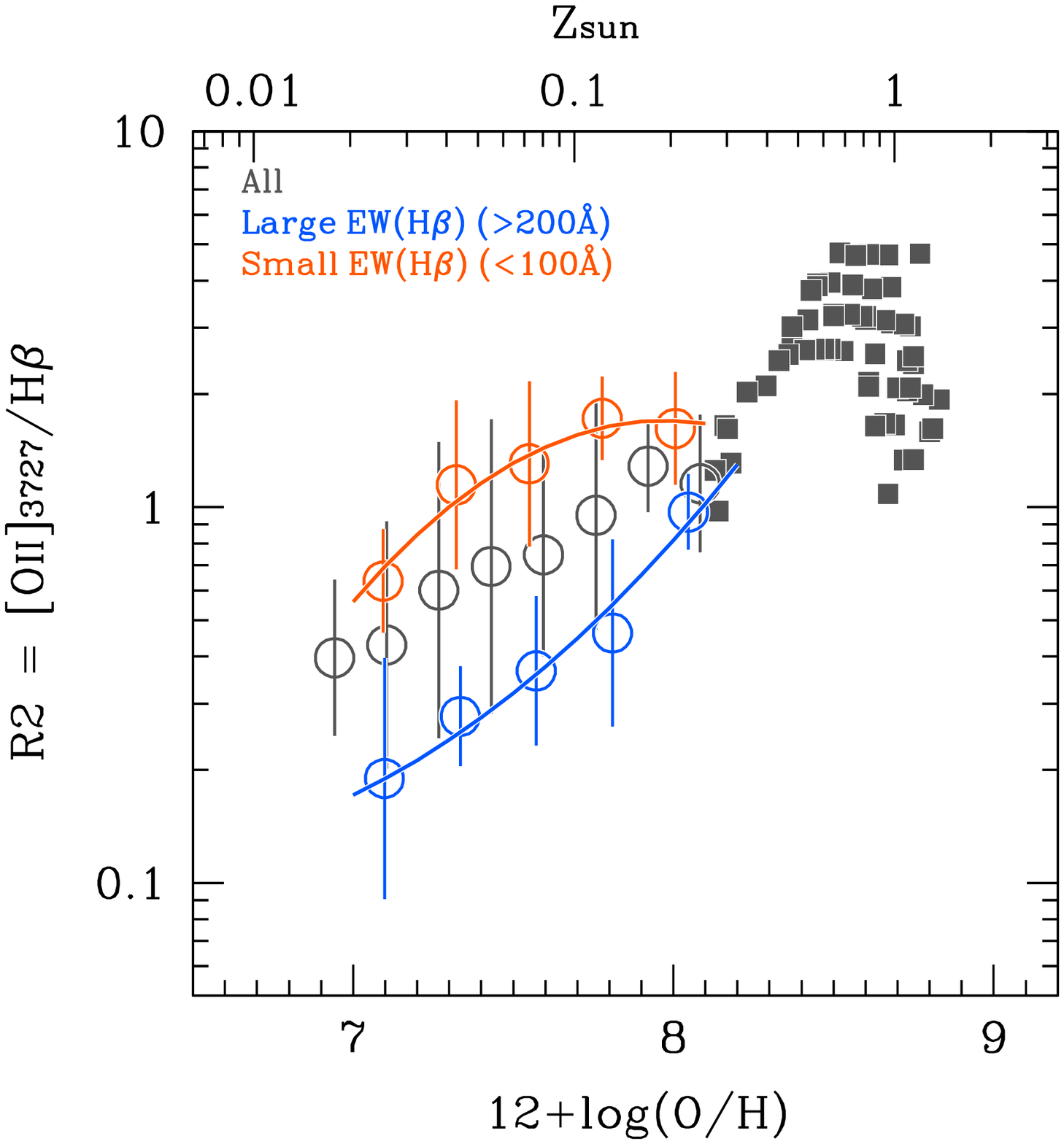}
        }
        
        \subfloat{
            \includegraphics[bb=23 161 522 698, width=0.3\textwidth]{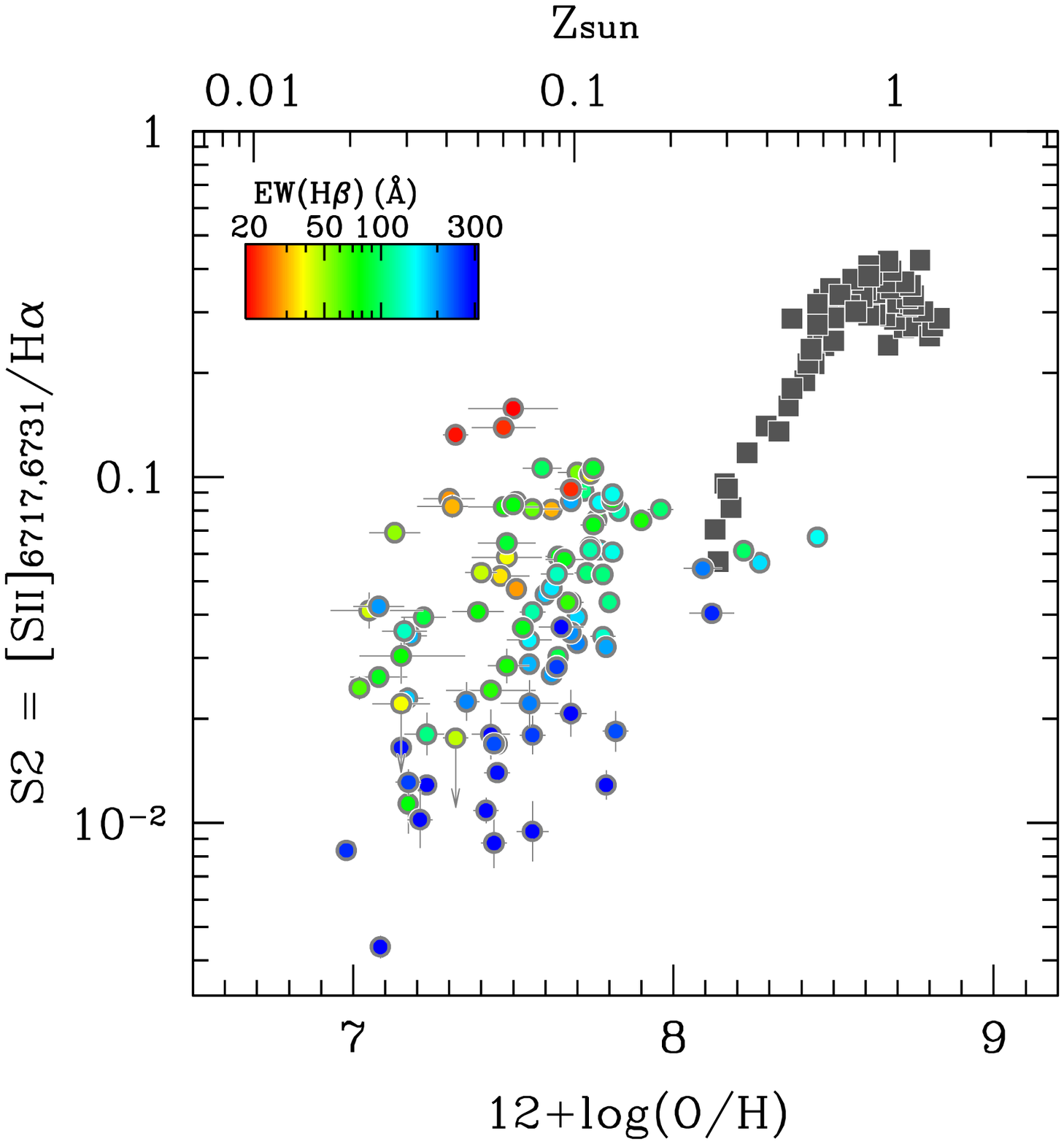}
        }
        \hspace{0.05\textwidth}
        \subfloat{
            \includegraphics[bb=23 161 522 698, width=0.3\textwidth]{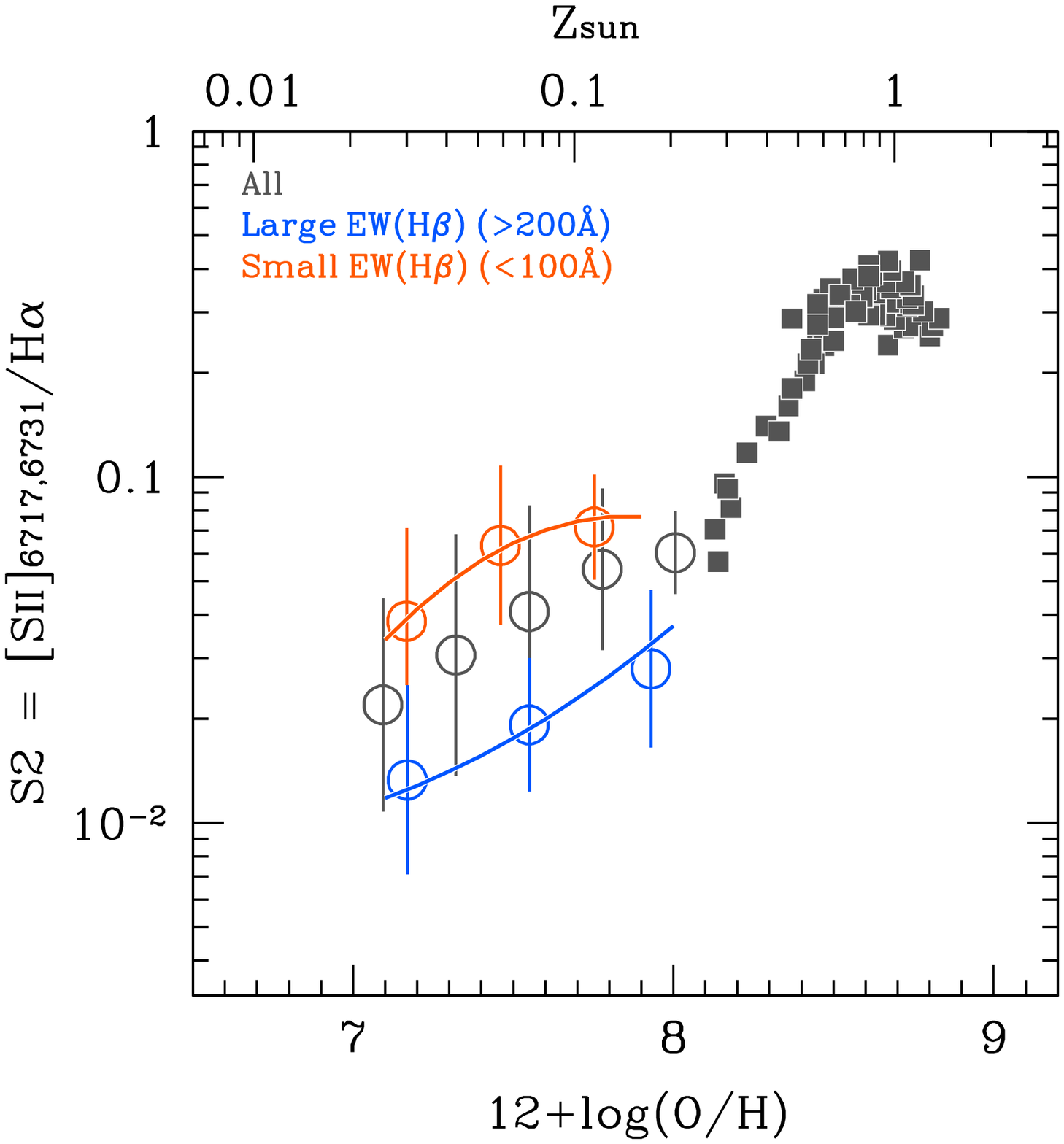}
        }
        \caption{%
		Same as Figure \ref{fig:Z_empirical_ewhb1} but for the line ratios of 
		R3, R2, and S2 (from top to bottom).
        }
        \label{fig:Z_empirical_ewhb2}
    \end{center} 
\end{figure*}


\begin{figure*}[t!]
    \begin{center}
        \subfloat{
            \includegraphics[bb=23 161 522 698, width=0.3\textwidth]{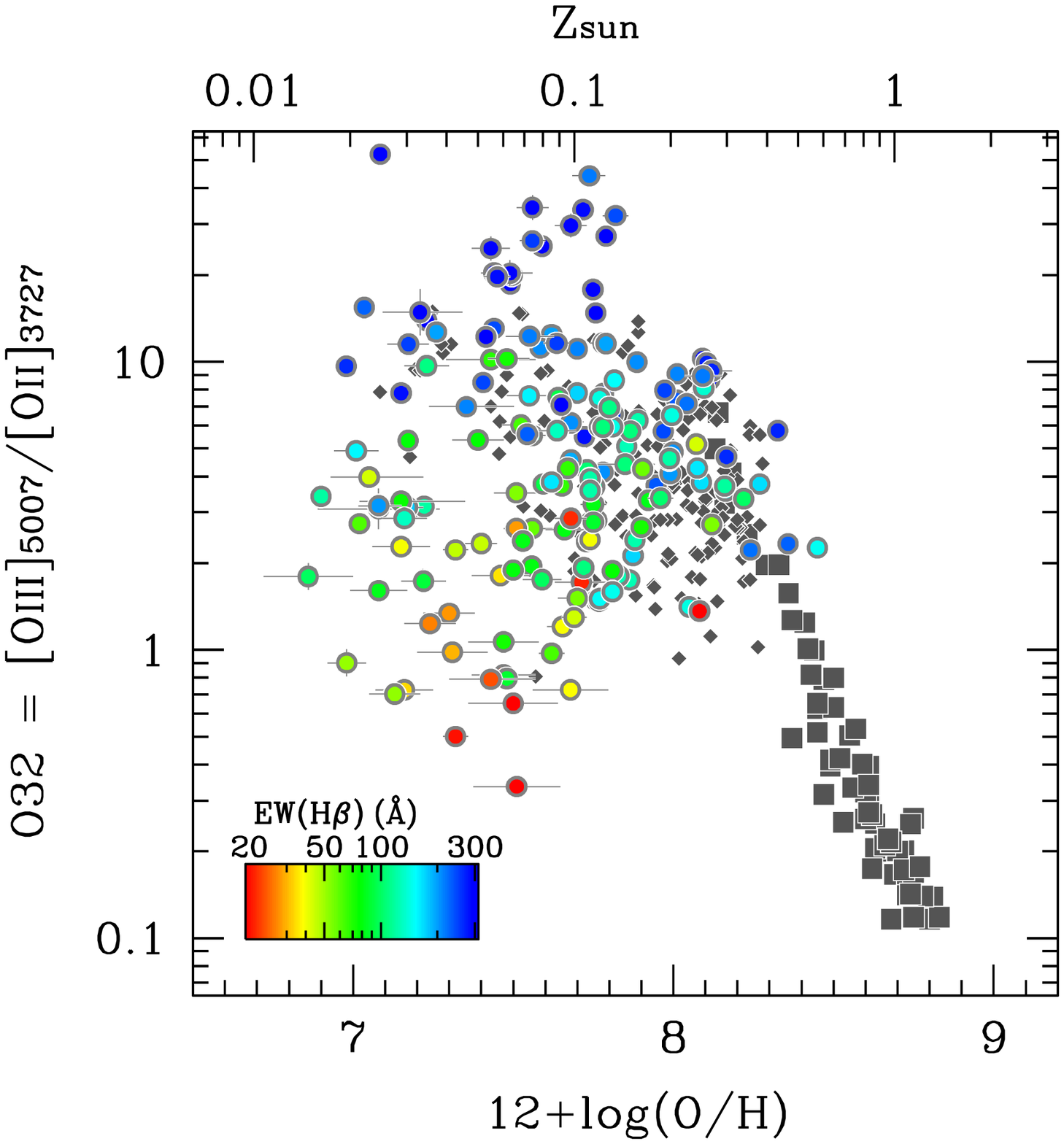}
        }
        \hspace{0.05\textwidth}
        \subfloat{
            \includegraphics[bb=23 161 522 698, width=0.3\textwidth]{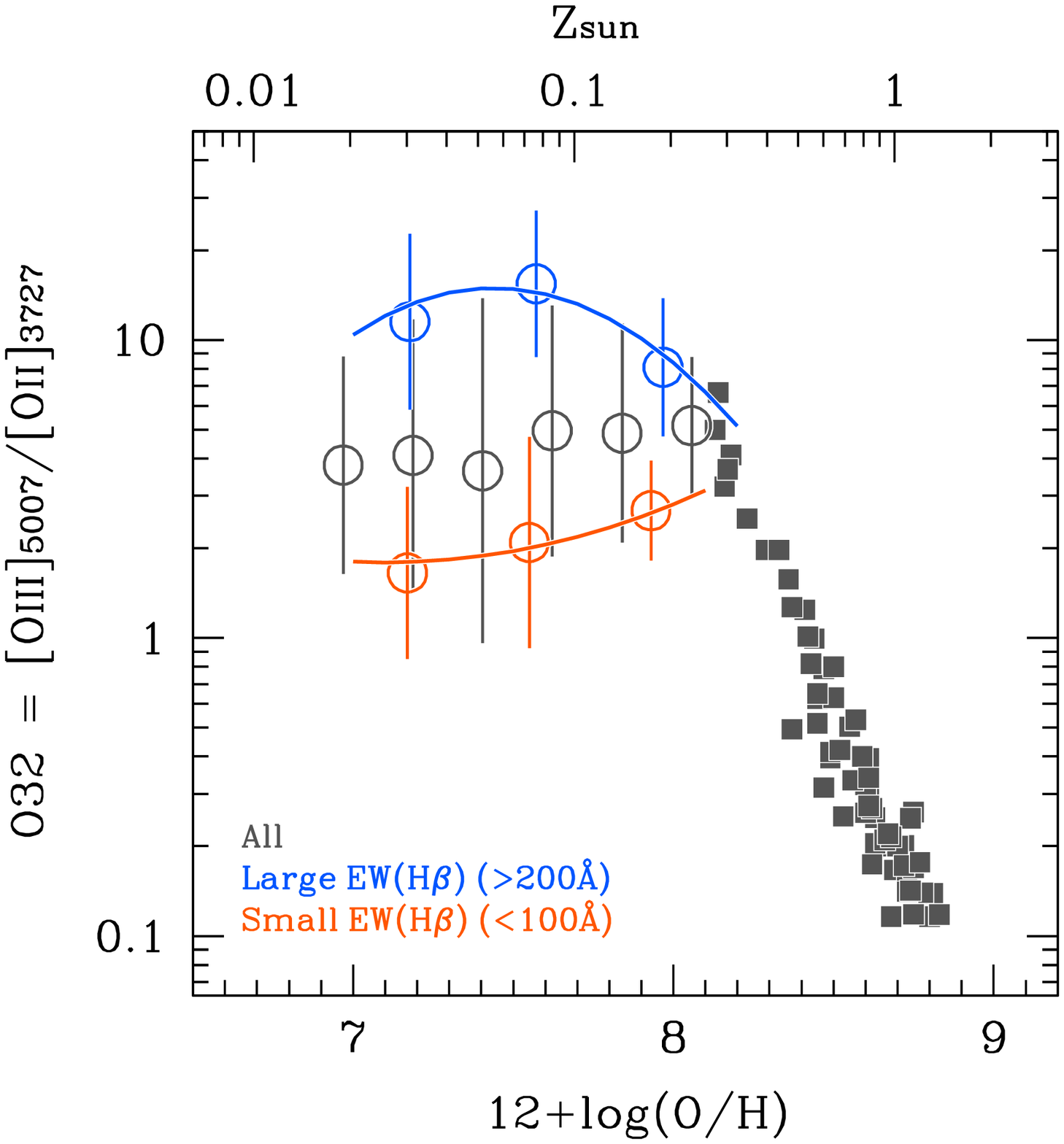}
        }
        
        \subfloat{
            \includegraphics[bb=23 161 522 698, width=0.3\textwidth]{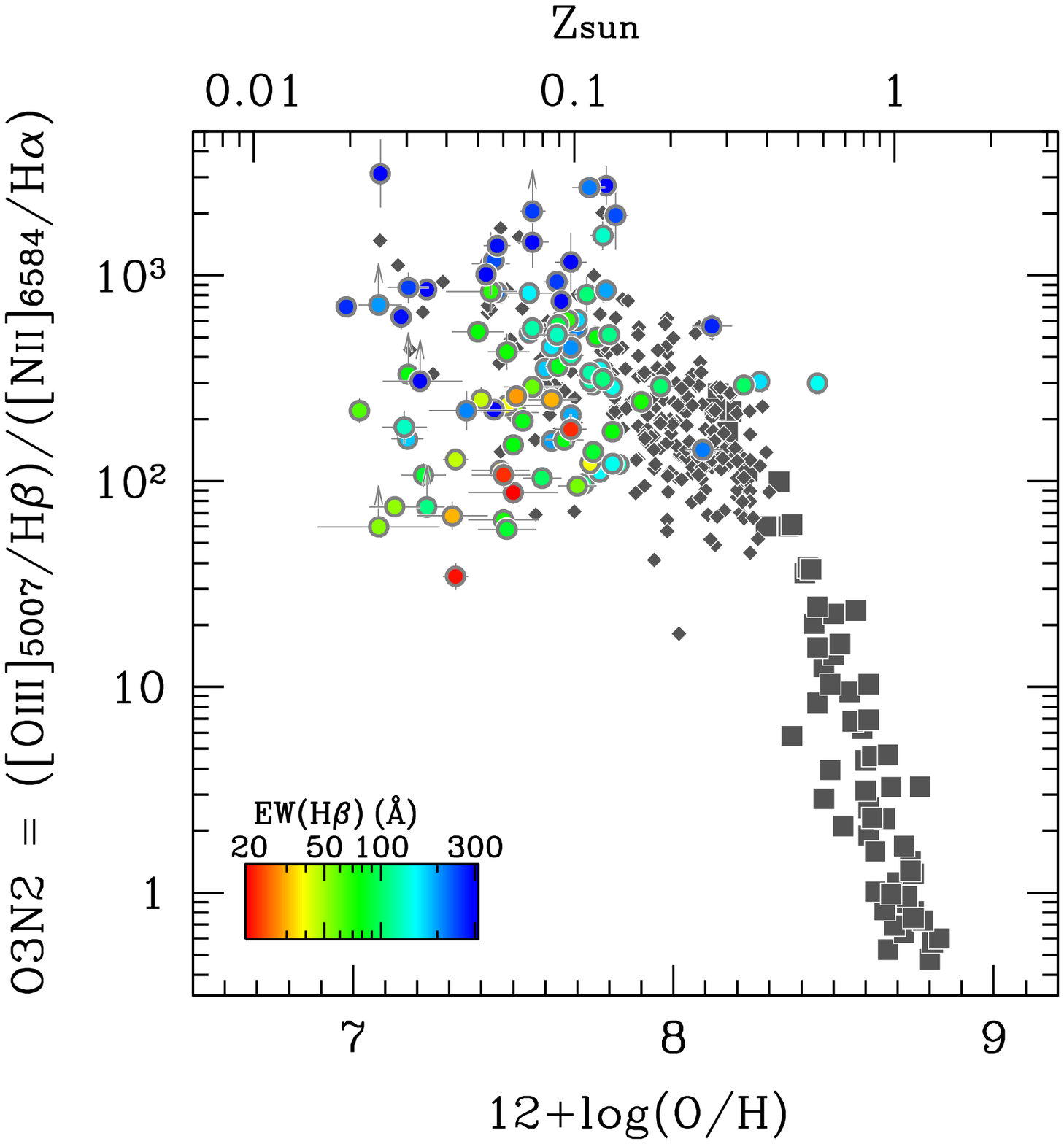}
        }
        \hspace{0.05\textwidth}
        \subfloat{
            \includegraphics[bb=23 161 522 698, width=0.3\textwidth]{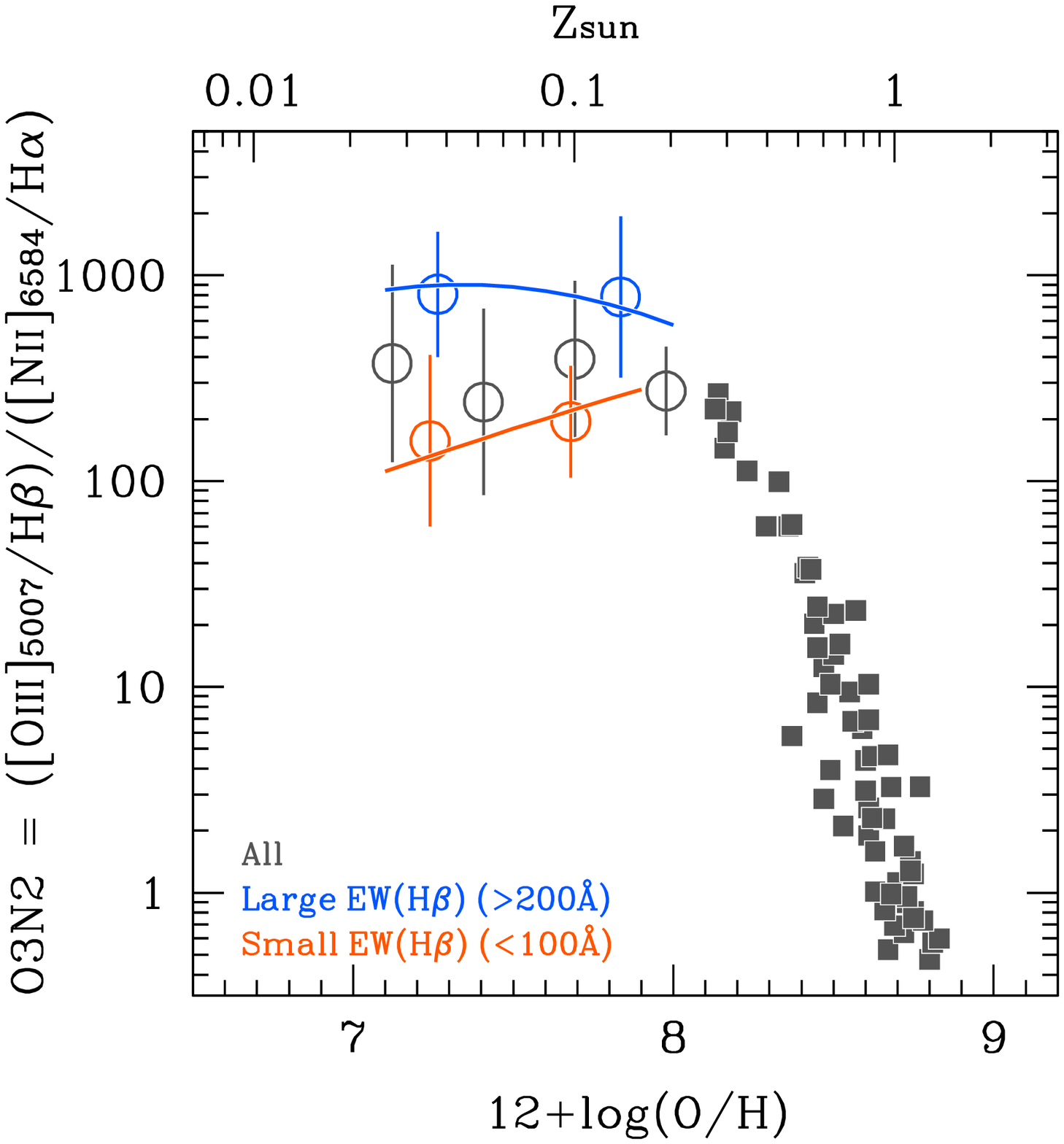}
        }
        
        \subfloat{
            \includegraphics[bb=23 161 522 698, width=0.3\textwidth]{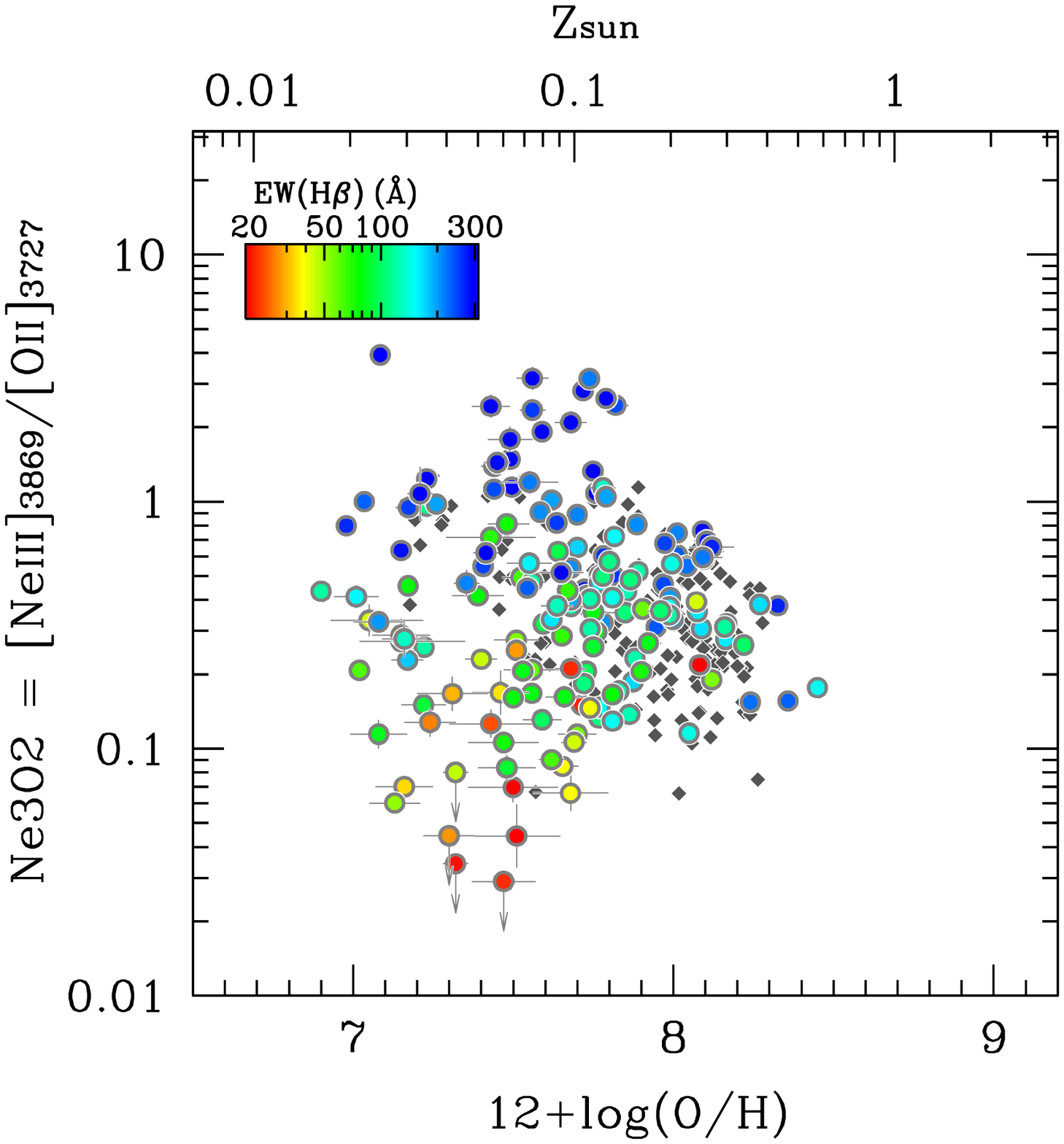}
        }
        \hspace{0.05\textwidth}
        \subfloat{
            \includegraphics[bb=23 161 522 698, width=0.3\textwidth]{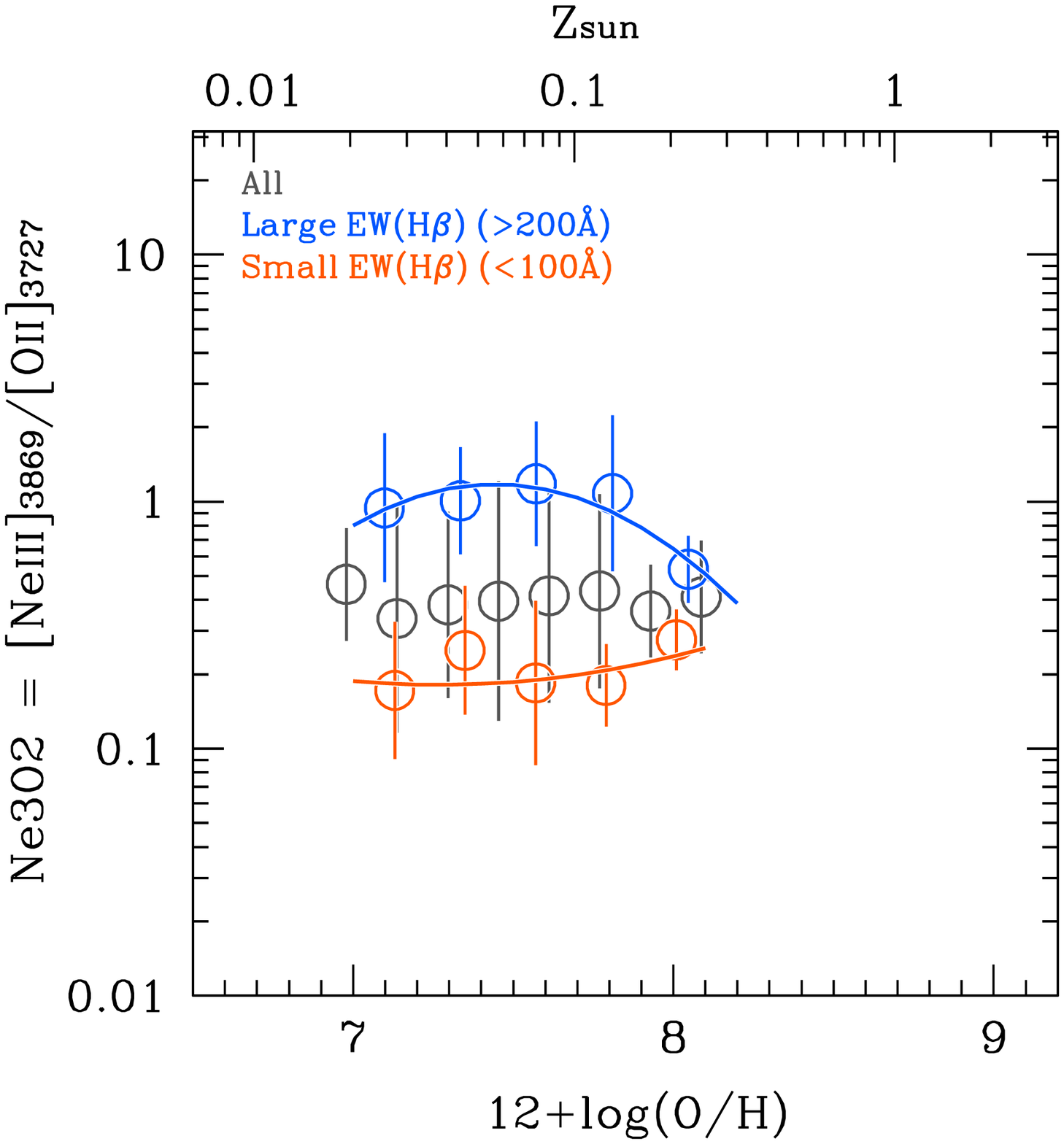}
        }
        \caption{%
		Same as Figure \ref{fig:Z_empirical_ewhb1} but for the line ratios of 
		O32, O3N2, and Ne3O2 (from top to bottom).
        }
        \label{fig:Z_empirical_ewhb3}
    \end{center} 
\end{figure*}


\subsection{Revisiting Diagnostics over \\ \Oabundance\ $\simeq 6.9-8.9$} 
\label{ssec:diagnostics_first}

Our compiled sample containing quite a few numbers of EMPGs ($\times 3$ 
larger than the previous work; e.g., \citealt{nagao2006_metallicity}) with an accurate
metallicity measurement serves as a good reference for checking and improving 
the empirical metallicity diagnostics at the low-metallicity regime.
Figure \ref{fig:Z_empirical_all1} illustrates the relationship of R23- and N2-index
as a function of metallicity. 
Our compiled sources (gray circles) shape a tight relationship 
for each of the indices, remarkably for R23-index, 
at the low metallicity regime.
The short-dashed and long-dashed curve is provided by \citet{maiolino2008} and 
\citet{curti2017,curti2020}, respectively, and the green curve shows our best-fit function
(as detailed below).
The two previous studies are complementary to each other; 
\citet{maiolino2008} make use of the individual low-metallicity galaxies but lack
data-points at high metallicity range with the direct $T_e$ method. 
On the other hand, \citet{curti2017,curti2020} obtain the relations at such 
a high-metallicity regime based on the direct $T_e$ method by stacking SDSS spectra (\S\ref{ssec:samples_sdss}), but lack enough metal-poor galaxies.

Therefore, as a first step in this section, we present the best-fit functions over 
the full metallicity range of \Oabundance\ $\simeq 6.9$ up to $8.9$
based fully on the direct $T_e$ method.
We use our compiled sample of EMPGs $+$ MPGs
(\Oabundance\ $=6.9$--$8.2$; \S\ref{ssec:samples_empgs})
in addition to the \citet{curti2017,curti2020}'s stacked result to compensate the 
high metallicity regime (\Oabundance\ $=8.1$--$8.9$; \S\ref{ssec:samples_sdss}).
We confirm in Figure \ref{fig:Z_empirical_all1} that the two independent samples 
are smoothly connected at the intersection of \Oabundance\ $\sim 8.0$.
Following the previous studies, we adopt the functional form:
\begin{equation}
\log  R = \Sigma_{n=0}^{N} c_nx^n
\label{eq:Z_empirical}
\end{equation}
to derive the best-fit,
where $\log R$ is the logarithm of the strong line ratio (e.g., R23 and N2-index),
and $x$ is the metallicity relative to solar 
(i.e., $x=\log(Z/Z_{\odot})=12+\log({\rm O}/{\rm H})-8.69$).
We bin the compiled (E)MPGs to obtain average line ratios for a given metallicity range 
(red open circles in Figure \ref{fig:Z_empirical_all1}; Table \ref{tbl:Z_empirical_obs}
for the ``All'' sample)
and use them to perform the least squares fitting with the stacked data-points of \citet{curti2017,curti2020}
without any weighting. 
For the binning we only use the individual (E)MPGs with a firm measurement of line ratio
and do not count those with a lower-/upper-limit of line ratio of interest. 
We employ the polynomial orders of $N=2$, $3$, and $4$ in Equation (\ref{eq:Z_empirical}), 
and choose the best-fit for each of the indices from the three functional forms 
which minimizes the dispersions along with both the \Oabundance\ and the line ratio directions. 
Note that we use the individual (E)MPGs rather than the binned average ones
for calculating the dispersions.
The coefficients as well as the $1\sigma$ uncertainties of the best-fit
are given in Table \ref{tbl:Z_empirical_bestfit} for each indices based on the ``All'' sample.
As expected, our best fit-functions show good agreements with 
those of \citet{maiolino2008} in the low-metallicity and 
\citet{curti2017,curti2020} at the high-metallicity regime.

\subsection{Scatters in the diagnostics} 
\label{ssec:diagnostics_scatters}

R23-index is confirmed to work as a good metallicity indicator with the metallicity 
uncertainty of $0.14$\,dex over the range of \Oabundance\ $=6.9-8.9$
(Table \ref{tbl:Z_empirical_bestfit}).
Although N2-index works as accurately as R23-index for galaxies in the high-metallicity 
range, we recognize large scatters as large as $\Delta\log$(O/H) $\sim 0.4$\,dex in the relation 
in the metal-poor regime.
A more or less similar scatter of the line ratio is also found 
in Figure \ref{fig:Z_empirical_all2}
when other famous indicators are 
chosen as a function of metallicity:  
\OIII$\lambda5007$$/$\Hb\ (R3),
\OII$\lambda3727$$/$\Hb\ (R2), 
\OIII$\lambda5007$$/$\OII$\lambda3727$ (O32),
R3$/$N2 (O3N2),
\SII$\lambda\lambda6717,6731$$/$\Ha\ (S2), and 
\NeIII$\lambda3869$$/$\OII$\lambda3727$ (Ne3O2).
Weird concave down features found in the best-fit in O32- and O3N2-index
as well as plateau-like behaviors in R2-, S2-, and Ne3O2-index
illustrate their weak dependence on metallicity in the low metallicity regime 
with the current function forms
(see also their large uncertainties of $\Delta\log$(O/H) in Table \ref{tbl:Z_empirical_bestfit}).
We note that this paper cannot fully address scatters in the high-metallicity range, 
as we only have the averaged line ratios based on the stacked SDSS spectra
(see also \S\ref{ssec:diagnostics_implications}). 
In the following, we explore and discuss the scatters in the low-metallicity regime
using our compiled (E)MPG sample.

As found in N2-index, a large scatter appears particularly associated
with the line ratios using singly-ionized, low-ionization lines.
We thus speculate this would be caused by a variation of ionization state
in the ISM for a given metallicity, 
such as ionization parameter and ionizing radiation field.
Indeed, \citet{pilyugin2016} demonstrate that, for a given metallicity, N2-index tends 
to become lower with the excitation parameter (=\OIII$/$(\OII$+$\OIII)).
The excitation parameter and O32-index are famous probes of the ionization parameter 
(e.g., \citealt{KD2002,NO2014}), and useful to correct for such an effect 
of ionization state on the metallicity indicators 
(see also \citealt{pilyugin2005, KD2002, izotov2019_lowZcandidates}). 
Although it is recommended to adopt these direct indicators of ionization parameter
as a correction factor, it will not be always possible to obtain such multiple emission lines
probing both high and low ionization gas from distant galaxies. 
Because our main purpose of this work is to provide metallicity diagnostics that will
be practical and useful even when limited sets of emission lines are available, 
we re-examine the scatters in the indices using a different observable quantity, 
which is easy-to-obtain and a good probe of the ionization state. 
Among several observables, we consider EW(\Hb) (or equivalently EW(\Ha)) is best-suited 
for this purpose for the following reasons.
First, O32-index is known to be positively correlated with specific star-formation rate
(e.g., \citealt{NO2014,sanders2016}), which should be proportional to EWs of \Hb\ 
and \Ha\ by definition.
Indeed, our (E)MPGs shape a tight positive correlation between O32-index and EW(\Hb).
Second, as will be illustrated in \S\ref{ssec:UV_properties_xiion}, 
\xiion, a gauge of the hardness of the ionizing radiation field, 
is primarily controlled by EW(\Hb), such that it becomes larger as EW(\Hb) increases.
Finally, as compared to the ionization/excitation parameters, O32-index, and \xiion, 
EW(\Hb) is a direct observable quantity and easy to get from high-redshift galaxies
with JWST, and reliably measured without being affected by dust extinction
and other assumptions. 
We expect at least either of \Hb\ and \Ha\ is available when using the metallicity
indicators, and EW(\Ha) can be translated into EW(\Hb), 
such that EW(\Ha) $=5.47\times$ EW(\Hb)
following a typical relationship \citep{kojima2020}.
Therefore, we expect we can use EW(\Hb) to probe the degree of ionization state in ISM 
of distant galaxies in a practical manner.
Moreover, a presence of diffuse ionized gas (DIG) can influence the line ratios, 
especially of the low-ionization lines \citep{zhang2017,sanders2017}.
This effect can also be quantified by EW(\Hb), as the fraction of emission from
DIG is thought to decrease with increasing star-formation intensity
\citep{oey2007}.

\begin{figure*}[t]
    \begin{center}
        \subfloat{
            \includegraphics[bb=23 161 522 698, width=0.40\textwidth]{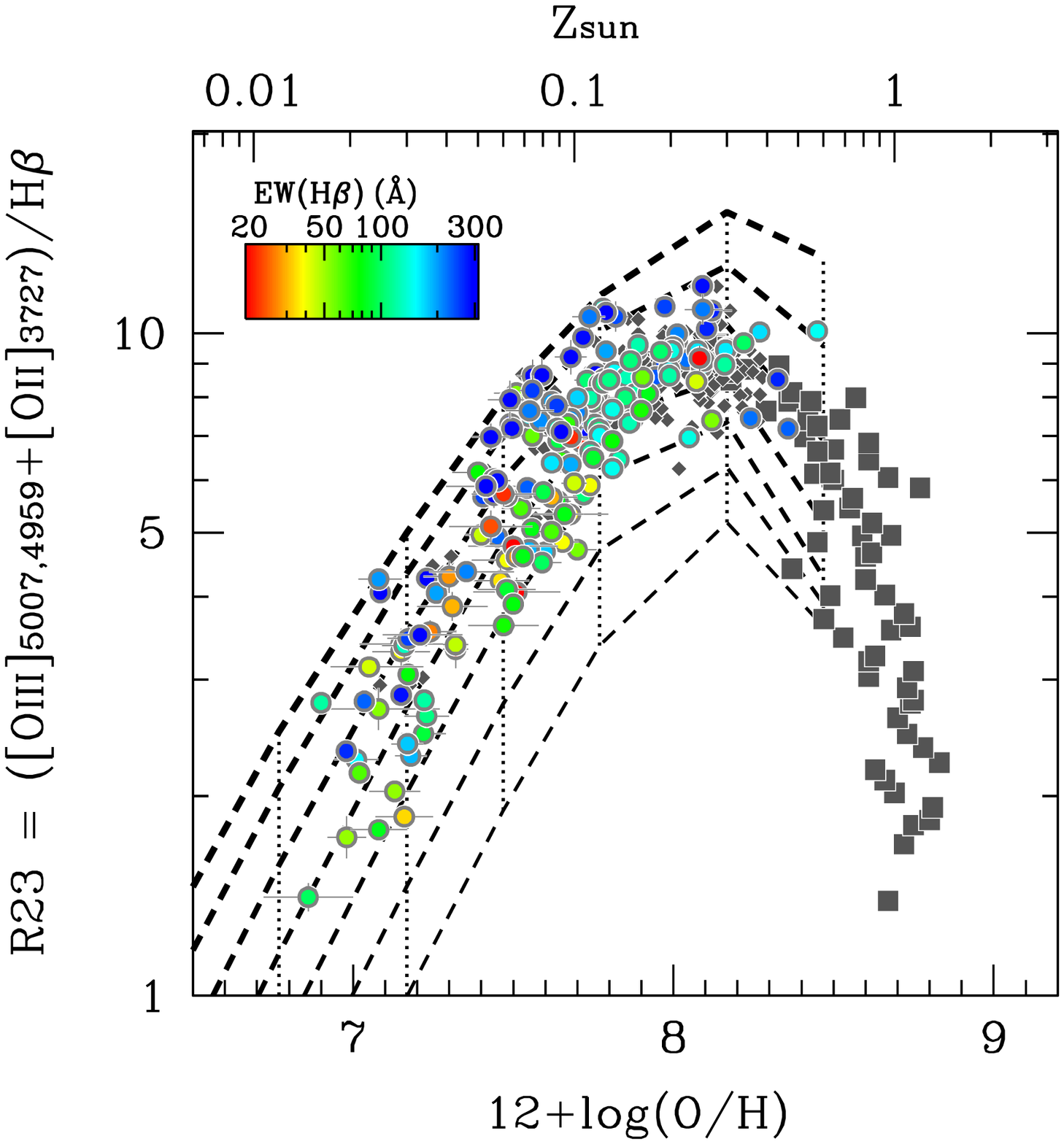}
        }
        \hspace{0.05\textwidth}
        \subfloat{
            \includegraphics[bb=23 161 522 698, width=0.40\textwidth]{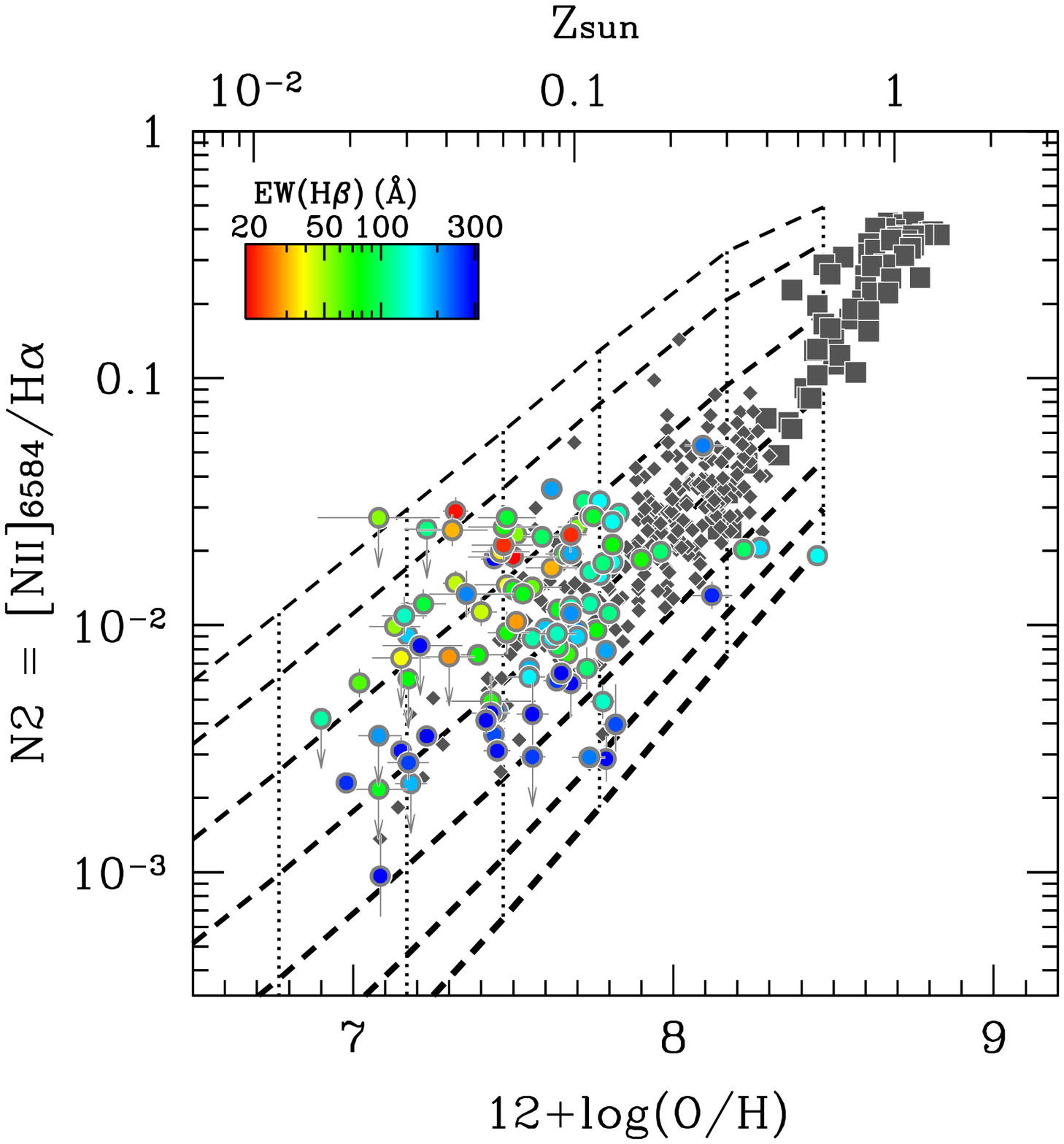}
        }
        
        \caption{%
        	Photoionization model predictions of 
		R23-index (left) and N2-index (right)
		and comparisons with the observations (Figure \ref{fig:Z_empirical_ewhb1}).
		The black grid illustrates the photoionization models 
		based on the binary evolution SEDs (BPASS-300bin; \citealt{nakajima2018_vuds}).
		The dashed-curves present model tracks with an ionization parameter
		$\log\,U$ from $-3.5$ (thin) to $-0.5$ (thick) with a step of $0.5$\,dex.
		Only the gas-phase oxygen abundance is counted for the models 
		(i.e., after removing the depleted component onto dust grains)
		in the abscissa axis
		to be directly compared with the observations. 
		The gas density is fixed to $100$\,cm$^{-3}$.
		The models explain the scatter and its correlation with EW(\Hb) 
		by changing the ionization parameter in a consistent manner
		for each of the indicators.
        }
        \label{fig:Z_empirical_model1}
    \end{center} 
\end{figure*}


\begin{figure*}[t]
    \begin{center}
        \subfloat{
            \includegraphics[bb=23 161 522 698, width=0.3\textwidth]{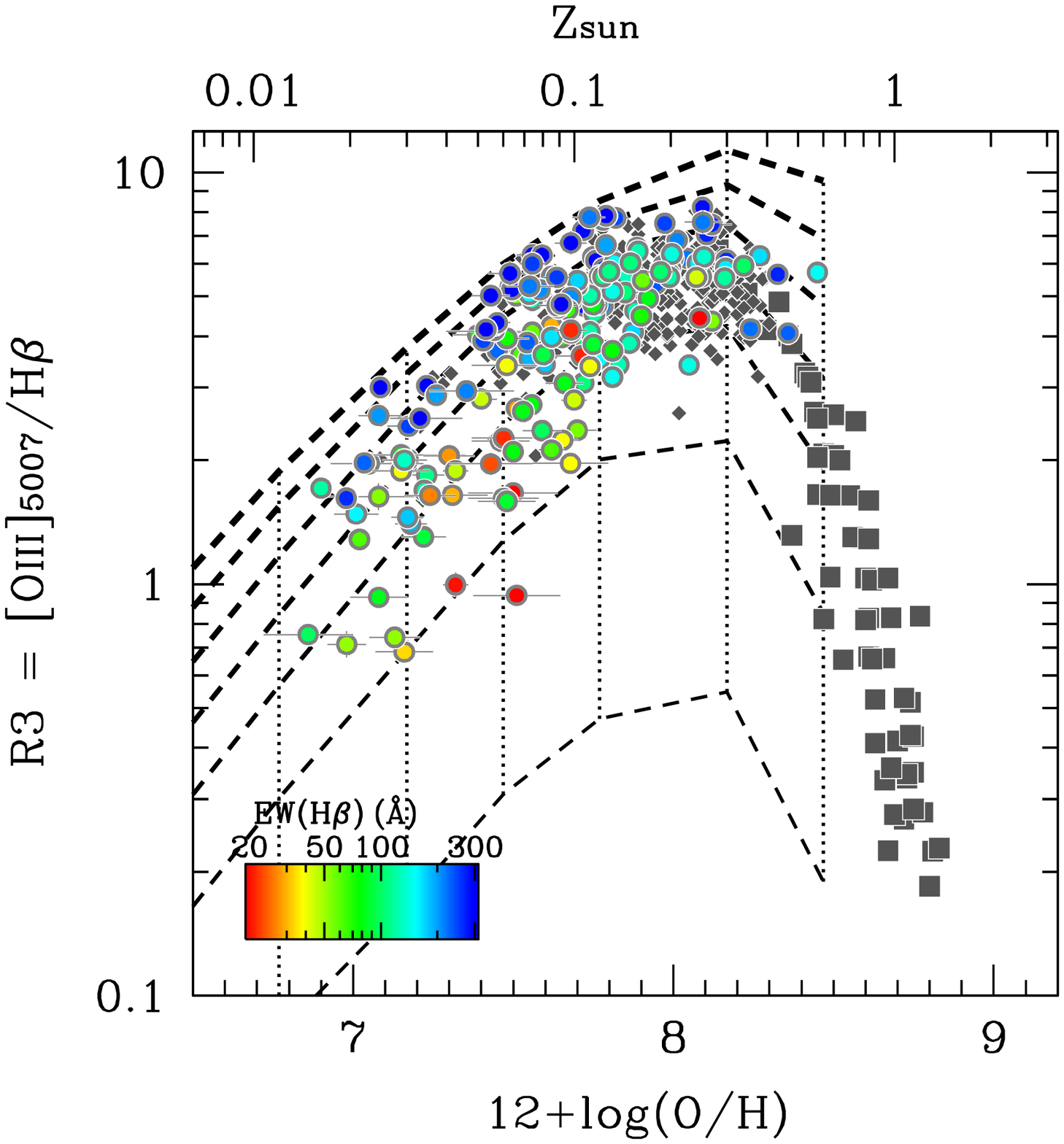}
        }
        \subfloat{
            \includegraphics[bb=23 161 522 698, width=0.3\textwidth]{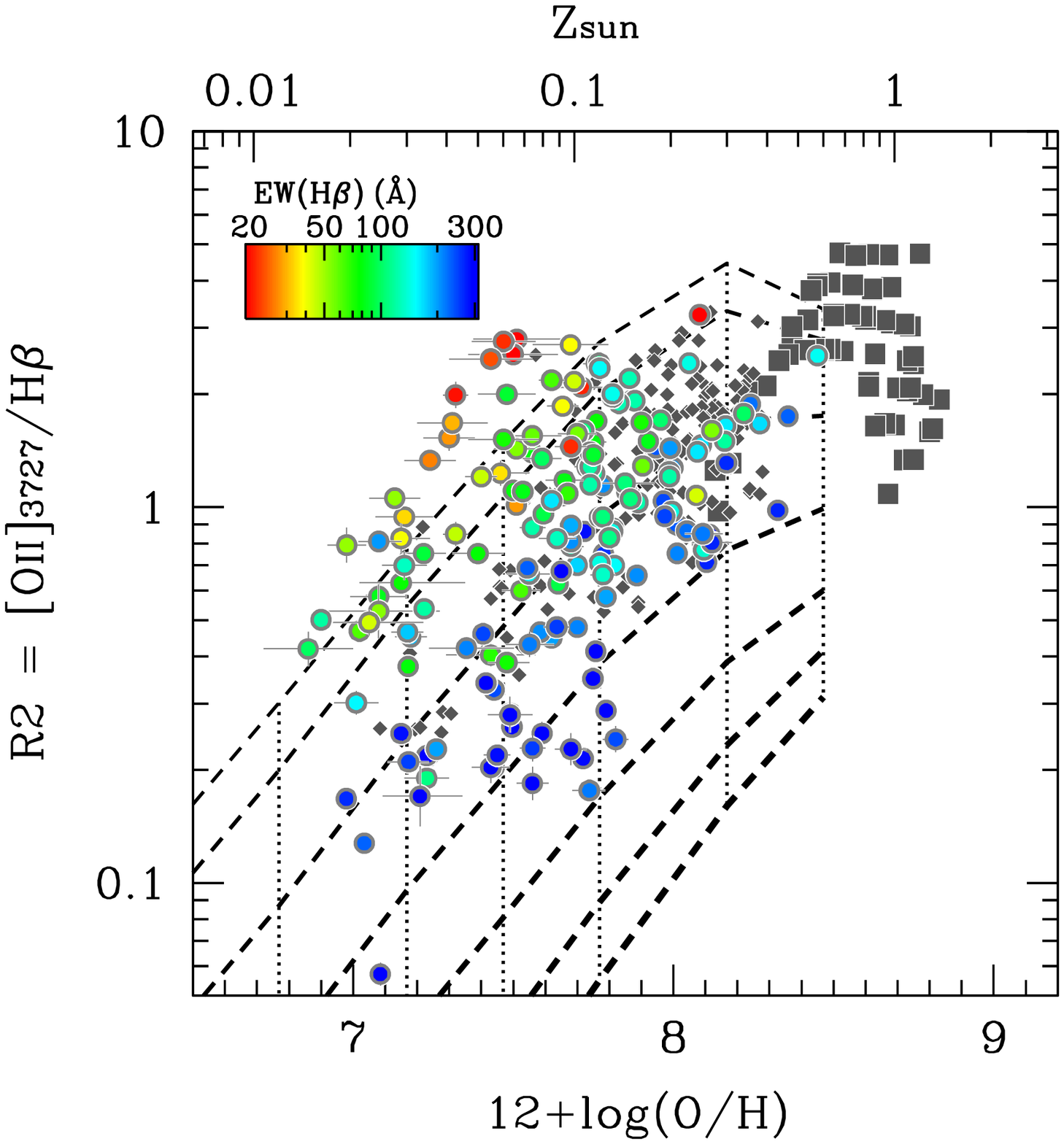}
        }
        \subfloat{
            \includegraphics[bb=23 161 522 698, width=0.3\textwidth]{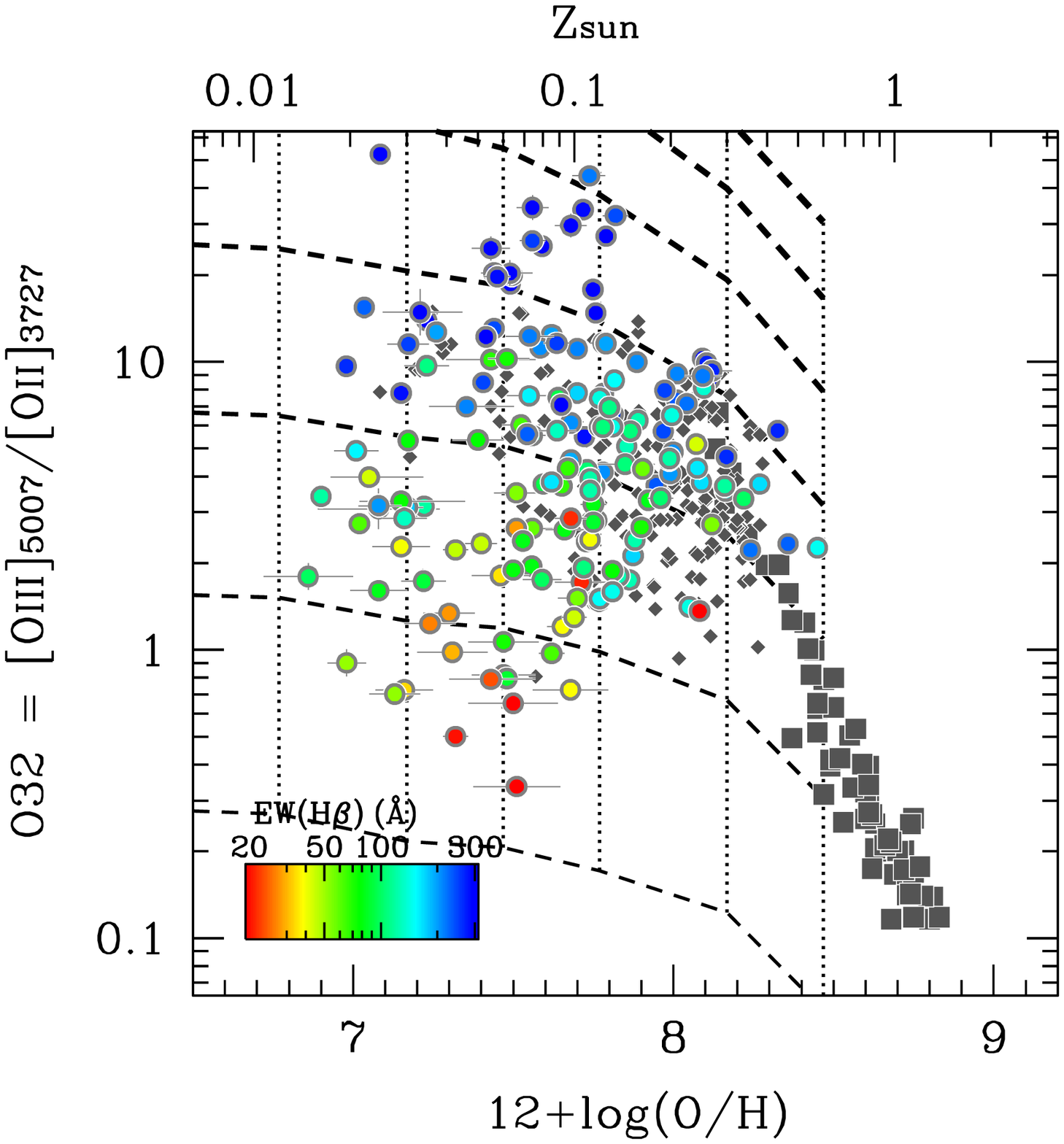}
        }
        
        \subfloat{
            \includegraphics[bb=23 161 522 698, width=0.3\textwidth]{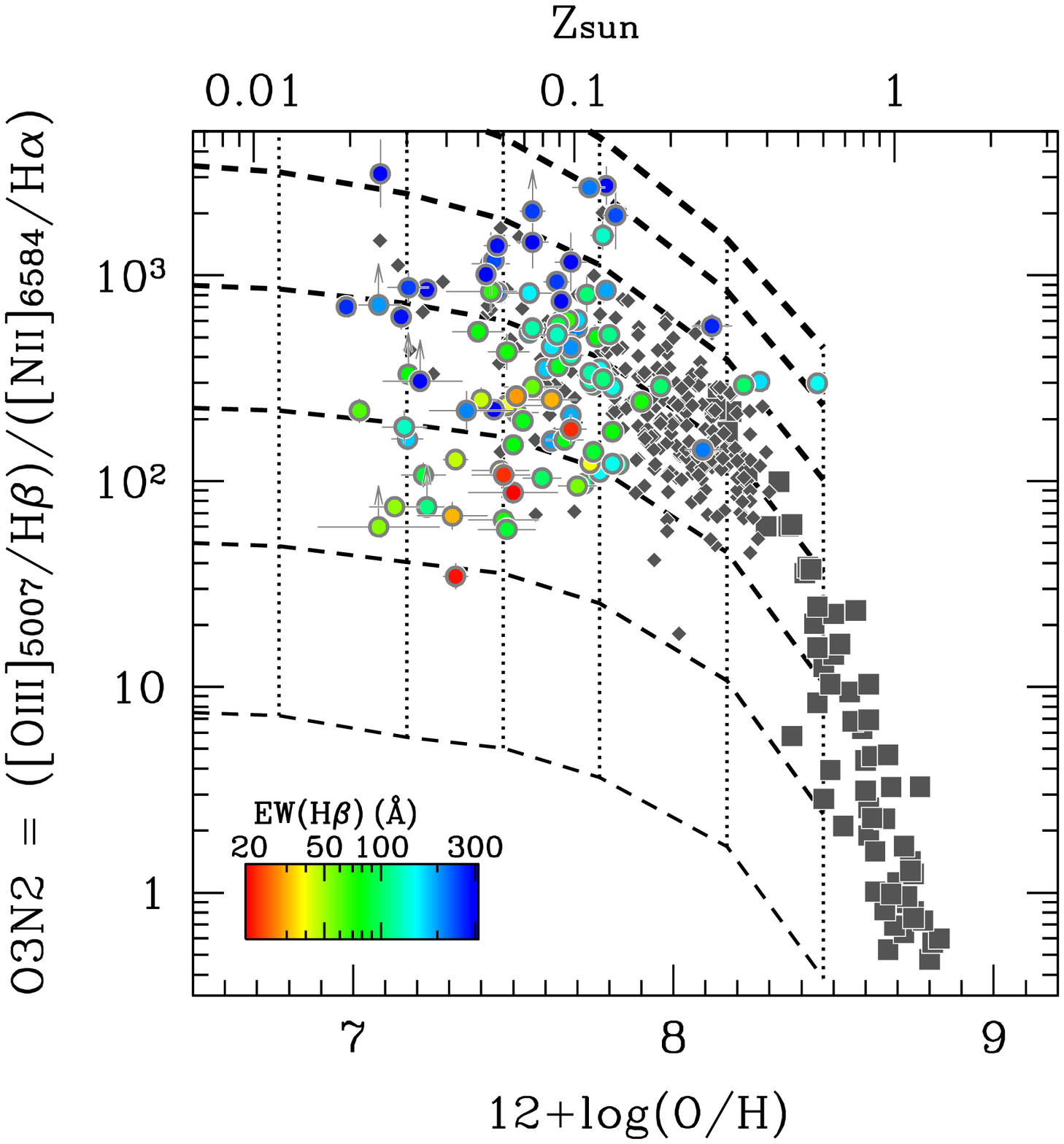}
        }
        \subfloat{
            \includegraphics[bb=23 161 522 698, width=0.3\textwidth]{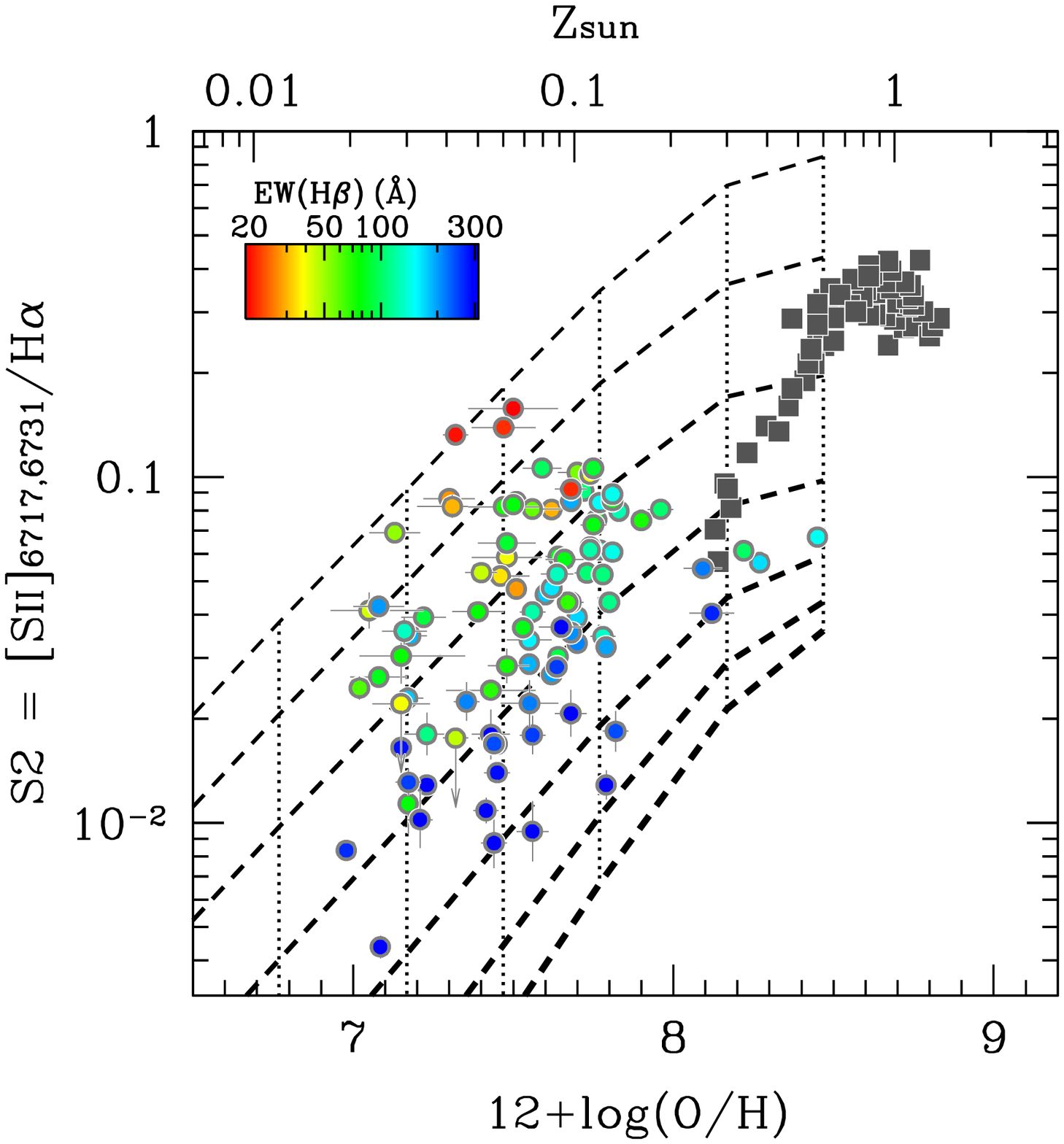}
        }
        \subfloat{
            \includegraphics[bb=23 161 522 698, width=0.3\textwidth]{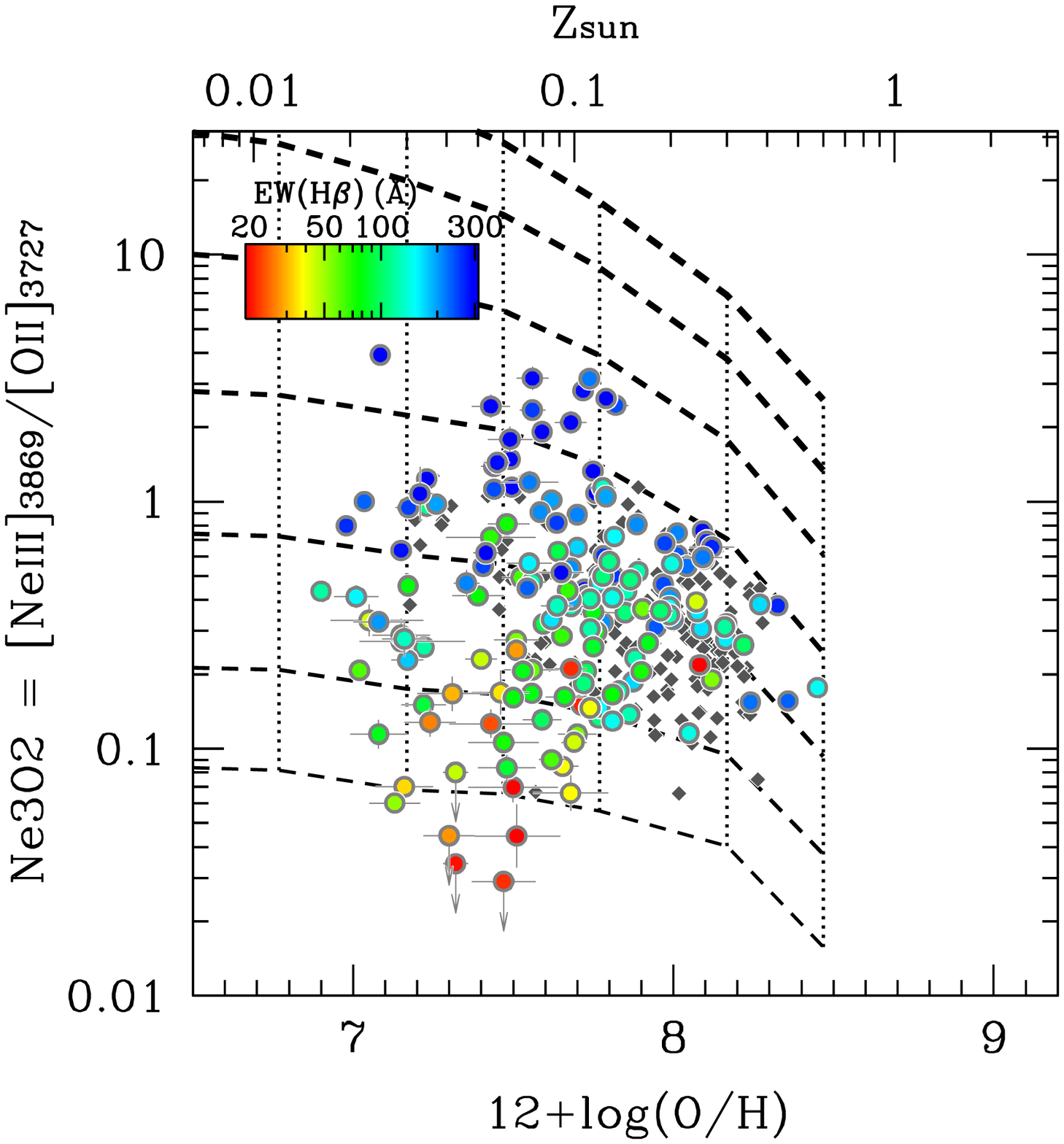}
        }
        \caption{%
        	Same as Figure \ref{fig:Z_empirical_model1} but for the line ratios of 
		R3, R2, O32, O3N2, S2, and Ne3O2 (from top left to bottom right).
        }
        \label{fig:Z_empirical_model2}
    \end{center} 
\end{figure*}


\begin{figure*}[t]
    \begin{center}
        \subfloat{
            \includegraphics[bb=23 161 522 698, width=0.27\textwidth]{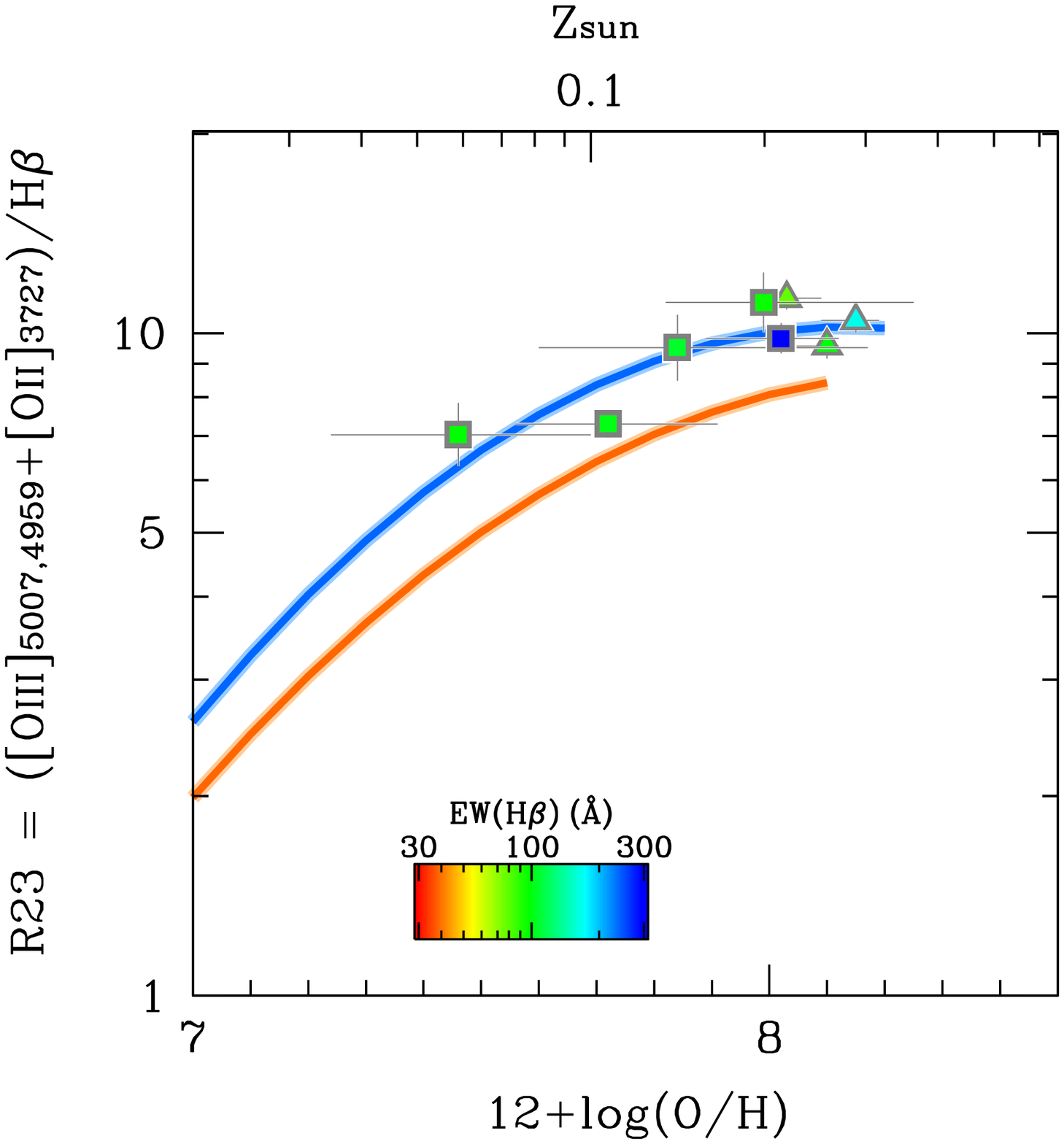}
        }
        \subfloat{
            \includegraphics[bb=23 161 522 698, width=0.27\textwidth]{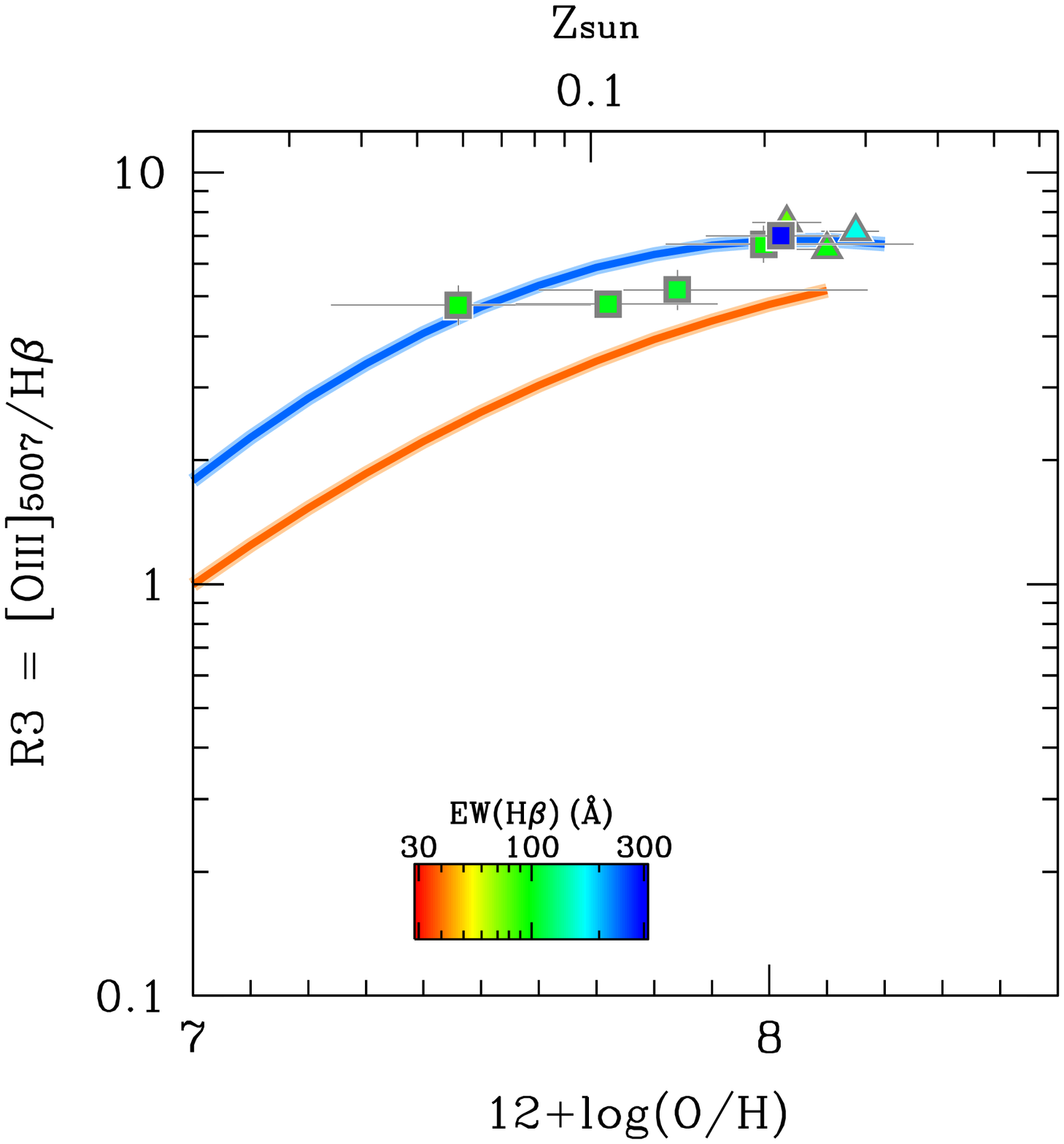}
        }      
        \subfloat{
            \includegraphics[bb=23 161 522 698, width=0.27\textwidth]{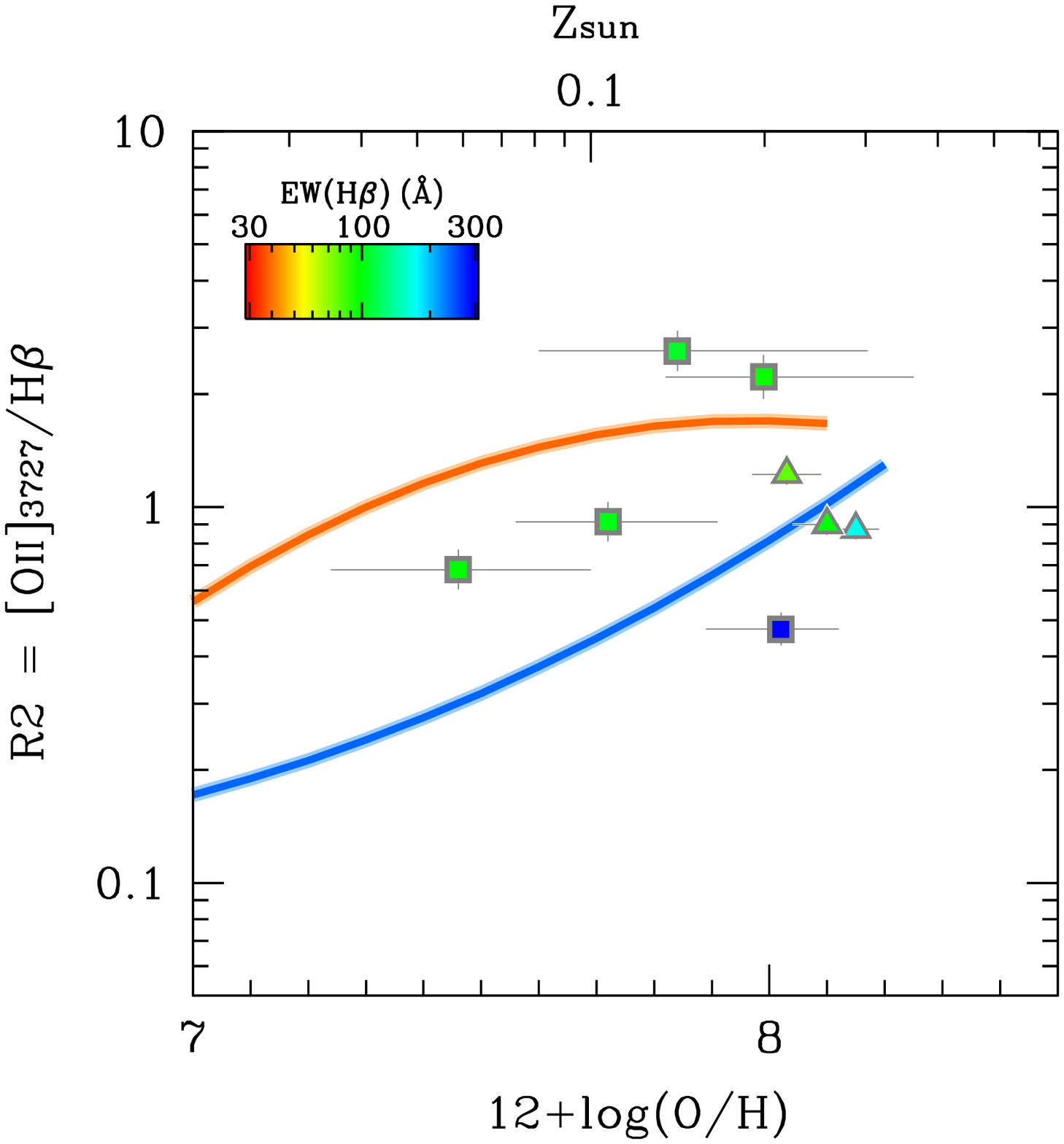}
        }
        
        \subfloat{
            \includegraphics[bb=23 161 522 698, width=0.27\textwidth]{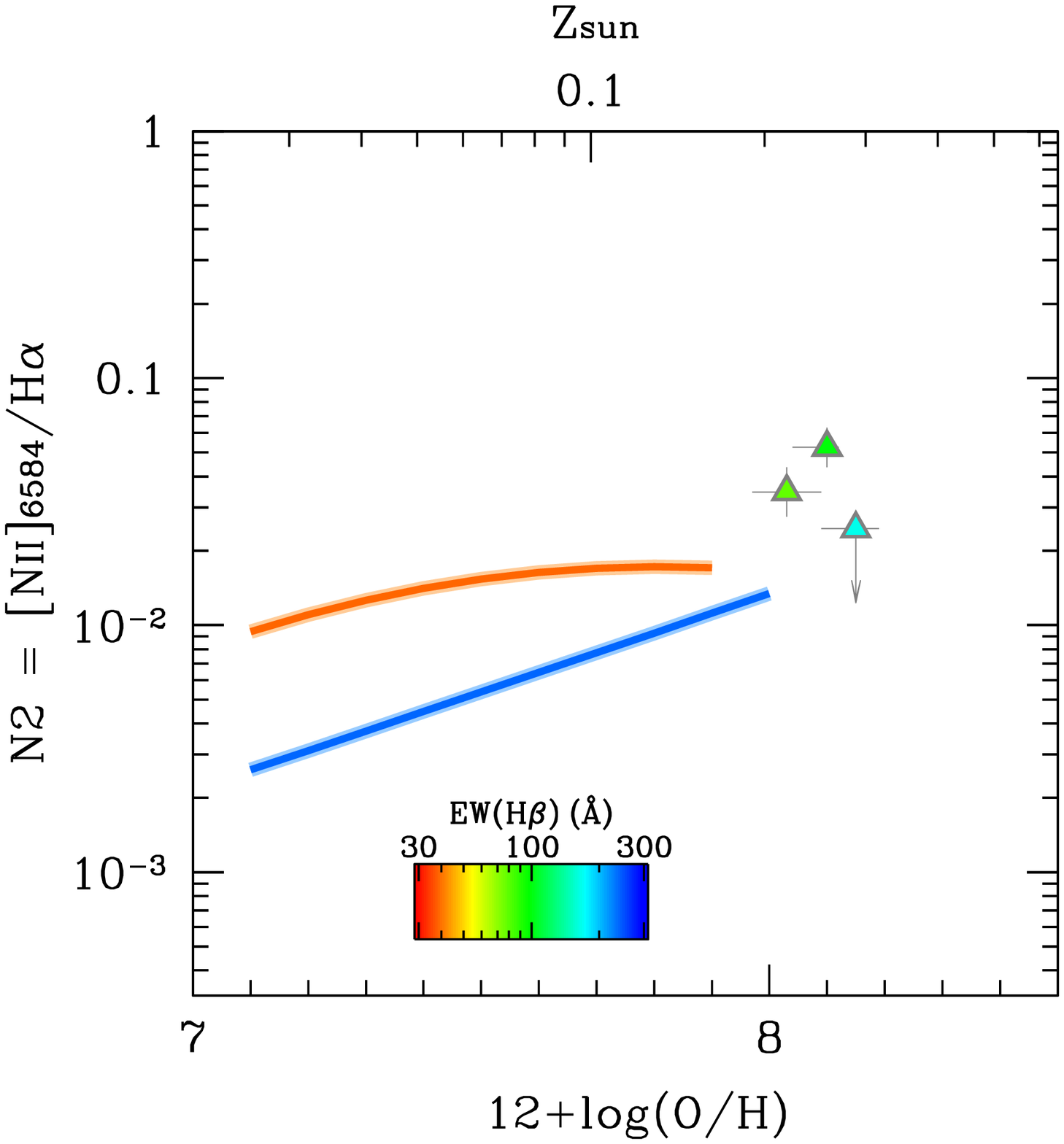}
        }
        \subfloat{
            \includegraphics[bb=23 161 522 698, width=0.27\textwidth]{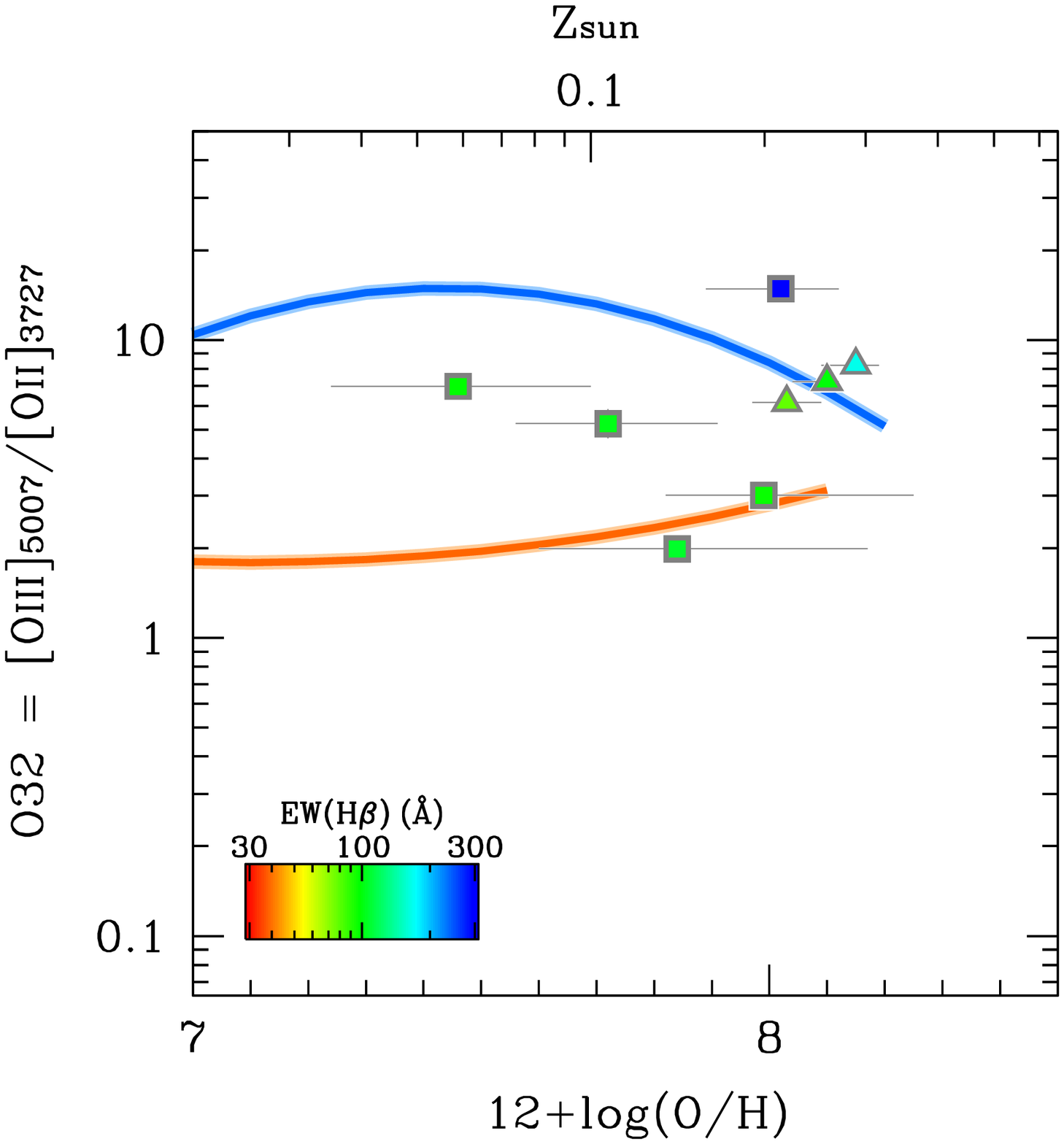}
        }
        \subfloat{
            \includegraphics[bb=23 161 522 698, width=0.27\textwidth]{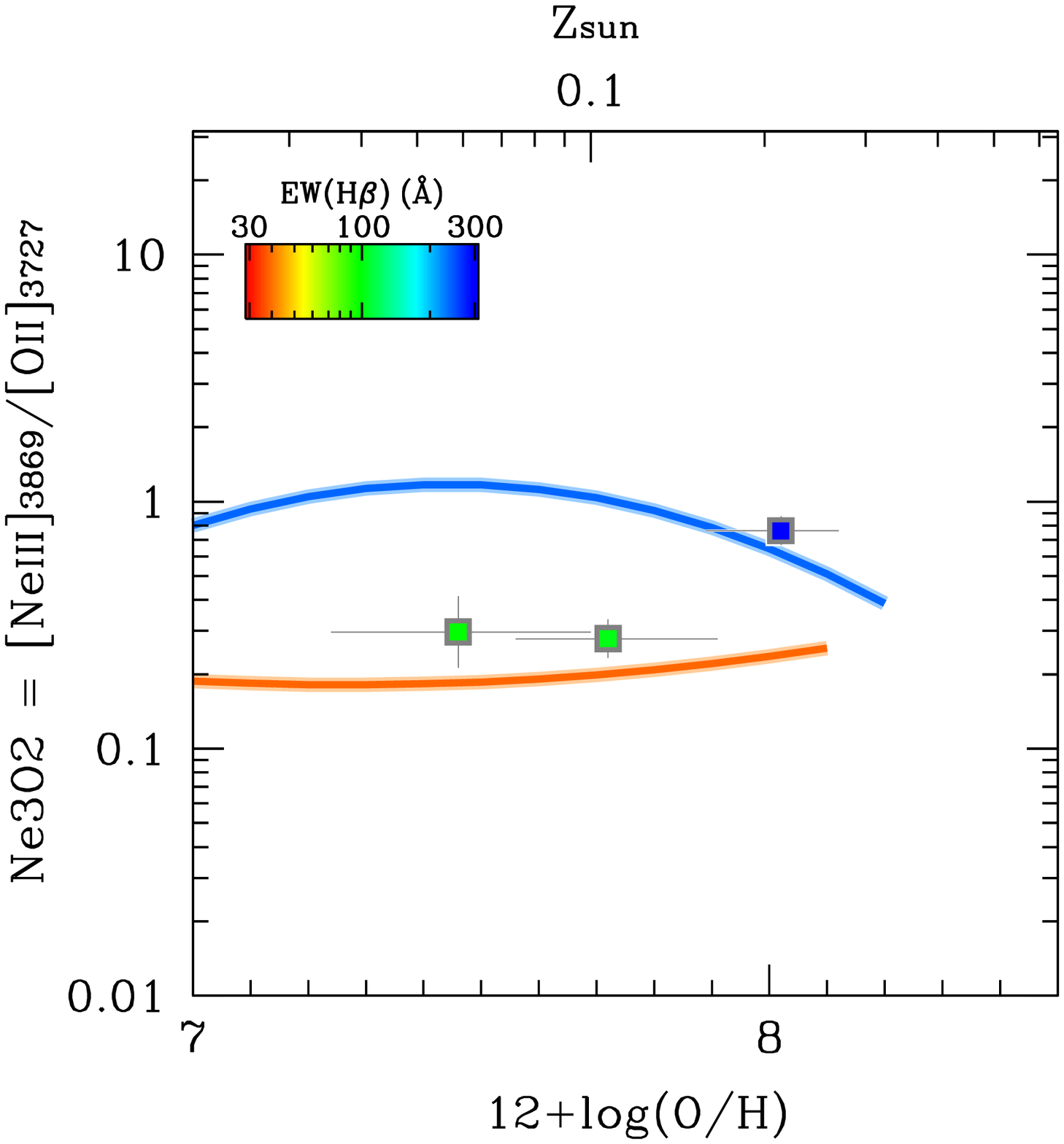}
        }

        \caption{%
		Relationships between metallicity and the strong line ratios
		of R23, R3, R2, N2, O32, and Ne3O2 (from top left to bottom right)
		and their dependence on EW(\Hb) 
		to be compared with observations of $z=2-3$ galaxies. 
		Squares show galaxies with a detection of \OIII$\lambda4363$
		\citep{christensen2012_laes,sanders2020}, and 
		triangles represent galaxies with \OIIIuv$\lambda\lambda1661,1666$
		\citep{erb2010,steidel2014}.
		The symbols are color-coded based on EW(\Hb) as shown in the legend.
		We assume the theoretical \OIII$\lambda5007/4959$ line ratio of $2.98$
		\citep{SZ2000} to correctly derive R23 and R3 as defined in this paper,
		if only the sum of the two lines or only the $5007$ component is listed 
		in the original literature. 
		The curves show the relationships derived in this paper; based on 
		the (E)MPG subsamples of Large EW(\Hb) (blue) and 
		Small EW(\Hb) (orange).
        }
        \label{fig:Z_empirical_highz}
    \end{center} 
\end{figure*}


Figure \ref{fig:Z_empirical_ewhb1} (Left) shows the same diagnostics of R23 and N2 
as in Figure \ref{fig:Z_empirical_all1} but our compiled EMPGs + MPGs are 
color-coded with EW(\Hb). 
To clarify the dependence on EW(\Hb), 
we split the sample into three subsamples, 
``Small EW'' ($<100$\,\AA),  
``Medium EW'' ($100$--$200$\,\AA), and 
``Large EW'' ($>200$\,\AA), 
and give the binned average results for the Small and Large EW subsamples 
in the right panels.
Note again that this paper explores the EW(\Hb) dependence only 
in the low metallicity regime (\Oabundance\ $\lesssim 8.0$).
We see a clear trend that the scatter in the indices is correlated with EW(\Hb).
For a given metallicity, sources with a larger EW(\Hb) tend to present a 
larger R23 and a smaller N2 value.
This can be interpreted as a result of large variation of ionization condition
in metal-poor clouds, 
i.e., the ionization parameter and/or the hardness of the ionizing spectrum
among galaxies with a comparable metallicity
\citep{pilyugin2016}.
The trend in N2-index is straightforward, while that of R23 needs more careful 
explanations because it uses a sum of singly and doubly ionized lines in the numerator.
Figures \ref{fig:Z_empirical_ewhb2} and \ref{fig:Z_empirical_ewhb3} summarize
the dependence of the other metallicity indicators on EW(\Hb).
We confirm the correlation between the scatters in the metallicity indicators 
and EW(\Hb), such that higher ionization lines (\OIII, \NeIII) tend to be stronger
with EW(\Hb) for a given metallicity than lower ionization lines (\OII, \NII, \SII) 
including the hydrogen recombination lines.
Because the binned average relationships of the Large EW sample are more similar 
to the earlier work showing relatively strong powers of these line ratios 
as metallicity indicators at the low metallicity range
(e.g., O32- and Ne3O2-index; \citealt{maiolino2008}), 
the weak metallicity-dependence we obtain for the ``ALL'' sample, 
in contrast to the earlier work, 
are thought to be due to the presence of objects with a low ionization condition.
The correlation between the metallicity and ionization state will be further 
discussed later (\S\ref{ssec:diagnostics_implications}).
As for R23-index, it is relatively strong against the variation of the 
ionization state by tracing both low and high ionization gas,
resulting in the smallest metallicity dispersion ($0.14$\,dex).
In the low metallicity regime, 
R23-index becomes dominated by R3-index as O32 becomes larger 
than unity. We thus see a similar trend of EW(\Hb) as seen in R3-index
but slightly milder in R23-index.

The binned average relationships of the metallicity indicators 
for each of the subsamples are summarized in Table \ref{tbl:Z_empirical_obs}, 
and their best-fit functions can be reproduced within the metallicity range 
and with the coefficients given in Table \ref{tbl:Z_empirical_bestfit}.
We fix $N=2$ for the best-fit functions for the subsamples.
Table \ref{tbl:Z_empirical_bestfit} also lists the standard deviations 
of \Oabundance\ as typical uncertainties for each of the indicators and subsamples.
For example, below \Oabundance\ $\leq 7.69$, the uncertainty of metallicity 
using R3-index (All) is 0.21\,dex, which becomes as small as 0.10\,dex 
(i.e., by a factor of 2) if the subsample's best-fits are employed. 
In a similar way, we confirm the accuracy of the metallicity indicators 
becomes significantly improved if the best-fit function 
depending on EW(\Hb) is used. 
Although the indicators using a high ionization line divided by a low ionization line,
O32, O3N2, and Ne3O2, show a smaller improvement in the standard deviation 
than the other indicators, 
it is mostly due to the larger impact of ionization condition on these line ratios
even within the subsamples of EW(\Hb) $>200$\,\AA\ and $<100$\,\AA\
(Fig.~\ref{fig:Z_empirical_ewhb3}). 
This is the current limitation.
Another practical note is that
we recommend to use the best-fit relationships of the Large and Small 
EW subsample for sources with EW(\Hb) $>200$\,\AA\ and $<100$\,\AA, respectively, 
and interpolate the two solutions for sources with EW(\Hb) in between $100$ and $200$\,\AA.

\subsection{Comparisons with photoionization models} 
\label{ssec:diagnostics_phoionizationmodels}

To confirm the variation of ionization condition on the strong line diagnostics,
we compare the results with photoionization models.
We exploit the \verb+CLOUDY+ photoionization modeling as detailed in 
\citet{nakajima2018_vuds}.
Briefly, we refer to the models of star-forming galaxies using binary evolution
SEDs of ``BPASS-300bin'' (v2; \citealt{eldridge2017}). 
For this comparison, we slightly expand the original BPASS model grid of 
\citet{nakajima2018_vuds} to cover the extremely low metallicity down below 
the $0.01$\,\Zsun\ adopting the same assumptions 
such as the abundance ratios and the relationship between stellar
and gas-phase metallicities. 
The ionization parameter is varied to assess the impact of changing 
the ionization condition on the emergent strong line ratios.
The gas density is fixed to a fiducial value of $100$\,cm$^{-3}$.
One caveat here is that the metallicity in the original \verb+CLOUDY+
models counts elements depleted onto dust grains in addition to those in gas-phase
in the ionized nebula. On the other hand, the metallicities observationally derived
based on the direct $T_e$ method trace the oxygen abundances only in gas-phase
(O$^{+}$/H$^{+}$ + O$^{2+}$/H$^{+}$ + O$^{3+}$/H$^{+}$; 
see \ref{sssec:samples_empress_new}).
To directly compare the models with the observations, we subtract the depleted 
component from the total oxygen abundance for each of the models 
using the adopted depletion factor of $0.6$ for oxygen.

In Figures \ref{fig:Z_empirical_model1} and \ref{fig:Z_empirical_model2}, 
we present the comparisons of the metallicity diagnostics 
between the photoionization predictions and the observations. 
The models reproduce the scatters as well as the trend observed with EW(\Hb)
for a given metallicity by changing the ionization parameter
within a reasonable range of $\log\,U$ from $-3.5$ to $-0.5$.
The comparison demonstrates that we can substitute EW(\Hb) as a gauge 
of the degree of ionization in ISM, and that the metallicity indicators that do not use
a sum of low and high ionization lines can be weak without any correction
for the ionization state, as they cannot trace both low and high ionization gas, 
latter of which particularly exist widely in metal-poor clouds.

\subsection{Implications} 
\label{ssec:diagnostics_implications}

The non-negligible correlations seen in the metallicity indicators in terms of EW(\Hb),
as a result of the difference in the degree of ionization state in the ISM, 
have some implications.
First, it would be important to adopt the appropriate relationship 
following the EW of \Hb\ (or \Ha), 
rather than using the best-fit for the ``All'' sample blindly, 
to improve the accuracy of the derived metallicity based on these
empirical relationships.
Particularly, this would influence discussions of mass-metallicity relation and 
its star-formation rate dependency (e.g., \citealt{mannucci2010,lara-lopez2010}), 
because EW(\Hb) is a strong function of stellar mass and star-formation rate. 
Although R23-index is the best indicator which is the least affected 
by the difference of the ionization state among various popular indicators, 
there still remains a factor of $\sim 1.5$ difference in R23-index
between the Large and Small EW subsamples
for a given metallicity below \Oabundance\ $\lesssim 8.0$.
This corresponds to a systematic offset of metallicity as large as 
$\Delta \log ({\rm O/H})$ $\sim 0.25$\,dex for a given R23 value 
between the large ($>200$\,\AA) and small ($<100$\,\AA) EW(\Hb) objects%
\footnote{
When using R23-index (i.e., both \OII\ and \OIII\ are available), 
the effect of ionization parameter can be more directly mitigated with 
the excitation parameter or O32-index \citep{pilyugin2016}.
}.
A worse case is to \textit{blindly} use a ratio of high to low ionization emission line
such as O32, O3N2, and Ne3O2 indicators which show a weak dependence
to estimate the metallicity below \Oabundance\ $\lesssim 8.0$
especially if EW(\Hb) is not available (Figure \ref{fig:Z_empirical_all2}).

Another important practical caveat when using R23-index and R3-index is that 
two solutions would be arithmetically obtained for a given index value.
One would need additional single-valued functions such as N2- and S2-index 
to resolve the degeneracy,
and be recommended to correct for the ionization condition using our prescription 
if the low-metallicity value (\Oabundance\ $\lesssim 8.0$) is the likely solution.
O32-index can also provide a discriminatory power to prefer the low-metallicity 
solution if O32-index is larger than $\sim 3$. 
Moreover, the two solution nature of R23-index and R3-index results in 
the plateau region around \Oabundance\ $\sim 8.0\pm 0.2$ 
which results in a difficulty in pinpointing the metallicity 
when the index value is close to the maximum value of $\sim 9$.
Indeed, the uncertainties of metallicity indicator of R23-index and R3-index
(EW-corrected) are $0.13-0.25$\,dex and $0.18-0.30$\,dex, respectively,
in the metallicity range of \Oabundance\ $\leq 8.1$ 
(Table \ref{tbl:Z_empirical_bestfit}).
These values get significantly smaller, $0.09-0.12$\,dex and $0.10-0.18$\,dex,
in the metallicity range of \Oabundance\ $\leq 7.69$, i.e., only considering
the low-metallicity branch apart from the plateau region.
The single-valued functions (e.g., N2- and S2-index), 
with a correction of ionization condition,
are suggested to be used together with R23-index and R3-index 
to help ease the difficulty around the plateau region.

A second implication is about the large variation found in the indicators 
using low-ionization lines and high-to-low ionization line ratios.
As clarified in the comparison plot of photoionization models and O32, O3N2, and Ne3O2-index
(Figure \ref{fig:Z_empirical_model2}),
the variation suggests a diverse ionization nature of ISM in metal-poor galaxies.
This appears in contrast to the SDSS's high-metallicity regime where 
a relatively tight anti-correlation exists between metallicity and ionization
parameter (e.g., \citealt{AM2013, sanders2020}).
Because the SDSS stacking in \citet{curti2017} is performed by dividing the 
sample based on the location on the R3 vs.~R2 plot, 
it implicitly follows the anti-correlation between the ionization 
parameter and metallicity typically found in the local universe,
resulting in a tight relationship between metallicity and O32, O3N2, and Ne3O2-index
\citep{curti2017}.
This is why the Large EW subsample looks more smoothly connected with
the SDSS sample at \Oabundance\ $\sim 8.0$, as the highly-ionized galaxies
are preferentially included in the lowest-metallicity bins of SDSS. 
If a similar anti-correlation typically exists between the metallicity and ionization parameter 
in the low-metallicity regime, 
the indicators that present weak dependences on metallicity
such as O32, O3N2, and Ne3O2-index
would become more helpful as seen in the 
Large EW subsample and as proposed in the earlier work \citep{maiolino2008}.
At the moment, however, our compilation demonstrates that outliers do exist 
as found in the Small EW subsample, such as EMPGs with a modestly ionized ISM 
having a low EW(\Hb).
A presence of DIG can also play a role in scattering the low EW(\Hb) objects 
toward larger values of the low-ionization line indices \citep{zhang2017,sanders2017}.
Our prescription will be useful to alleviate these uncertainties and derive  
metallicities for objects including such outliers.
Because the Small EW subsample tends to contain individual 
stellar clumps and \HII-regions resolved in near-by galaxies, 
our current compilation may bias the sample following the detection of 
\OIII$\lambda 4363$. 
We would need a larger and more well-constructed (e.g., mass-limited) EMPG sample 
to address the typical ionization condition as a function of metallicity.
At the high-metallicity regime, we also expect some deviations from 
the typical ionization parameter -- metallicity relationship and hence
some improvements of diagnostics by using EW(\Hb)
(cf. \citealt{brown2016,cowie2016,sanders2017}).
However, this is beyond this paper's scope as we would need different analyses
including how the SDSS sample is binned and stacked.

Finally, the metallicity prescriptions with the EW(\Hb) dependence 
would be essential for high-redshift galaxy studies.
Because higher-$z$ galaxies are thought to present a higher ionization parameter
(e.g., \citealt{NO2014}), the typical relationships (i.e., without the EW(\Hb) correction)
constructed based on the local galaxies' ionization parameter -- metallicity relation
may cause a systematic uncertainty of metallicity at high-redshift.
To test the utility of the new metallicity diagnostics for high-redshift galaxies, 
we use the $z=2-3$ galaxies whose metallicity is determined with 
the electron temperature. 
As compiled in \citet{sanders2020}, we collect $8$ galaxies in total at $z=1.7$--$3.6$,
five of which present a detection of \OIII$\lambda 4363$ \citep{christensen2012_laes,sanders2020},
and the remaining three's electron temperatures are determined with
\OIIIuv$\lambda\lambda 1661,1666$ \citep{erb2010,steidel2014}.
Figure \ref{fig:Z_empirical_highz} shows the $z=2-3$ galaxies,
color-coded with EW(\Hb), superposed on the metallicity indicators 
found in the local universe.
Albeit with the small sample size, the $z=2-3$ galaxies appear to support 
the same dependence of the relationships on EW(\Hb).
The single galaxy having a very large EW(\Hb) ($340$\,\AA) notably follows the 
functions constructed with the Large EW(\Hb) subsample in the local universe
($>200$\,\AA). 
The other $7$ galaxies have modest values of EW(\Hb) ($80-170$\,\AA), and indeed
fall in between the Large and Small EW subsamples on the metallicity indicators' 
plots, with a tendency that the second largest EW(\Hb) galaxy ($170$\,\AA)
prefers the relationship of the Large EW(\Hb) subsample.
We admit the current sample size and the individual metallicity measurement uncertainties 
for the existing data-points do not permit a conclusive discussion. 
Still, we can argue
that we do not see any clear contradiction of the 
metallicity indicators and the EW(\Hb) dependence at different redshifts up to 
$z\sim 3$.

The result is also consistent with the tendency found in \citet{bian2018}.
The authors derive the empirical relationships between metallicity and strong line ratios 
for the typical local star-forming galaxies of SDSS as well as 
for analogs of $z\sim 2$ galaxies which are selected in the local universe 
but based on the offset location on the N2 vs. R3 plot 
(i.e., \NII\ BPT diagram)
as typically seen at $z\sim 2$ (see also \citealt{steidel2014,shapley2015}). 
Using stacked spectra for each of the $z=0$ and $2$ samples, the authors suggest
similar systematic offsets in the metallicity indicators between $z=0$ and $\sim 2$
as identified in this study. This makes sense as the $z=2$ analogous sample is 
constructed based on the elevated R3 value for a given N2-index, 
and preferentially contain galaxies with a higher ionization parameter and/or a harder 
ionizing spectrum for a given metallicity which can be characterized by a large EW(\Hb).
The evolution of galaxies on the \NII\ BPT diagram would thus be largely due to
the evolution of ionized ISM conditions and caused by the evolution of EW(\Hb)
with redshift
(see also the discussion of \citealt{reddy2018_mosdef} based on EWs of \OIII).

In brief, we can argue that a correction of ionization condition would be important 
in determining metallicities based on the strong line ratios, and the new metallicity 
prescriptions we develop using the local galaxies can be applicable even for 
high-redshift galaxies.
We believe our prescriptions using EW(\Hb) are practically useful as it is easy to get
as compared to ionization parameter and \xiion.
This work demonstrates careful applications of the strong line ratios are necessary to
discuss the chemical evolution of galaxies as a function of stellar mass and
star-formation activity.
Nevertheless, we understand the current sample size is
too small to confirm the applicability of the indicators at high-redshift.
The relationships at high-redshift as well as local universe need to be further tested 
and improved, if necessary, with the forthcoming large and sensitive spectroscopic surveys 
such as Subaru/PFS \citep{takada2014} and VLT/MOONS \citep{cirasuolo2014}.

\setcounter{table}{2}
\startlongtable
\begin{deluxetable}{llcr}
\tablecaption{Binned average relationships of line ratio as a function of 
metallicity based on Compiled (E)MPGs
\label{tbl:Z_empirical_obs}}
\tablewidth{0.99\columnwidth}
\renewcommand{\arraystretch}{1.4}
\tabletypesize{\scriptsize}
\tablehead{
\colhead{Flux ratio}
& \colhead{Sample}
& \colhead{12+log(O/H)}
& \colhead{log R}
}
\startdata
R23 & All & $6.94$ & $0.32 \pm 0.10$ \\
 & & $7.11$ & $0.45 \pm 0.11$ \\
 & & $7.27$ & $0.53 \pm 0.08$ \\
 & & $7.43$ & $0.73 \pm 0.10$ \\
 & & $7.59$ & $0.80 \pm 0.09$ \\
 & & $7.76$ & $0.88 \pm 0.08$ \\
 & & $7.92$ & $0.94 \pm 0.04$ \\
 & & $8.08$ & $0.97 \pm 0.06$ \\
\cline{2-4}
 & Large EW & $7.10$ & $0.51 \pm 0.09$ \\
 & & $7.33$ & $0.72 \pm 0.07$ \\
 & & $7.57$ & $0.89 \pm 0.05$ \\
 & & $7.81$ & $0.95 \pm 0.06$ \\
 & & $8.05$ & $1.00 \pm 0.04$ \\
\cline{2-4}
 & Medium EW & $6.99$ & $0.32 \pm 0.12$ \\
 & & $7.25$ & $0.43 \pm 0.06$ \\
 & & $7.51$ & $0.77 \pm 0.10$ \\
 & & $7.77$ & $0.89 \pm 0.06$ \\
 & & $8.03$ & $0.95 \pm 0.04$ \\
\cline{2-4}
 & Small EW & $7.09$ & $0.39 \pm 0.11$ \\
 & & $7.32$ & $0.62 \pm 0.12$ \\
 & & $7.55$ & $0.71 \pm 0.09$ \\
 & & $7.78$ & $0.82 \pm 0.07$ \\
 & & $8.01$ & $0.91 \pm 0.03$ \\
\cline{1-4}
R2 & All & $6.94$ & $-0.40 \pm 0.21$ \\
 & & $7.11$ & $-0.37 \pm 0.33$ \\
 & & $7.27$ & $-0.22 \pm 0.39$ \\
 & & $7.43$ & $-0.16 \pm 0.39$ \\
 & & $7.59$ & $-0.13 \pm 0.27$ \\
 & & $7.76$ & $-0.02 \pm 0.30$ \\
 & & $7.92$ & $0.11 \pm 0.12$ \\
 & & $8.08$ & $0.06 \pm 0.18$ \\
\cline{2-4}
 & Large EW & $7.10$ & $-0.72 \pm 0.32$ \\
 & & $7.33$ & $-0.56 \pm 0.13$ \\
 & & $7.57$ & $-0.44 \pm 0.20$ \\
 & & $7.81$ & $-0.34 \pm 0.25$ \\
 & & $8.05$ & $-0.01 \pm 0.10$ \\
\cline{2-4}
 & Medium EW & $6.99$ & $-0.40 \pm 0.09$ \\
 & & $7.25$ & $-0.36 \pm 0.19$ \\
 & & $7.51$ & $-0.03 \pm 0.10$ \\
 & & $7.77$ & $0.08 \pm 0.17$ \\
 & & $8.03$ & $0.14 \pm 0.13$ \\
\cline{2-4}
 & Small EW & $7.09$ & $-0.20 \pm 0.14$ \\
 & & $7.32$ & $0.06 \pm 0.23$ \\
 & & $7.55$ & $0.12 \pm 0.22$ \\
 & & $7.78$ & $0.24 \pm 0.11$ \\
 & & $8.01$ & $0.21 \pm 0.15$ \\
\cline{1-4}
R3 & All & $6.94$ & $0.07 \pm 0.15$ \\
 & & $7.11$ & $0.22 \pm 0.18$ \\
 & & $7.27$ & $0.27 \pm 0.14$ \\
 & & $7.43$ & $0.49 \pm 0.20$ \\
 & & $7.59$ & $0.61 \pm 0.13$ \\
 & & $7.76$ & $0.67 \pm 0.14$ \\
 & & $7.92$ & $0.74 \pm 0.07$ \\
 & & $8.08$ & $0.78 \pm 0.09$ \\
\cline{2-4}
 & Large EW & $7.10$ & $0.35 \pm 0.09$ \\
 & & $7.33$ & $0.58 \pm 0.08$ \\
 & & $7.57$ & $0.73 \pm 0.06$ \\
 & & $7.81$ & $0.80 \pm 0.07$ \\
 & & $8.05$ & $0.82 \pm 0.05$ \\
\cline{2-4}
 & Medium EW & $6.99$ & $0.09 \pm 0.15$ \\
 & & $7.25$ & $0.22 \pm 0.06$ \\
 & & $7.51$ & $0.59 \pm 0.11$ \\
 & & $7.77$ & $0.67 \pm 0.10$ \\
 & & $8.03$ & $0.74 \pm 0.07$ \\
\cline{2-4}
 & Small EW & $7.09$ & $0.10 \pm 0.19$ \\
 & & $7.32$ & $0.31 \pm 0.19$ \\
 & & $7.55$ & $0.44 \pm 0.18$ \\
 & & $7.78$ & $0.55 \pm 0.12$ \\
 & & $8.01$ & $0.69 \pm 0.04$ \\
\cline{1-4}
O32 & All & $6.97$ & $0.58 \pm 0.36$ \\
 & & $7.19$ & $0.61 \pm 0.46$ \\
 & & $7.40$ & $0.56 \pm 0.58$ \\
 & & $7.62$ & $0.69 \pm 0.42$ \\
 & & $7.84$ & $0.69 \pm 0.37$ \\
 & & $8.06$ & $0.71 \pm 0.23$ \\
\cline{2-4}
 & Large EW & $7.18$ & $1.06 \pm 0.29$ \\
 & & $7.57$ & $1.19 \pm 0.24$ \\
 & & $7.97$ & $0.91 \pm 0.23$ \\
\cline{2-4}
 & Medium EW & $7.08$ & $0.55 \pm 0.20$ \\
 & & $7.51$ & $0.65 \pm 0.23$ \\
 & & $7.95$ & $0.59 \pm 0.24$ \\
\cline{2-4}
 & Small EW & $7.17$ & $0.22 \pm 0.29$ \\
 & & $7.55$ & $0.32 \pm 0.35$ \\
 & & $7.93$ & $0.43 \pm 0.17$ \\
\cline{1-4}
N2 & All & $7.09$ & $-2.37 \pm 0.35$ \\
 & & $7.32$ & $-2.03 \pm 0.30$ \\
 & & $7.55$ & $-1.94 \pm 0.27$ \\
 & & $7.78$ & $-1.90 \pm 0.30$ \\
 & & $8.01$ & $-1.65 \pm 0.23$ \\
\cline{2-4}
 & Large EW & $7.17$ & $-2.51 \pm 0.34$ \\
 & & $7.55$ & $-2.20 \pm 0.30$ \\
 & & $7.93$ & $-2.04 \pm 0.45$ \\
\cline{2-4}
 & Medium EW & $7.29$ & $-2.00 \pm 0.04$ \\
 & & $7.56$ & $-1.98 \pm 0.18$ \\
 & & $7.83$ & $-1.81 \pm 0.24$ \\
\cline{2-4}
 & Small EW & $7.17$ & $-1.94 \pm 0.22$ \\
 & & $7.46$ & $-1.83 \pm 0.20$ \\
 & & $7.75$ & $-1.78 \pm 0.19$ \\
\cline{1-4}
O3N2 & All & $7.12$ & $2.57 \pm 0.48$ \\
 & & $7.41$ & $2.38 \pm 0.45$ \\
 & & $7.69$ & $2.59 \pm 0.38$ \\
 & & $7.98$ & $2.44 \pm 0.21$ \\
\cline{2-4}
 & Large EW & $7.26$ & $2.91 \pm 0.30$ \\
 & & $7.83$ & $2.90 \pm 0.39$ \\
\cline{2-4}
 & Medium EW & $7.36$ & $2.57 \pm 0.28$ \\
 & & $7.76$ & $2.48 \pm 0.31$ \\
\cline{2-4}
 & Small EW & $7.24$ & $2.20 \pm 0.42$ \\
 & & $7.68$ & $2.29 \pm 0.27$ \\
\cline{1-4}
S2 & All & $7.09$ & $-1.66 \pm 0.31$ \\
 & & $7.32$ & $-1.51 \pm 0.35$ \\
 & & $7.55$ & $-1.39 \pm 0.31$ \\
 & & $7.78$ & $-1.27 \pm 0.23$ \\
 & & $8.01$ & $-1.22 \pm 0.12$ \\
\cline{2-4}
 & Large EW & $7.17$ & $-1.88 \pm 0.27$ \\
 & & $7.55$ & $-1.72 \pm 0.19$ \\
 & & $7.93$ & $-1.55 \pm 0.23$ \\
\cline{2-4}
 & Medium EW & $7.29$ & $-1.57 \pm 0.12$ \\
 & & $7.56$ & $-1.31 \pm 0.17$ \\
 & & $7.83$ & $-1.21 \pm 0.13$ \\
\cline{2-4}
 & Small EW & $7.17$ & $-1.42 \pm 0.27$ \\
 & & $7.46$ & $-1.20 \pm 0.23$ \\
 & & $7.75$ & $-1.15 \pm 0.15$ \\
\cline{1-4}
Ne3O2 & All & $6.98$ & $-0.34 \pm 0.23$ \\
 & & $7.14$ & $-0.47 \pm 0.46$ \\
 & & $7.30$ & $-0.42 \pm 0.38$ \\
 & & $7.45$ & $-0.40 \pm 0.48$ \\
 & & $7.61$ & $-0.38 \pm 0.43$ \\
 & & $7.77$ & $-0.36 \pm 0.39$ \\
 & & $7.93$ & $-0.44 \pm 0.19$ \\
 & & $8.09$ & $-0.39 \pm 0.23$ \\
\cline{2-4}
 & Large EW & $7.10$ & $-0.03 \pm 0.30$ \\
 & & $7.33$ & $0.00 \pm 0.22$ \\
 & & $7.57$ & $0.07 \pm 0.25$ \\
 & & $7.81$ & $0.03 \pm 0.32$ \\
 & & $8.05$ & $-0.27 \pm 0.14$ \\
\cline{2-4}
 & Medium EW & $7.03$ & $-0.37 \pm 0.01$ \\
 & & $7.28$ & $-0.48 \pm 0.23$ \\
 & & $7.53$ & $-0.47 \pm 0.22$ \\
 & & $7.78$ & $-0.48 \pm 0.24$ \\
 & & $8.04$ & $-0.48 \pm 0.18$ \\
\cline{2-4}
 & Small EW & $7.13$ & $-0.76 \pm 0.28$ \\
 & & $7.35$ & $-0.60 \pm 0.26$ \\
 & & $7.57$ & $-0.73 \pm 0.33$ \\
 & & $7.79$ & $-0.74 \pm 0.17$ \\
 & & $8.01$ & $-0.56 \pm 0.12$ \\
\cline{1-4}
\enddata
\end{deluxetable}

\newpage

\begin{deluxetable*}{llcccccccc}[h]
\tablecaption{Coefficients for empirical metallicity diagnostics (Eq.~\ref{eq:Z_empirical})
\label{tbl:Z_empirical_bestfit}}
\tablewidth{0.99\columnwidth}
\tabletypesize{\scriptsize}
\tablehead{
\colhead{Flux ratio} &
\colhead{Sample}&
\colhead{$c_0$} &
\colhead{$c_1$} &
\colhead{$c_2$} &
\colhead{$c_3$} &
\colhead{$c_4$} &
\colhead{Range$^{(\dag)}$} &
\colhead{$\Delta\log R$$^{(\ddag)}$} &
\colhead{$\Delta\log({\rm O}/{\rm H})$$^{(\ddag)}$}
} 
\startdata
R23 & All & $0.515$ & $-1.474$ & $-1.392$ & $-0.274$ & -- & [6.9:8.9] & $0.10$ & $0.14$ \\
 & & & & & & & & $0.08$ ($0.10$) & $0.14$ ($0.13$) \\
 & Large EW & $0.866$ & $-0.515$ & $-0.463$ & -- & -- & [7.1:8.1] & $0.05$ ($0.06$) & $0.25$ ($0.09$) \\
 & Medium EW$^{(\star)}$ & $0.986$ & $-0.178$ & $-0.335$ & -- & -- & [7.0:8.0] & $0.07$ ($0.10$) & $0.14$ ($0.12$) \\
 & Small EW & $0.875$ & $-0.313$ & $-0.387$ & -- & -- & [7.1:8.0] & $0.08$ ($0.08$) & $0.13$ ($0.10$) \\
\hline
R2 & All & $0.429$ & $-1.044$ & $-4.586$ & $-4.117$ & $-1.145$ & [6.9:8.9] & $0.27$ & $0.38$ \\
 & & & & & & & & $0.30$ ($0.34$) & $0.43$ ($0.45$) \\
 & Large EW & $0.697$ & $1.327$ & $0.273$ & -- & -- & [7.1:8.1] & $0.21$ ($0.21$) & $0.28$ ($0.30$) \\
 & Medium EW$^{(\star)}$ & $0.362$ & $0.204$ & $-0.147$ & -- & -- & [7.0:8.0] & $0.16$ ($0.16$) & $0.23$ ($0.21$) \\
 & Small EW & $-0.033$ & $-0.732$ & $-0.510$ & -- & -- & [7.1:8.0] & $0.19$ ($0.21$) & $0.31$ ($0.26$) \\
\hline
R3 & All & $-0.277$ & $-3.182$ & $-2.832$ & $-0.637$ & -- & [6.9:8.9] & $0.16$ & $0.23$ \\
 & & & & & & & & $0.15$ ($0.18$) & $0.27$ ($0.21$) \\
 & Large EW & $0.628$ & $-0.660$ & $-0.522$ & -- & -- & [7.1:8.1] & $0.06$ ($0.06$) & $0.30$ ($0.10$) \\
 & Medium EW$^{(\star)}$ & $0.718$ & $-0.297$ & $-0.387$ & -- & -- & [7.0:8.0] & $0.10$ ($0.11$) & $0.22$ ($0.13$) \\
 & Small EW & $0.780$ & $-0.072$ & $-0.316$ & -- & -- & [7.1:8.0] & $0.15$ ($0.17$) & $0.18$ ($0.18$) \\
\hline
O32 & All & $-0.693$ & $-2.722$ & $-1.201$ & -- & -- & [7.0:8.9] & $0.39$ & $0.39$ \\
 & & & & & & & & $0.45$ ($0.51$) & $0.45$ ($0.45$) \\
 & Large EW & $-0.080$ & $-2.008$ & $-0.804$ & -- & -- & [7.2:8.0] & $0.25$ ($0.25$) & $0.38$ ($0.42$) \\
 & Medium EW$^{(\star)}$ & $0.344$ & $-0.525$ & $-0.250$ & -- & -- & [7.1:8.0] & $0.23$ ($0.19$) & $0.36$ ($0.35$) \\
 & Small EW & $0.865$ & $0.771$ & $0.243$ & -- & -- & [7.2:7.9] & $0.31$ ($0.35$) & $0.48$ ($0.56$) \\
\hline
N2 & All & $-0.482$ & $1.052$ & $-3.979$ & $-5.479$ & $-1.904$ & [7.1:8.9] & $0.24$ & $0.40$ \\
 & & & & & & & & $0.29$ ($0.28$) & $0.52$ ($0.50$) \\
 & Large EW & $-1.309$ & $0.826$ & $0.014$ & -- & -- & [7.2:7.9] & $0.31$ ($0.28$) & $0.38$ ($0.35$) \\
 & Medium EW$^{(\star)}$ & $0.375$ & $3.642$ & $1.358$ & -- & -- & [7.3:7.8] & $0.20$ ($0.15$) & $0.15$ ($0.15$) \\
 & Small EW & $-2.181$ & $-0.943$ & $-0.532$ & -- & -- & [7.2:7.8] & $0.19$ ($0.20$) & $0.35$ ($0.33$) \\
\hline
O3N2 & All & $0.226$ & $-4.710$ & $-2.138$ & -- & -- & [7.1:8.9] & $0.42$ & $0.34$ \\
 & & & & & & & & $0.49$ ($0.51$) & $0.44$ ($0.43$) \\
 & Large EW & $2.126$ & $-1.229$ & $-0.456$ & -- & -- & [7.3:7.8] & $0.34$ ($0.32$) & $0.38$ ($0.39$) \\
 & Medium EW$^{(\star)}$ & $0.033$ & $-4.540$ & $-2.020$ & -- & -- & [7.4:7.8] & $0.29$ ($0.23$) & $0.37$ ($0.37$) \\
 & Small EW & $2.786$ & $0.396$ & $-0.043$ & -- & -- & [7.2:7.7] & $0.30$ ($0.32$) & $0.40$ ($0.38$) \\
\hline
S2 & All & $-0.452$ & $-0.297$ & $-5.262$ & $-5.881$ & $-1.882$ & [7.1:8.9] & $0.24$ & $0.40$ \\
 & & & & & & & & $0.30$ ($0.31$) & $0.50$ ($0.48$) \\
 & Large EW & $-0.788$ & $1.097$ & $0.239$ & -- & -- & [7.2:7.9] & $0.20$ ($0.20$) & $0.35$ ($0.37$) \\
 & Medium EW$^{(\star)}$ & $-0.456$ & $0.959$ & $0.148$ & -- & -- & [7.3:7.8] & $0.14$ ($0.15$) & $0.22$ ($0.25$) \\
 & Small EW & $-1.561$ & $-1.066$ & $-0.635$ & -- & -- & [7.2:7.8] & $0.21$ ($0.23$) & $0.29$ ($0.29$) \\
\hline
Ne3O2 & All & $-0.317$ & $0.161$ & $0.070$ & -- & -- & [7.0:8.1] & $0.39$ & $0.66$ \\
 & & & & & & & & $0.39$ ($0.44$) & $0.66$ ($0.60$) \\
 & Large EW & $-1.240$ & $-2.106$ & $-0.846$ & -- & -- & [7.1:8.1] & $0.25$ ($0.26$) & $0.36$ ($0.40$) \\
 & Medium EW$^{(\star)}$ & $-0.527$ & $-0.059$ & $-0.002$ & -- & -- & [7.0:8.0] & $0.23$ ($0.19$) & $0.58$ ($0.54$) \\
 & Small EW & $-0.306$ & $0.613$ & $0.216$ & -- & -- & [7.1:8.0] & $0.28$ ($0.31$) & $0.56$ ($0.63$) \\
\hline
\enddata
\tablecomments{%
The polynomial order is either $N=4$, $3$, or $2$ which is determined 
for each of the indices to minimize the dispersions. 
The subsamples (\Oabundance\ $\lesssim 8.0$) adopt a fixed order $N=2$.
No value is listed in the coefficient(s) of $c_4$ ($c_4$ and $c_3$) 
if the order $N=3$ ($N=2$) is chosen.
$(\dag)$ The range of \Oabundance\ used for deriving the best-fit.
$(\ddag)$ The $1\sigma$ logarithmic uncertainty (standard deviation) of line ratio for a given \Oabundance\ ($\Delta\log R$),
and that of \Oabundance\ for a given line ratio ($\Delta\log({\rm O}/{\rm H})$) 
calculated over the range: $(\dag)$.
For the All sample of each of the indices, 
the second row lists the uncertainties calculated solely with 
the MPGs + EMPGs (i.e., omitting the high-metallicity SDSS stacks
as done for the subsamples' best-fit).
The values in the round brackets give the uncertainties 
calculated only with EMPGs (\Oabundance\ $\leq 7.69$).
$(\star)$ These indicators' functions for the Medium EW subsample are not well behaved,
particularly for N2-, O3N2-, and S2-index, 
due to the small sample size at the moment. 
It would be rather recommended to interpolate the best-fits of the Large and the Small EW 
subsamples if these indicators are to be used for sources with EW(\Hb) $=100-200$\,\AA.
}
\end{deluxetable*}

\begin{figure*}[t]
    \begin{center}
        \subfloat{
            \includegraphics[bb=38 163 412 496, width=0.45\textwidth]{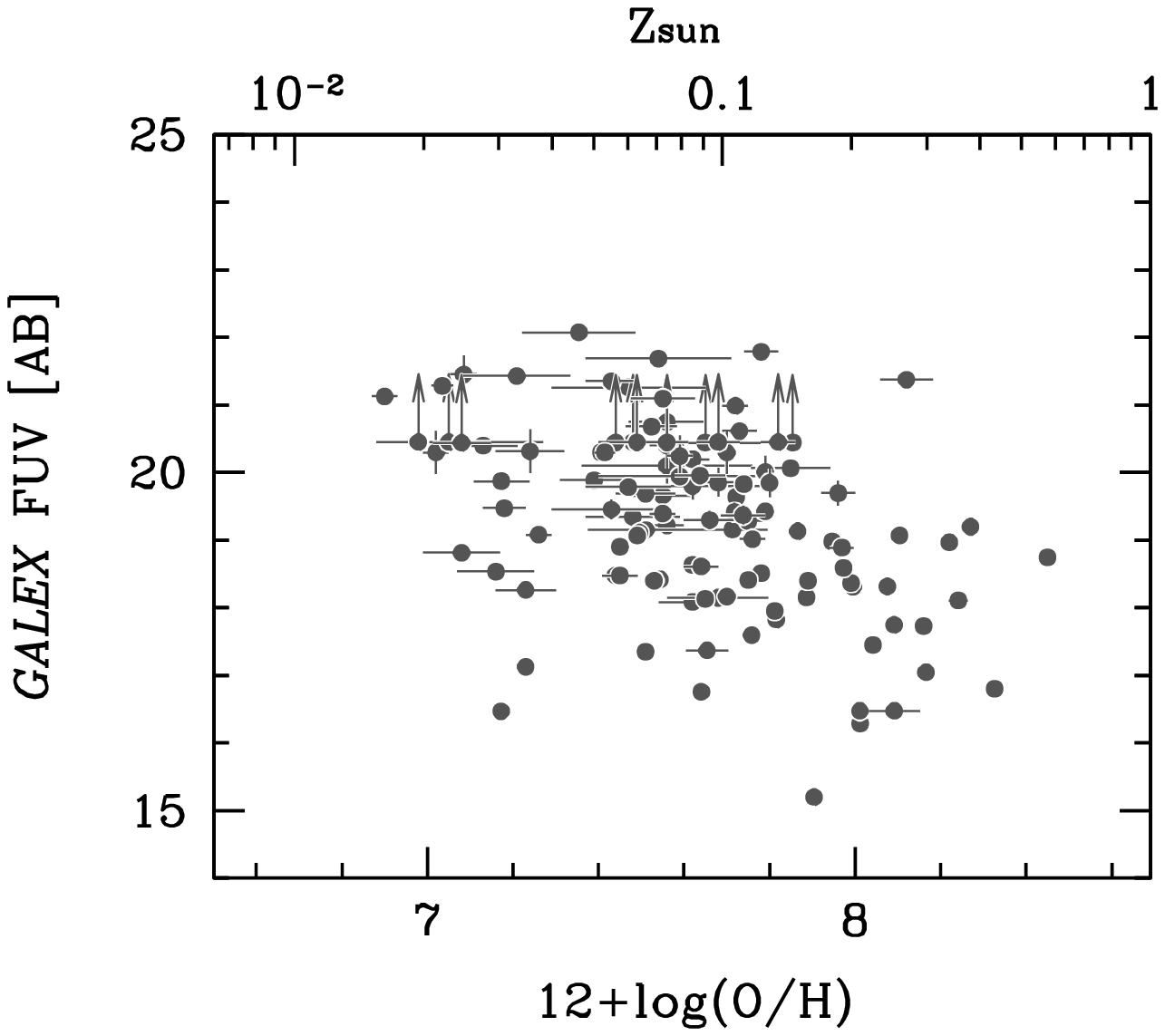}
        }
        \hspace{0.05\textwidth}
        \subfloat{
            \includegraphics[bb=38 163 412 496, width=0.45\textwidth]{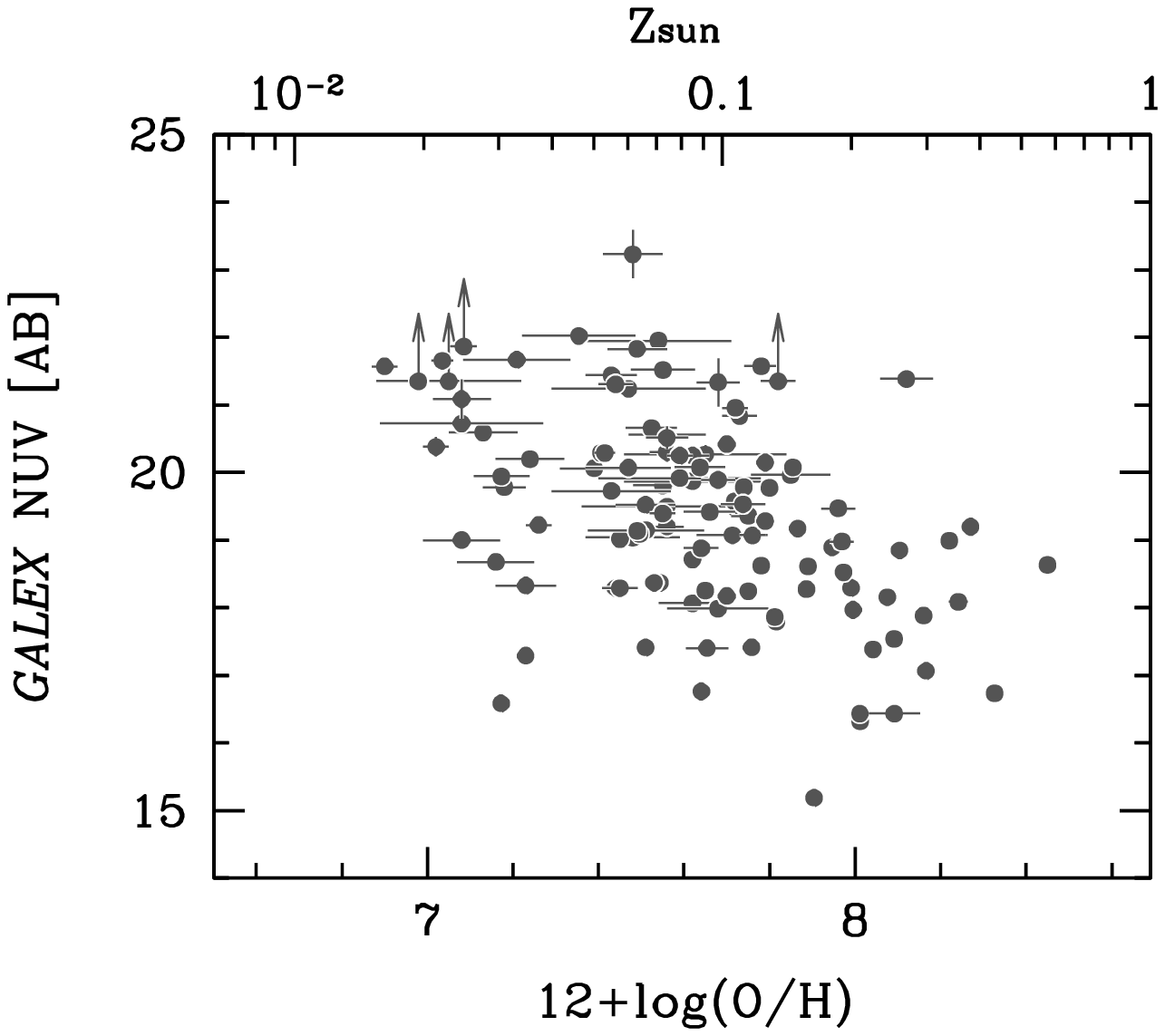}
        }
        \caption{%
		The distributions of \textsl{GALEX} FUV- (left) and NUV-band (right) photometry 
		for the compiled EMPGs and MPGs as a function of metallicity.
		The $3\sigma$ upper-limits are adopted on the photometry for the sources
		without a significant detection.
        }
        \label{fig:galex_Z}
    \end{center} 
\end{figure*}


\section{UV Properties of EMPGs} \label{sec:UV_properties}

The EMPGs compiled in this study also provide important anchors of properties 
in the early phase of galaxy evolution
that will be useful for high-redshift galaxy studies. 
In this section, we present the properties of EMPGs in the rest-frame
UV wavelength to gain insights into the early star-formation and 
production of ionizing photons.

\subsection{GALEX data} \label{ssec:UV_properties_data}

In order to characterize the UV properties of (E)MPGs in the local universe, 
we utilize the FUV ($\lambda_{\rm eff} \sim 1540$\,\AA) 
and NUV ($\lambda_{\rm eff} \sim 2320$\,\AA) 
photometric data taken by \textsl{GALEX} \citep{morrissey2007}.
The data are collected from the Mikulski Archive for Space Telescopes (MAST) portal%
\footnote{
\url{https://mast.stsci.edu}
}.
For each of the objects in our compiled sample (\S\ref{ssec:samples_empgs}) 
we search for the deepest NUV and FUV imaging data 
that are available at the spatial position of the object. 
Eight objects are not covered with any pointings of \textsl{GALEX}.
For the remaining $177$ objects, we carefully check the downloaded 
\textsl{GALEX} images to remove sources 
that are highly contaminated by the neighboring objects 
due to the low image resolutions 
(FWHM $\sim 4^{\prime\prime}$ in FUV and $\sim 5^{\prime\prime}$ in NUV).
This procedure is particularly important for the nearby stellar clumps/\HII\ regions 
where multiple clumps are found in a galaxy, as well as
EMPGs that are associated with a bright extended tail (see e.g., \citealt{isobe2021_tail}). 
By comparing with the higher-resolution optical images, we label $62$ objects,
all of which are nearby stellar clumps, as blended in the \textsl{GALEX} images. 
The $70$ ($=8+62$) objects are not used in the analyses of UV properties
(but used in the analyses of emission line ratios; \S\ref{sec:diagnostics}).

\begin{deluxetable}{lcccccc}[t]
\tablecaption{\textsl{GALEX} Imaging Surveys 
\label{tbl:galex_details}}
\tabletypesize{\scriptsize}
\tablehead{
\colhead{} &
\colhead{} &
\multicolumn{2}{c}{FUV} &
\colhead{} &
\multicolumn{2}{c}{NUV} \\
\cline{3-4} 
\cline{6-7}
\colhead{Survey} &
\colhead{} &
\colhead{$N^o$} &
\colhead{Depth$^{(\dag)}$} &
\colhead{} &
\colhead{$N^o$} &
\colhead{Depth$^{(\dag)}$} 
} 
\startdata
AIS & & $53$ & $20.45$ & & $40$ & $21.35$  \\
MIS & & $34$ & $23.25$ & & $42$ & $23.25$  \\
DIS & & $3$ & $25.35$ & & $3$ & $24.95$  \\
GII & & $17$ & $23.4$ & & $22$ & $23.5$  \\
NGS & & $8$ & $23.6$ & & $8$ & $23.7$  \\
\enddata
\tablecomments{%
($\dag$) Limiting magnitude at the $3\sigma$ level.
The depths for AIS, MIS, and DIS are given in \citep{morrissey2007},
and those for GII and NGS are estimated by random aperture photometry.
}
\end{deluxetable}

The \textsl{GALEX} imaging observations from five types of surveys are used
for the $115$ objects. These surveys are the All-sky Imaging Survey (AIS), 
the Medium Imaging Survey (MIS), the Deep Imaging Survey (DIS), 
the Guest Investigators Survey (GII), and the Nearby Galaxy Survey (NGS).
The numbers of sources and the typical depths for each surveys and for each bands 
are given in Table \ref{tbl:galex_details}.
We use the SExtractor software \citep{BA1996} to perform source detection and 
photometry. 
We identify sources with five adjoining pixels and brightness above $>2\sigma$
of the background, and then cross-match with each of our sources to find 
a counterpart in the FUV/NUV images within $5^{\prime\prime}$ ($\simeq$ FWHM) 
from the source position.
We adopt \verb+MAG_AUTO+ for the total magnitude if the cross-matched object
is brighter than the $3\sigma$ limiting magnitude.
We find a high detection rate ($>3\sigma$) in the \textsl{GALEX} images
($103/115$ in FUV  and $111/115$ in NUV) for the compiled (E)MPGs.
The magnitudes are corrected for Galactic extinction in the same way as detailed
in \S\ref{sssec:samples_empress_new}.
Figure \ref{fig:galex_Z} presents the distributions of \textsl{GALEX} FUV and NUV
photometry for our sources ($N^o=115$) as a function of metallicity.

In the \textsl{GALEX} FUV and NUV photometric bands, the UV emission lines 
such as \CIV$\lambda 1549$, \HeII$\lambda 1640$, and \CIII$\lambda 1909$
stay and can contribute to the observed \textsl{GALEX} photometry.
Nevertheless, the UV emission lines would not have a significant impact on the 
photometry and the resulting UV properties below. 
Even the strongest UV emission lines present the maximum EWs as large as 
$\sim 20$\,\AA\ for \CIII\ and $\sim 10$\,\AA\ for \CIV\ 
for star-formation dominated systems \citep{nakajima2018_vuds}.
Even with such extreme EWs, the photometry would be boosted 
by only a negligibly small amount
($\lesssim 0.03$\,mag for an average brightness galaxy ($\sim 19.3$\,mag) 
in the sample at $z=0.03$).
We therefore do not correct for any possible contribution of the UV emission lines
to the observed \textsl{GALEX} photometry.

\begin{figure*}[t]
    \begin{flushleft}
        \subfloat{
            \includegraphics[bb=40 163 567 494, width=0.55\textwidth]{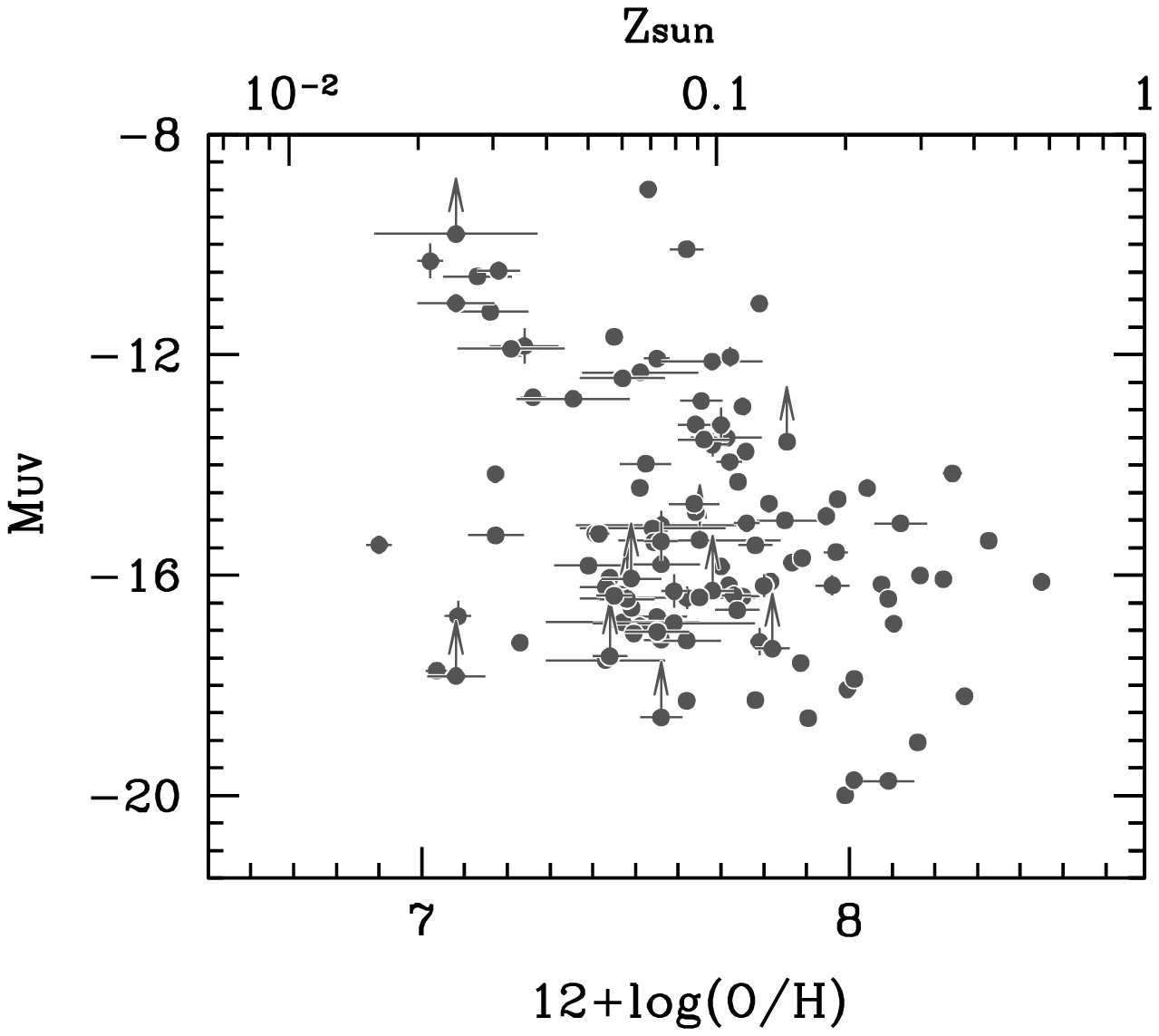}
        }
        \subfloat{
            \includegraphics[bb=190 163 717 494, width=0.55\textwidth]{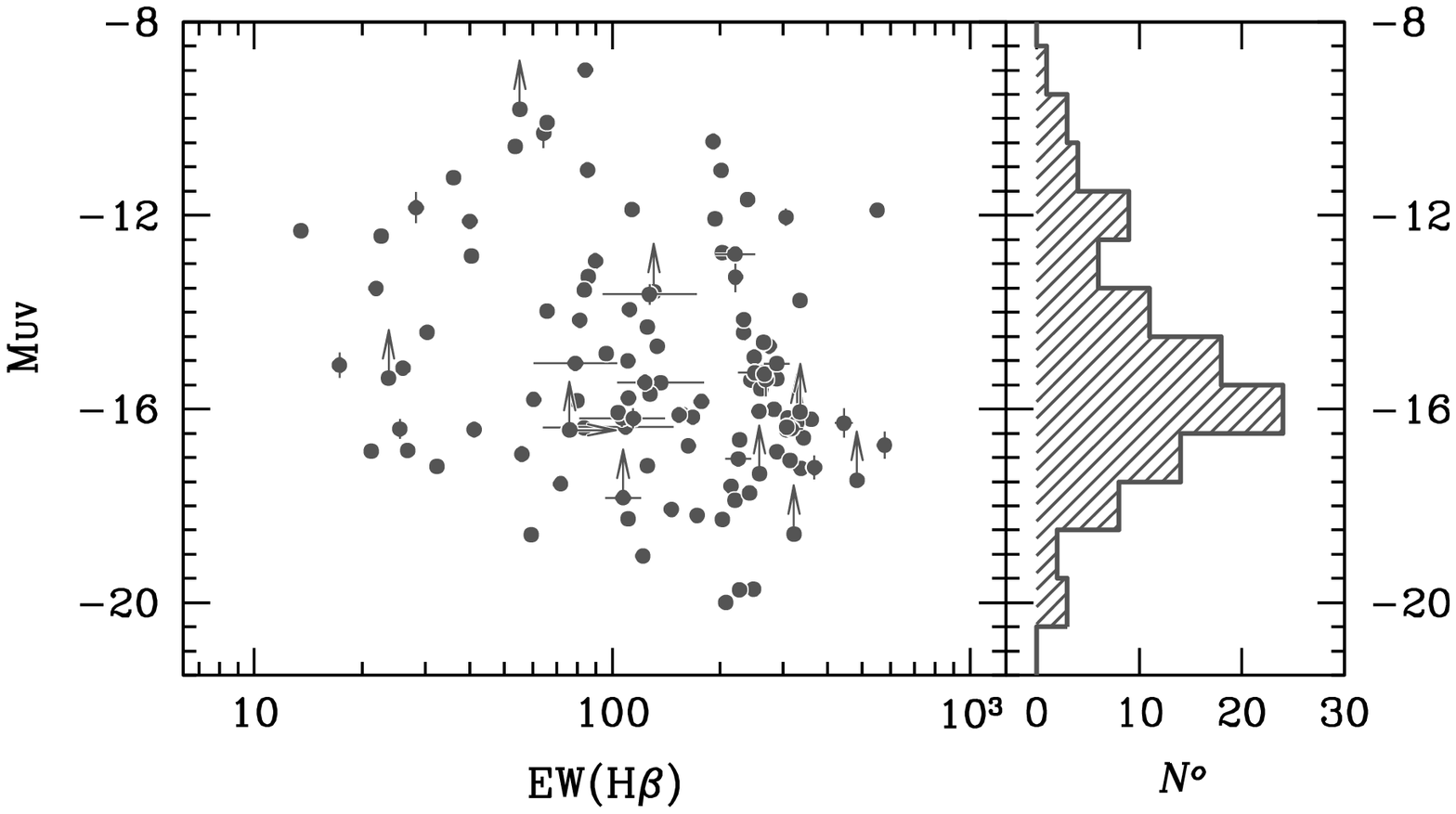}
        }
        
        \caption{%
        	UV absolute magnitude \Muv\ as a function of 
		metallicity (left) and EW(\Hb) (right),
	 	and its distribution (at the most right end; 
		the undetected sources are not included in the histogram)
		for the compiled (E)MPGs in this paper.
        }
        \label{fig:dist_Muv_ewhb_Z}
    \end{flushleft} 
\end{figure*}


\subsection{UV absolute magnitude \Muv} \label{ssec:UV_properties_Muv}

We start with deriving a key fundamental property of UV absolute magnitude, \Muv\
for the compiled metal-poor objects.
Here the absolute UV magnitudes are derived from the \textsl{GALEX} FUV band 
photometry, which probes the rest-frame $\sim 1500$\,\AA\ emission,
in addition to the luminosity distance.
We do not take into account any k-correction for the \Muv\ estimations 
as the sample is almost built at the similar redshift of $z< 0.05$ 
(Figure \ref{fig:dist_Mi_ewhb_Z_redshift}b).
Moreover, the errors in the luminosity distance (or redshift) are not available 
for the compiled sample, and not included in the \Muv\ calculation.
A correction for dust reddening is applied according to the degree of attenuation 
for nebular emission. We utilize the Balmer decrements and the \citet{calzetti2000}'s
attenuation curve to obtain \ebv\ for the nebular emission, 
divide it by $0.44$ for stellar emission \citep{calzetti2000}, 
and correct for the reddening of the FUV stellar light.
The corrections are generally small for the metal-poor objects studied in this paper,
and our results are not significantly affected by 
the choice of the attenuation curve and the relationship of \ebv\  
between stellar and nebular emission.
Figure \ref{fig:dist_Muv_ewhb_Z} shows the distribution of \Muv\ for the compiled objects
as functions of metallicity and EW(\Hb).
The compiled sample extraordinarily reaches the faintest UV magnitude of \Muv\ $\sim -9$.

\begin{figure*}[t]
    \begin{center}     
        \subfloat{
            \includegraphics[bb=18 159 565 550, width=0.46\textwidth]{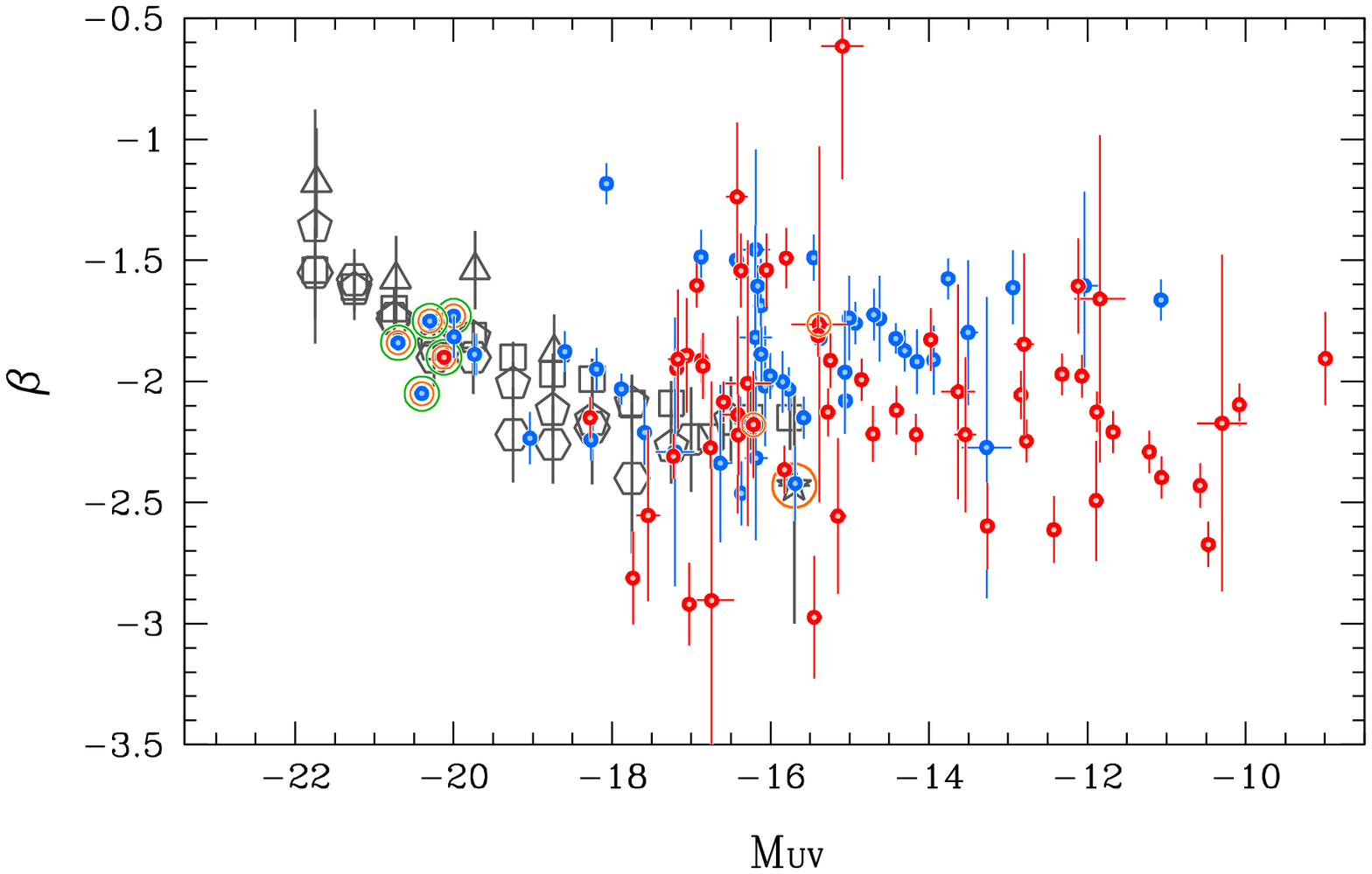}
        }
        \hspace{0.03\textwidth}
        \subfloat{
            \includegraphics[bb=18 159 565 550, width=0.46\textwidth]{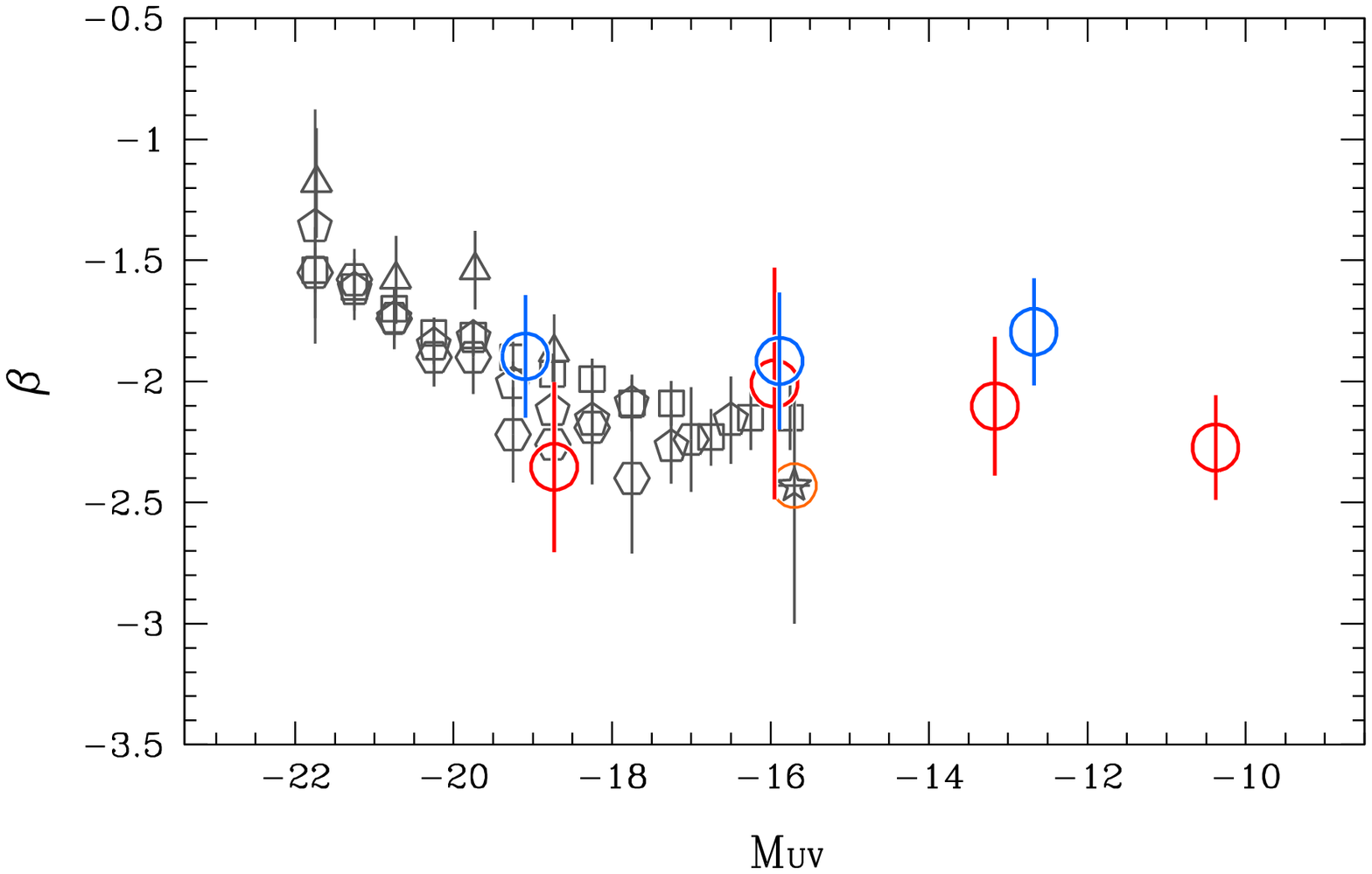}
        }
        
        \subfloat{
            \includegraphics[bb=18 159 565 550, width=0.46\textwidth]{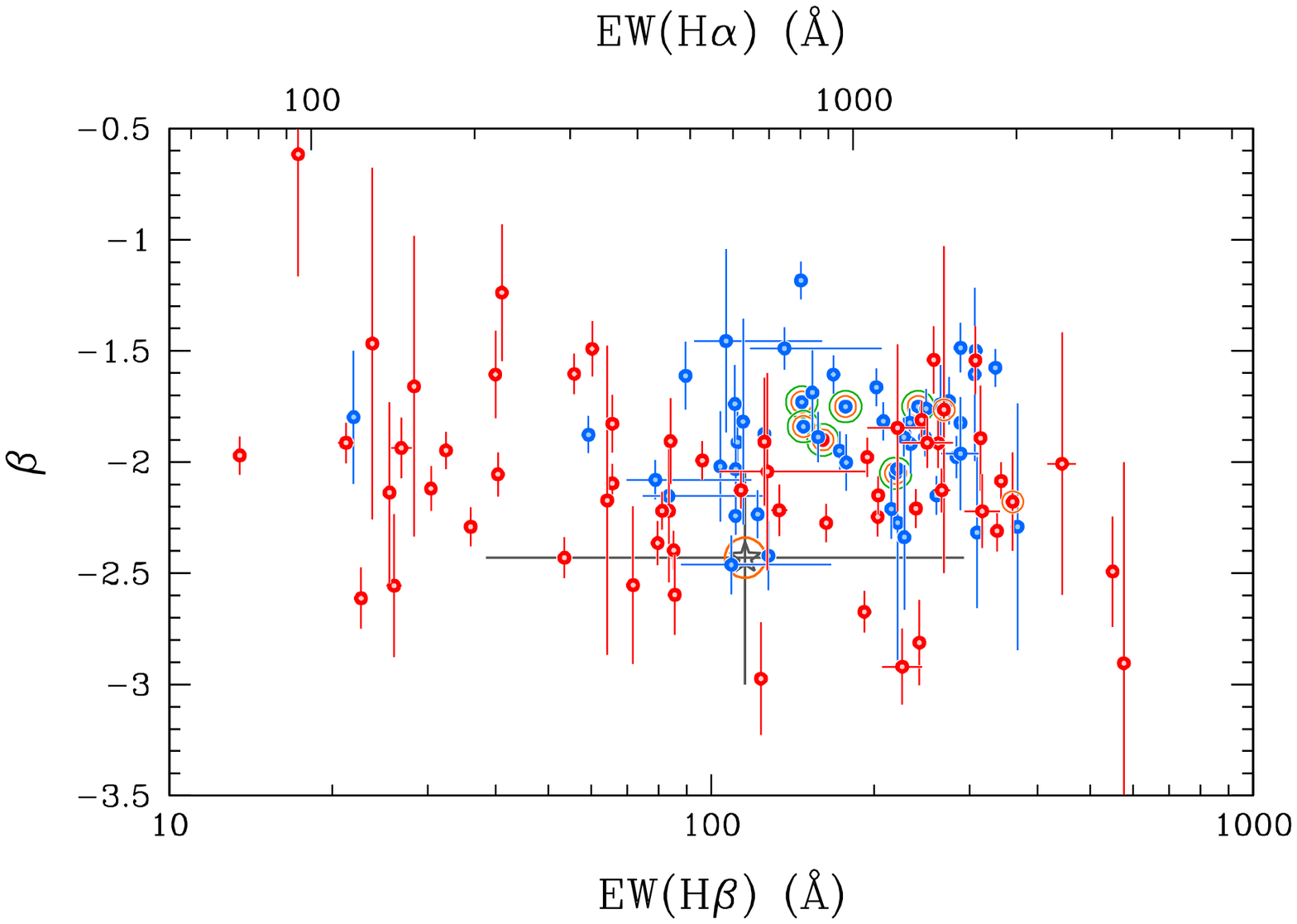}
        }
        \hspace{0.03\textwidth}
        \subfloat{
            \includegraphics[bb=18 159 565 550, width=0.46\textwidth]{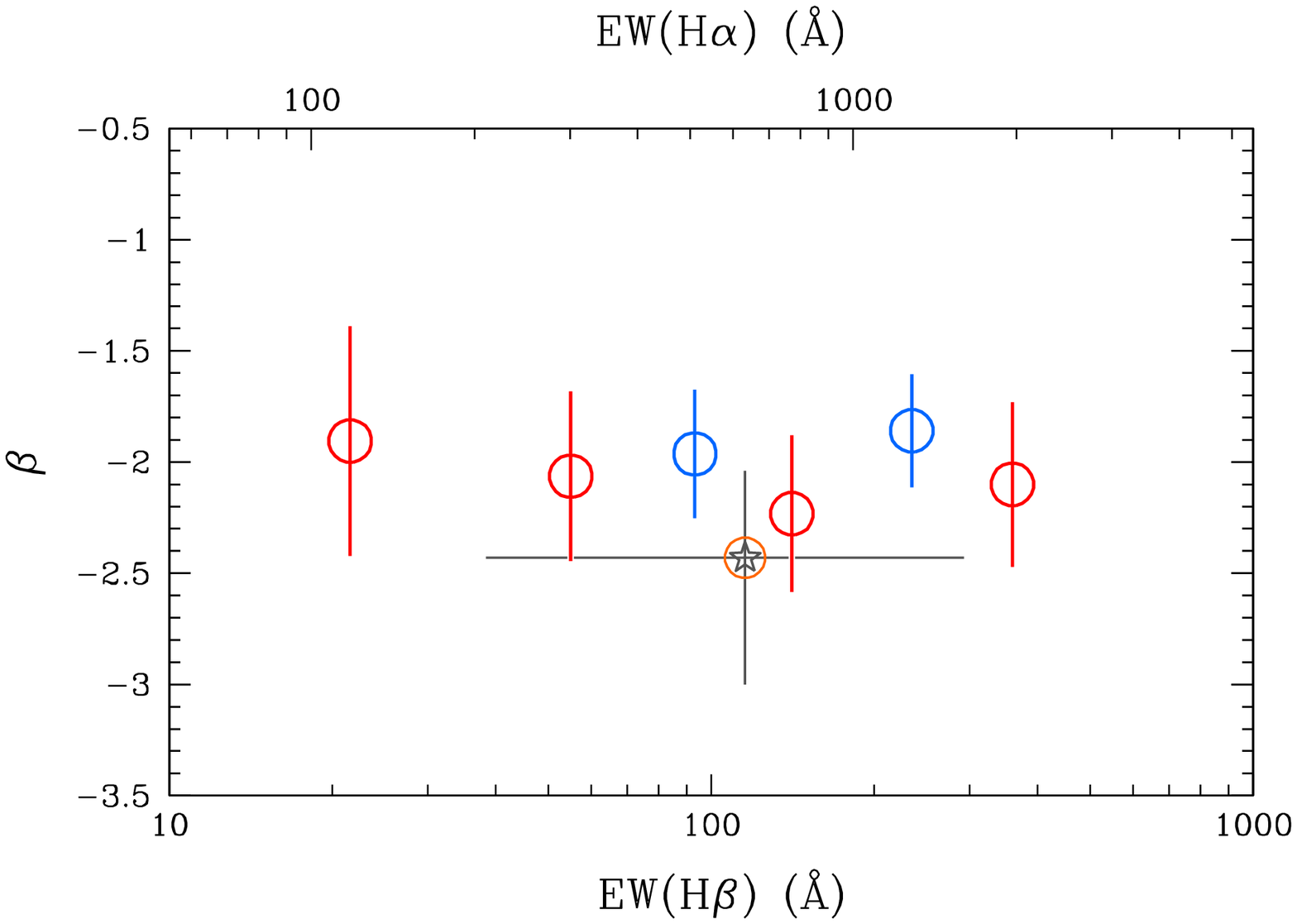}
        }
        
        \subfloat{
            \includegraphics[bb=18 159 565 550, width=0.46\textwidth]{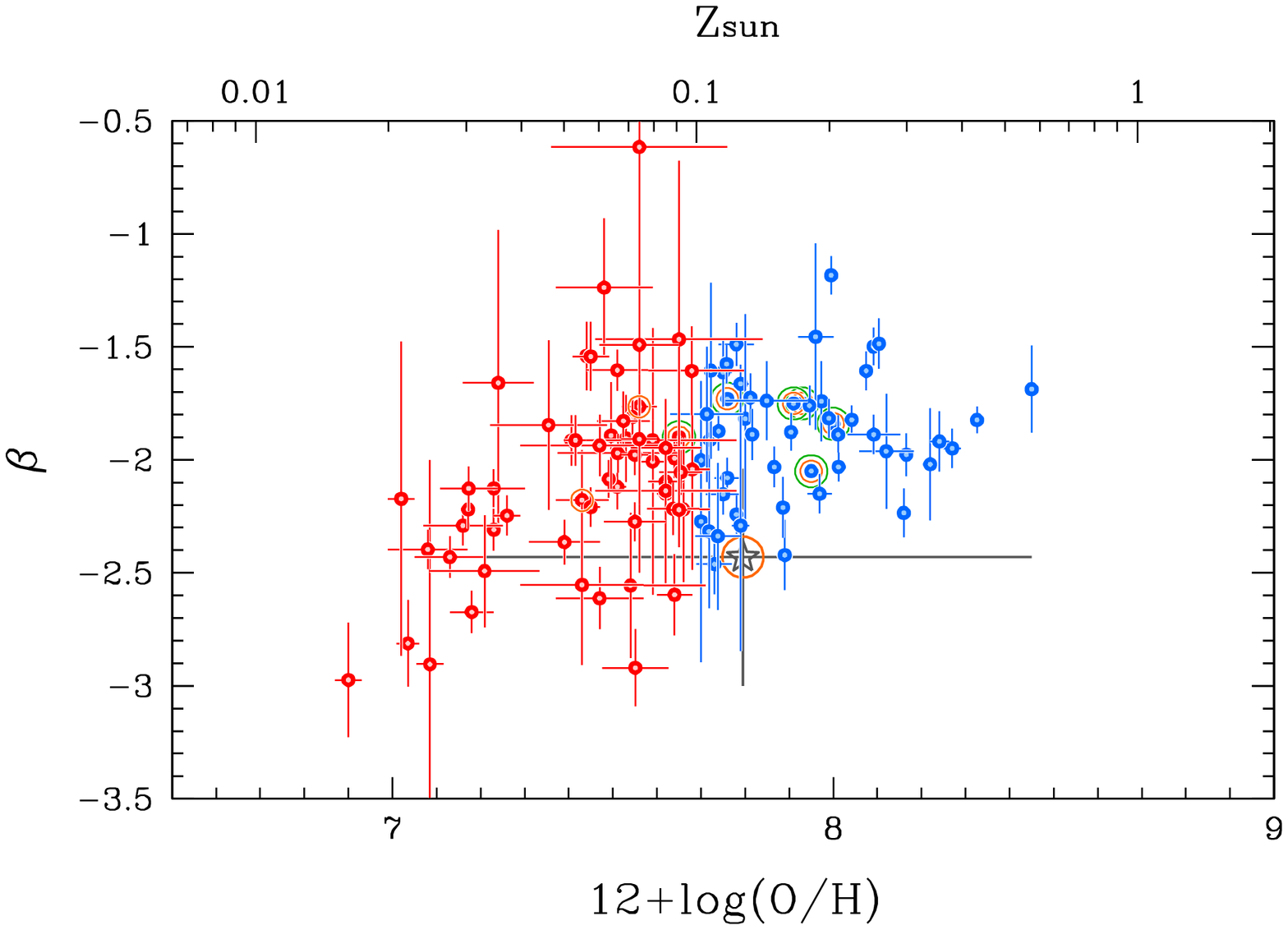}
        }
        \hspace{0.03\textwidth}
        \subfloat{
            \includegraphics[bb=18 159 565 550, width=0.46\textwidth]{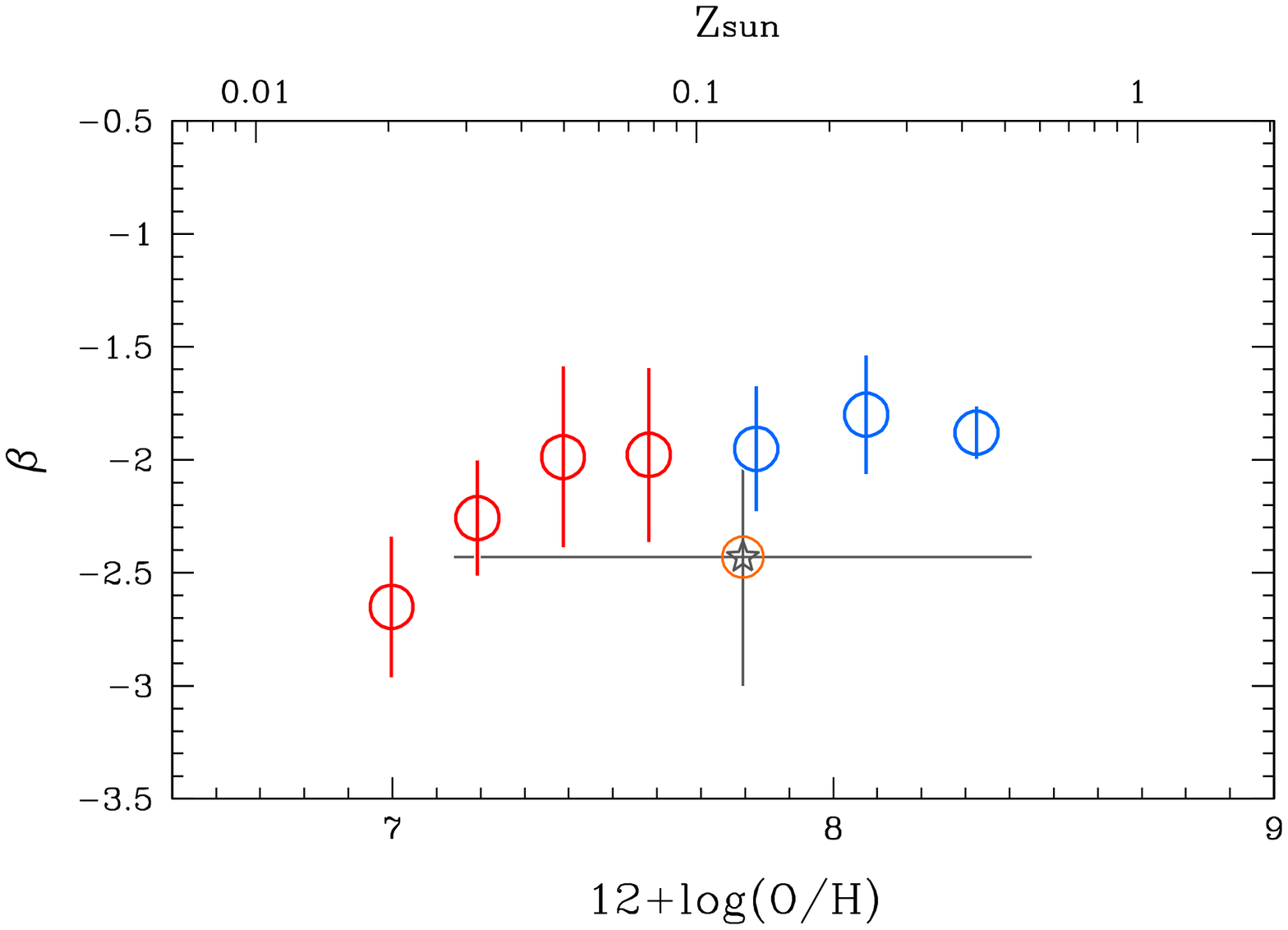}
        }
        
        \caption{%
		Relationships of UV slope $\beta$ as a function of 
		\Muv\ (top), EW(\Hb) (middle), and metallicity (bottom). 
		In the left panels, our compiled EMPGs (red small circles) and MPGs (blue small circles) 
		are presented on an individual basis,
		while the binned average relationships for EMPGs (red open circles) and MPGs 
		(blue open circles) are presented with larger symbols in the right panels
		using the individual data points shown in the left-hand panels.
		The errorbar represents the standard deviation of each of 
		the binned distribution.
		The points enclosed with a green circle 
		are $z\simeq 0.3$ green pea galaxies whose \fesc\ is directly measured 
		\citep{izotov2016_nature,izotov2016_4more}.
		The other large open symbols denote average relationships at higher-redshifts
		of $z\sim 2.5$ (triangles), $4$ (squares), $5$ (pentagons), and $6$ (hexagons)
		\citep{bouwens2009,bouwens2014}.
		The open star shows a stacked result of faint 
		LAEs at $z=4-5$ \citep{maseda2020}.
		Like the \citet{maseda2020}'s data-point, 
		the orange-markers represent galaxies whose \Lya\
		is observed to be strong (LAEs with EW(\Lya) $>20$\,\AA; see also \citealt{izotov2020}).
		For high-redshift galaxies whose EW(\Hb) is unknown, 
		we substitute EW(\Ha) using the empirical conversion of 
		EW(\Ha)$/$EW(\Hb) $=5.47$ \citep{kojima2020}.
        }
        \label{fig:beta_Muv_ewhb_Z}
    \end{center} 
\end{figure*}


\subsection{UV slope $\beta$} \label{ssec:UV_properties_beta}

Our next interest is to determine the rest-frame UV continuum slope $\beta$
($f_{\lambda}\propto\lambda ^{\beta}$) for EMPGs and examine the distribution 
of $\beta$ toward the lowest metallicity and faintest UV luminosity.
We estimate the UV continuum slope $\beta$ using 
two wavelength photometric points of \textsl{GALEX} FUV and NUV band.
We note again that the k-correction would be 
negligible in our $\beta$ measurements (\S\ref{ssec:UV_properties_Muv}).
The wavelength range of $\sim 1500$--$2300$\,\AA\ is consistent with those 
used for $\beta$ measurements in higher-redshift galaxies 
(e.g., \citealt{bouwens2009,bouwens2012}).
For each object we fit the FUV and NUV magnitudes with a power-law function 
at the measured redshift and determine $\beta$. 
We also repeat the fitting to randomly fluctuated photometry following the errors 
as Gaussian distributions and obtain the $1\sigma$ uncertainty of $\beta$ which 
encompasses $68$ percent of $\beta$ values drawn from the procedure.

Figure \ref{fig:beta_Muv_ewhb_Z} presents how the UV slopes of (E)MPGs are 
distributed as a function of absolute UV magnitude, EW(\Hb), and metallicity.
The left panels show the individual objects, while the right panels present
an average value of $\beta$ for a bin of the abscissa quantity,
with red symbols for EMPGs and blue for MPGs.
The relationship between $\beta$ and \Muv\ (top panels)
reveals that EMPGs overall present a small value of $\beta$ from $\sim -1.9$
to $\sim -2.3$ in the range of \Muv\ $= -18$ to $-10$ on average, 
mostly independent of the UV luminosity.
The reference sample of MPGs in the local universe covers a slightly bright but 
still faint range of \Muv\ $=-20$ to $-12$, showing a similarly flat relationship of $\beta$
as a function of \Muv\ and showing a similarly small value of $\beta$ from $\sim -1.8$ 
to $\sim -2.0$. The difference between the EMPG and MPG samples is therefore
not significant on the $\beta$ vs. \Muv\ plane.
At higher-redshifts ($z\gtrsim 3$), 
using continuum-selected galaxies such as Lyman break galaxies (LBGs),
earlier studies have indicated a decreasing trend of $\beta$ toward a fainter \Muv\ 
on average in the bright end of \Muv\ $=-22$ to $-17$ 
(e.g., \citealt{bouwens2009,bouwens2014}). 
On the other hand, some studies find a rather flat trend of $\beta \sim -2$ 
irrespective of \Muv\ \citep{finkelstein2012,dunlop2013} as similarly seen
in the local (E)MPG sample.
The relationship between $\beta$ and \Muv\ remains controversial at high-redshift,
and will be revisited below.

The middle panels of Figure \ref{fig:beta_Muv_ewhb_Z} show the distribution of $\beta$
as a function of EW(\Hb). We are interested in EW(\Hb) to use it as a proxy for stellar age 
as it corresponds to the current star-formation activity divided by the stellar mass 
(i.e., specific star-formation rate; sSFR).
We find a very weak trend of decreasing $\beta$ toward a larger EW(\Hb) in the EMPG sample.
Even the largest EW(\Hb) objects ($> 200$\,\AA) present a large scatter of $\beta$.
At first sight, it is weird to find such a weak correlation 
because a dust content is thought to be tightly linked with the past star-formation history and 
hence the stellar age.
We speculate this would be due to the fact that EW(\Hb) is a measure of the stellar age 
based on the current star-formation activity. 
Past star-formation episodes would play a role in dust production even for the largest
EW(\Hb) galaxies that are experiencing another phase of active star-formation.
Accordingly, EW(\Hb) would not be the primary governing the UV continuum slope
especially in the local universe.

At the bottom, the relationship between $\beta$ and metallicity is presented.
We clearly see a trend such that the UV continuum slope gets smaller at a lower metallicity, 
particularly at \Oabundance\ $\lesssim 7.3$.
At the lowest metallicity, 
the $\beta$ value reaches $\beta\sim -2.6$, 
which is as small as the intrinsic slope $\beta_0$ theoretically predicted
\citep{reddy2018_dust}.
This supports that EMPGs, especially with \Oabundance\ $\simeq 7.0$ or less metallicity, 
are almost dust-free systems as the productions and abundances of dust and metals 
are closely related to each other, both following the past star-formation history.
A relatively large scatter of metallicity for a given \Muv\ and EW(\Hb)
(Figures \ref{fig:dist_Mi_ewhb_Z_redshift}a(top) and \ref{fig:dist_Muv_ewhb_Z}(right))
would weaken the relationship and result in a flat trend of $\beta$ as seen in the 
top and middle panels.
On the other hand, the high-redshift samples are usually continuum-selected and 
hence magnitude-limited. 
Following a relatively tight luminosity--metallicity relationship and its redshift evolution 
(e.g., \citealt{zahid2011}),
fainter and higher-redshift galaxies tend to be metal-poorer.
The negative relationship between $\beta$ and \Muv\ observed in high-redshift UV-selected LBGs
and its possible redshift evolution toward smaller $\beta$ as seen in the top panels 
would thus be caused by such a luminosity--metallicity relation.

\begin{figure*}[t]
    \begin{center}
        \subfloat{
            \includegraphics[bb=18 159 565 550, width=0.46\textwidth]{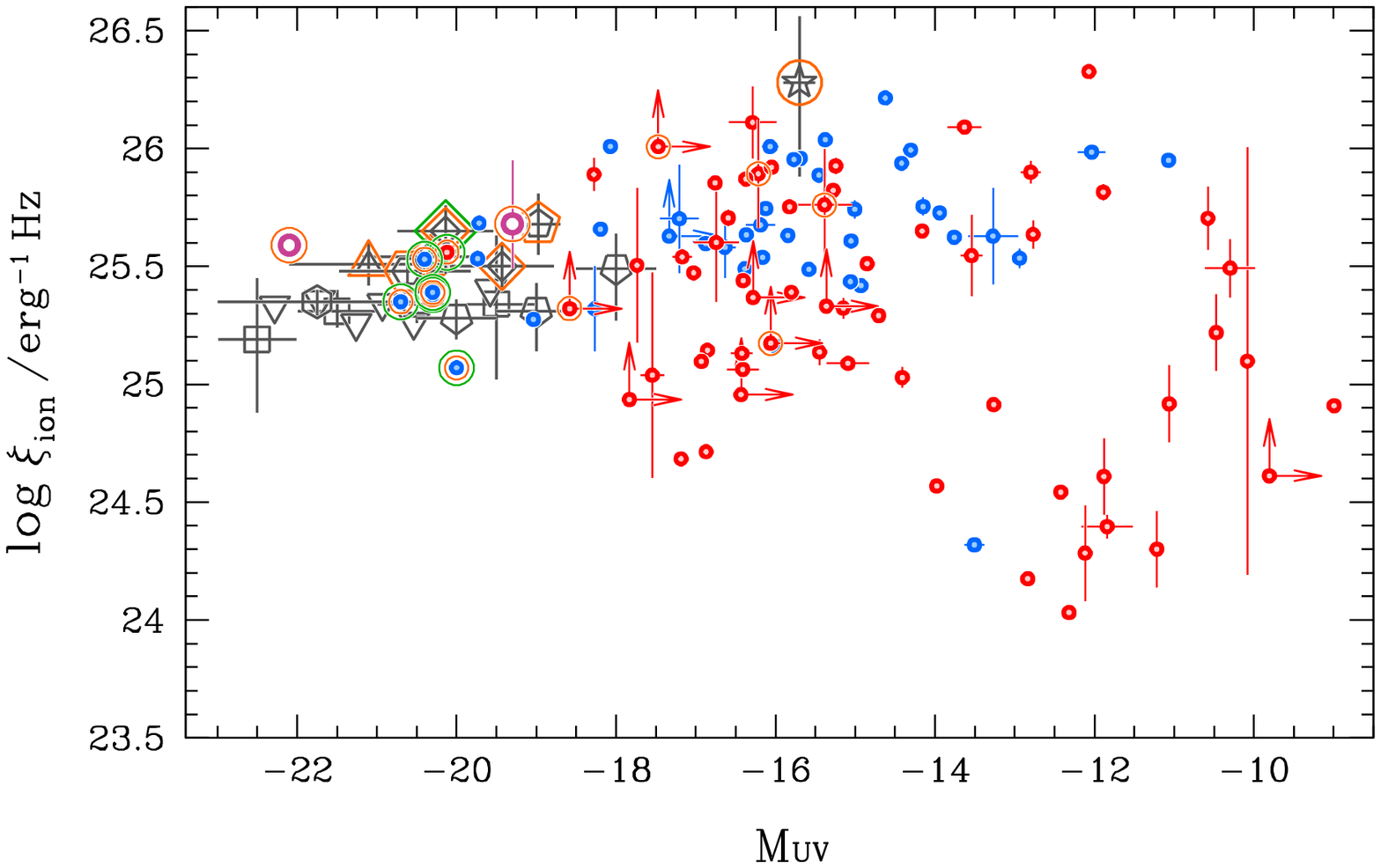}
        }
        \hspace{0.03\textwidth}
        \subfloat{
            \includegraphics[bb=18 159 565 550, width=0.46\textwidth]{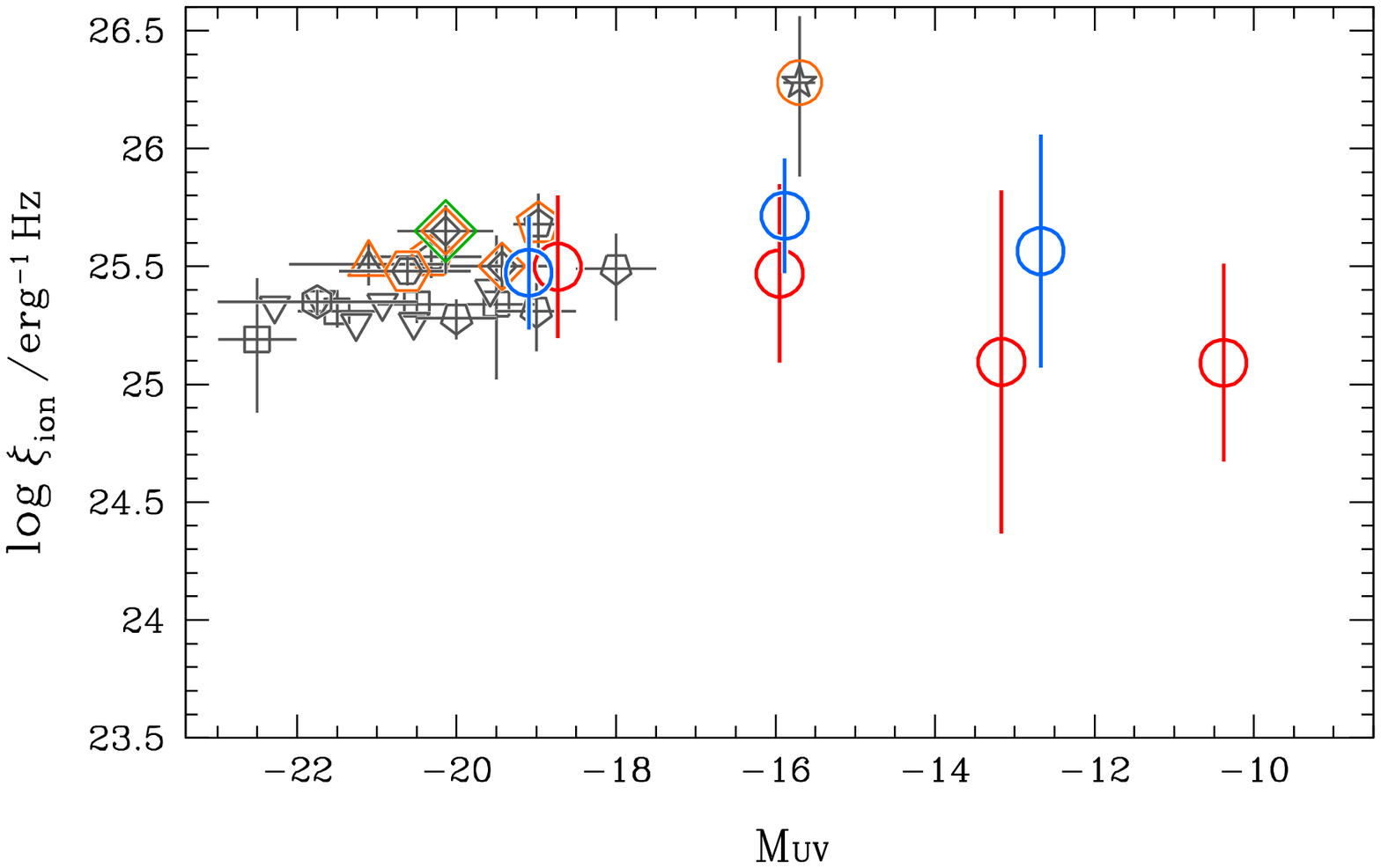}
        }
    
        \subfloat{
            \includegraphics[bb=18 159 565 550, width=0.46\textwidth]{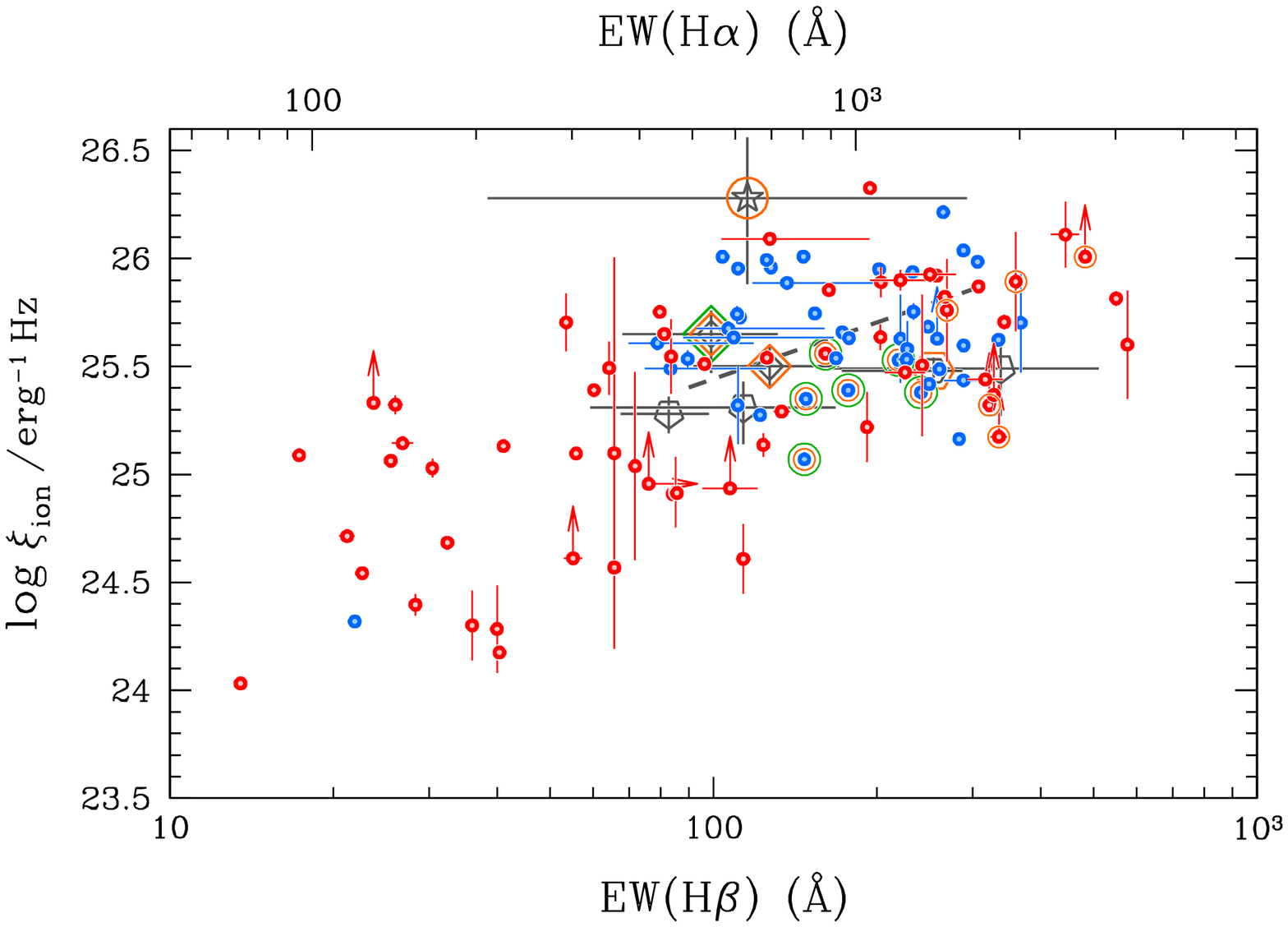}
        }
        \hspace{0.03\textwidth}
        \subfloat{
            \includegraphics[bb=18 159 565 550, width=0.46\textwidth]{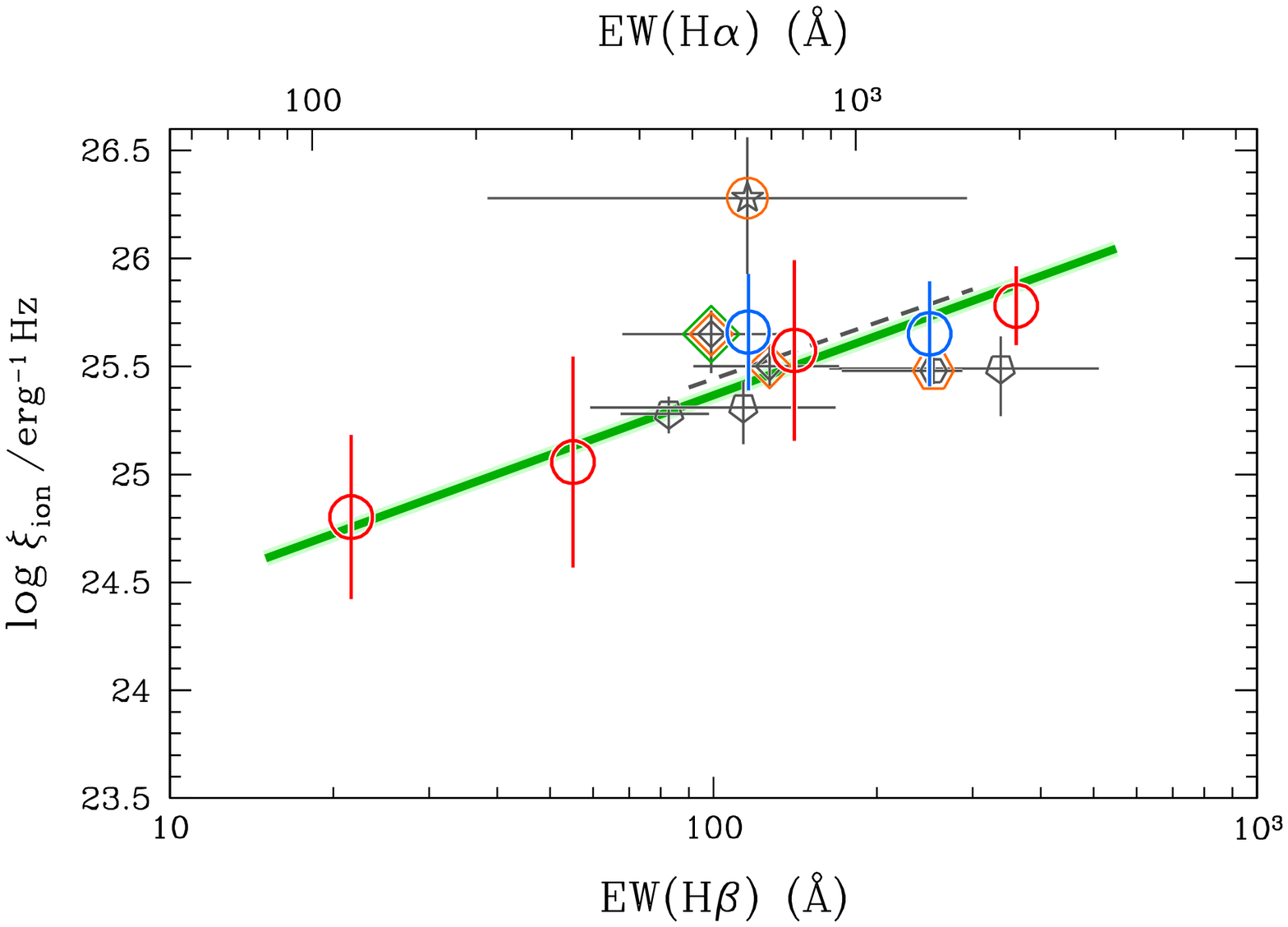}
        }
        
        \subfloat{
            \includegraphics[bb=18 159 565 550, width=0.46\textwidth]{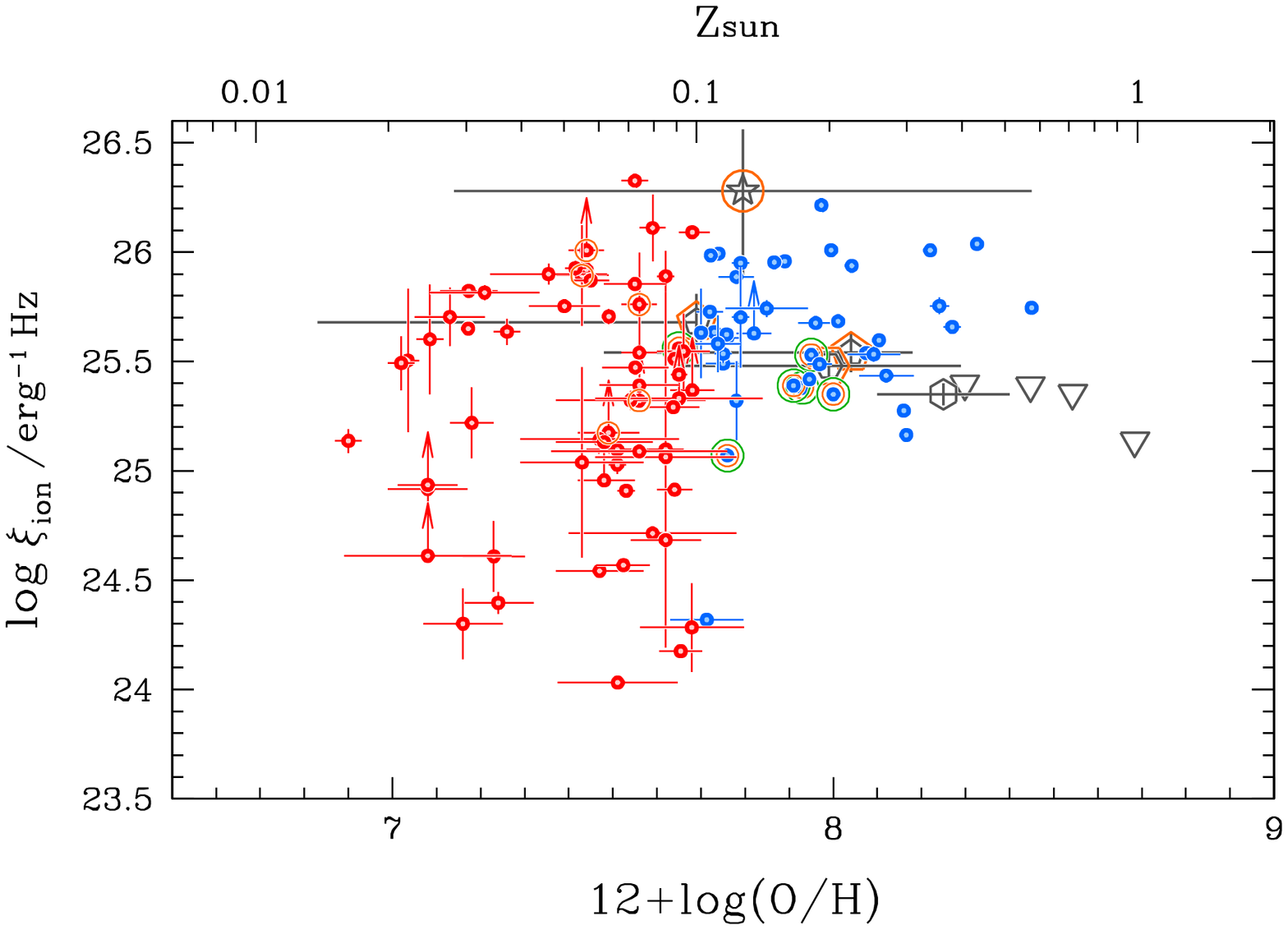}
        }
        \hspace{0.03\textwidth}
        \subfloat{
            \includegraphics[bb=18 159 565 550, width=0.46\textwidth]{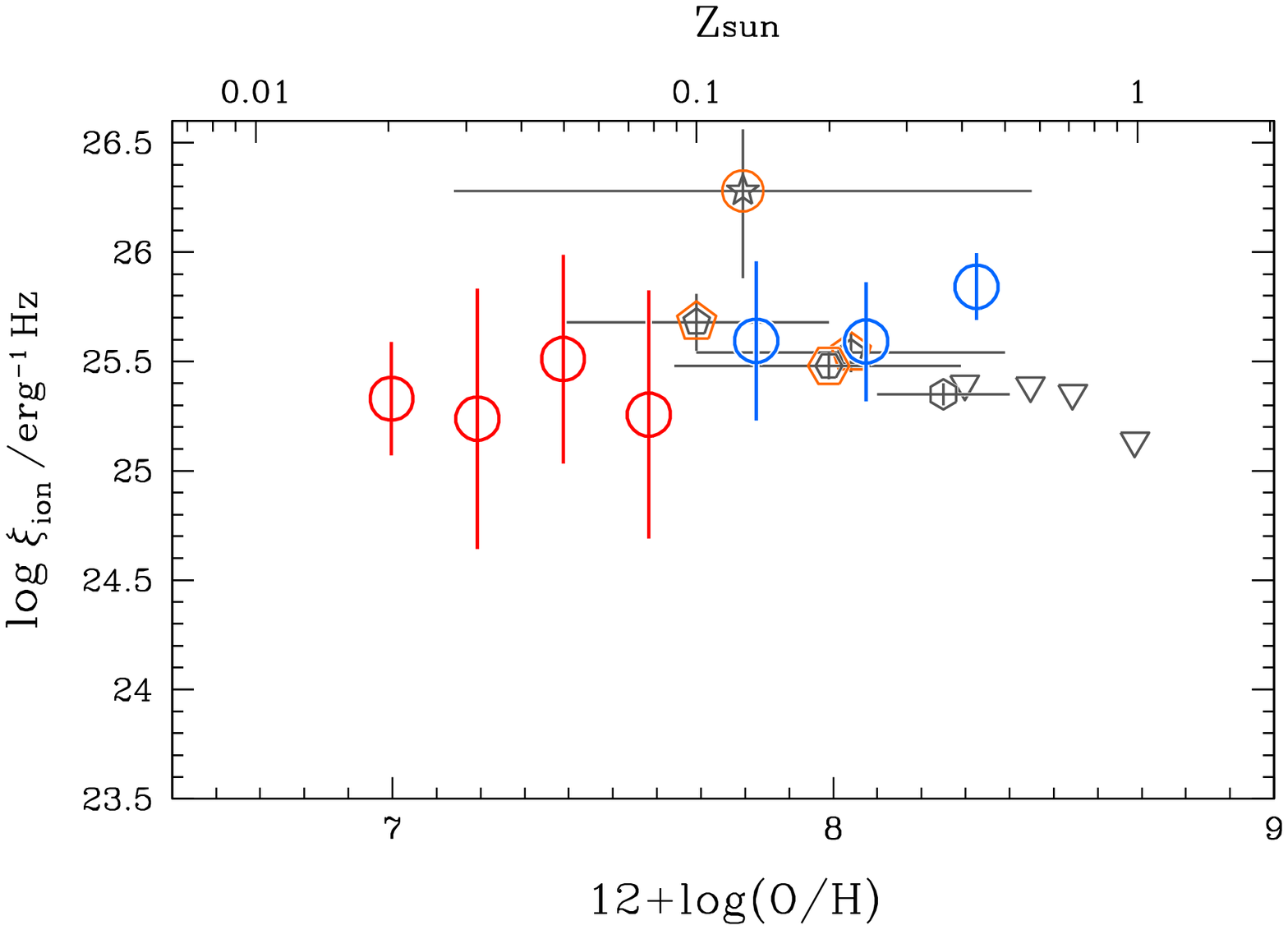}
        }
        \caption{%
        	Same as Figure \ref{fig:beta_Muv_ewhb_Z} but for the 
		relationships of ionizing photon production efficiency \xiion.
		Low-$z$ galaxies are as plotted in Figure \ref{fig:beta_Muv_ewhb_Z}.
		The other large open symbols denote average relationships at higher-redshifts;
		continuum-selected LBGs at $z=1.4-2.6$ (inverted open triangles; \citealt{shivaei2018}), 
		at $z=2-4$ ($30$\,deg-rotated hexagon; \citealt{nakajima2018_vuds,lefevre2019}), 
		at $z=3.8-5.0$ (open squares; \citealt{bouwens2016_xiion}),
		and at $z=4-5$ (inverted open pentagons; \citealt{lam2019}).
		Orange-marked symbols are LAEs
		at $z=3$ (open pentagons from \citealt{nakajima2018_z3laes} based on the UV emission lines, 
		open diamonds from \citealt{nakajima2020} based on the \Hb\ spectroscopy), 	
		at $z=4-5$ (star; \citealt{maseda2020}, 
		at $z=4.9$ (hexagon; \citealt{harikane2018}), and 
		at $z=6-7$ (triangle; \citealt{matthee2017}).
		We note that the \xiion\ is corrected for the escaping ionizing photons
		for the sources with an \fesc-measurement 
		(green-enclosed;\citep{izotov2016_nature,izotov2016_4more,schaerer2016,izotov2018_46per,schaerer2018,nakajima2020}).
		Two individual galaxies at $z=7-8$ are also presented with magenta circles
		whose \Lya\ and UV emission lines are detected \citep{stark2015_c4,stark2017}.
		In the middle panels, an average EW(\Hb)--\xiion\ relationship seen in extreme 
		emission line galaxies at $z=1.3-2.4$ is illustrated with a gray dashed line
		\citep{tang2019}.
		Our best-fit relationship of Equation (\ref{eq:xiion_ewhb}) is presented with
		a green solid line, confirming the previously known relationship but over a 
		wider range of EW(\Hb) of $\sim 10-600$\,\AA.
        }
        \label{fig:xiion_Muv_ewhb_Z}
    \end{center} 
\end{figure*}


\subsection{Ionizing photon production efficiency \xiion} \label{ssec:UV_properties_xiion}

Another key UV property is the ionizing photon production efficiency, \xiion.
This quantity represents the number of hydrogen-ionizing photons per UV luminosity; 
\begin{equation}
\xi_{\rm ion} = \frac{Q_{{\rm H}^0}}{L_{\rm UV}}.
\end{equation}
where $Q_{{\rm H}^0}$ is the ionizing photon production rate below $912$\,\AA,
and $L_{\rm UV}$ is the intrinsic UV continuum luminosity typically at around $1500$\,\AA.
The number of ionizing photons, $Q_{{\rm H}^0}$, can be determined via 
hydrogen recombination lines such as \Ha\ and \Hb\ (e.g \citealt{LH1995}), 
and the UV luminosity, $L_{\rm UV}$, is now derived from the \textsl{GALEX} 
FUV band photometry for the nearby galaxies
(as \Muv\ estimated; see \S\ref{ssec:UV_properties_Muv}).
Note that the conversion of $Q_{{\rm H}^0}$ assumes no escaping ionizing photons, 
i.e., all are converted into recombination radiation unless otherwise specified.
We will revisit the assumption later.

A crucial step is to match the apertures for the measurements of 
hydrogen recombination lines and FUV band photometry.
This is complicated by the spectroscopic measurements which were taken 
heterogeneously. 
We then use the total \Ha\ flux estimated from the excess in the broadband photometry,
which are measured in a consistent way as the total magnitude of \textsl{GALEX} FUV.
As illustrated in Figure \ref{fig:dist_Muv_ewhb_Z}, our EMPG sample present an intense 
hydrogen recombination lines of EW(\Hb) ($\sim 10$--$600$\,\AA\ at the rest-frame, 
$\sim 130$\,\AA\ median).
Given the typical EW ratio of EW(\Ha)$/$EW(\Hb) $=5.47$ \citep{kojima2020},
\Ha\ EWs are supposed to be very large on average for the EMPGs ($\sim 700$\,\AA)
and boost the photometry of the broadband where the \Ha\ is redshifted and falls in.
We can estimate the strength of \Ha\ by evaluating the excess in the photometric
broadband SED.
We retrieve the broadband photometric data from the archives of the 
SDSS Data Release 16
and the HSC-SSP S20A Data Release.
We use the HSC $riz$ band photometry for those in the HSC-SSP footprint.
Otherwise we use the SDSS $riz$ band photometry.
Fifteen sources, most of which are in the MPG sample from
\citet{guseva2007}, are not cross-matched with anything on the two catalogs, 
and thus not used in the following \xiion\ analyses.
We adopt the photometry of \verb+cmodel+ (HSC) and \verb+ModelMag+ (SDSS)
which delivers the total magnitude for each of the broadbands.
At $z\lesssim0.055$ with HSC, the $r$-band captures the \Ha\ line, and the $i$- and 
$z$-bands are used for the continuum estimation.
At a higher-redshift (up to the highest redshift of $z=0.13$ in our sample), we use
the $i$-band to probe the \Ha\ (+continuum) and the $z$-band for the continuum.
We use a BPASS SED with a stellar metallicity of $Z\sim0.07$\,\Zsun, constant star-formation,
and with an age of $50$\,Myr to assume the shape of the continuum around $iz$-bands.
A different demarcation redshift of $0.040$ is adopted for the SDSS photometry.
For each of the objects we use the systemic redshift determined from the 
optical spectroscopy as well as the actual transmission curves of the broadbands
to translate the broadband excess into the flux of \Ha.
We ignore the small contributions ($\lesssim10$\,\%) of the weaker emission lines 
such as \NII$\lambda 6584$ and \SII$\lambda \lambda 6717,6731$ 
for the EMPG sample.
The flux of \Ha\ is then reddening-corrected with the \ebv\ value estimated from
the Balmer decrements of the spectroscopy data with the \citet{calzetti2000}'s
attenuation law.
Finally, as a sanity check, we compare the dust-corrected fluxes of \Ha\ derived 
from the above method (i.e., the broadband excess) and spectroscopy. 
If the spectroscopic measure of \Ha\ is not available, 
we substitute the dust-corrected \Hb\ multiplied by the intrinsic \Ha$/$\Hb\ ratio 
($2.86$ in the Case B recombination) to be compared with the \Ha\ strength 
from the broadband excess.
The difference corresponds to the aperture correction for the slit- or fiber-loss 
in the spectroscopic observation.
The difference typically varies from $1.0$ to $2.2$ ($1.4$ median), 
which are reasonable as aperture corrections for EMPGs (e.g., \citealt{izotov2011}).

Figure \ref{fig:xiion_Muv_ewhb_Z} provides the distribution of \xiion\ 
as a function of various key properties of \Muv, EW(\Hb), and metallicity.
We have now explored the \xiion\ for the EMPGs over the wide ranges of 
properties, down to \Muv\ $\sim -10$, EW(\Hb) $\sim 10$--$600$\,\AA,
and \Oabundance\ down to $\sim 6.9$.
At the faint end of the UV luminosity \Muv\ $\gtrsim -15$,
a large variation of \xiion\ is interestingly recognized especially for the EMPG sample
(top panels of Figure \ref{fig:xiion_Muv_ewhb_Z}).
This is in contrast to the previous studies at high-redshift where an almost flat, 
or a weakly increasing trend of \xiion\ is suggested with UV luminosity decreasing 
in the range of \Muv\ from $-22$ to $\sim -19$
(e.g., \citealt{bouwens2016_xiion,shivaei2018,nakajima2018_z3laes}).
A stacking of faint, strong \Lya\ emitters at $z=4-5$ infers 
a very large value of $\log$ \xiion\ of $26.3$ albeit with a large uncertainty,
and supports an elevated ionizing photon production efficiency in faint galaxies 
\citep{maseda2020}.
LAEs are generally thought to be more efficient producers of ionizing photons 
at a given UV luminosity compared to continuum-selected LBGs
\citep{trainor2016,nakajima2016,matthee2017,harikane2018,nakajima2018_z3laes,nakajima2020}.
The EMPGs examined in this study do not apparently follow the trend.
Although the typical \xiion\ matches with those reported in high-redshift galaxies 
at \Muv\ $\sim -19$, it rather decreases with \Muv\ with a tendency that 
MPGs have a more or less larger \xiion\ value than EMPGs 
for a given UV luminosity.
The bottom panels of Figure \ref{fig:xiion_Muv_ewhb_Z} clarify the metallicity
dependence of \xiion.
A large variation remains at the lowest metallicity of \Oabundance\ $\sim 7.0$
whose \xiion\ varies from $log$ \xiion\ $\sim 24.0$ to $26.0$.
Faint UV luminosity and/or low metallicity in gas-phase would not be the primary properties 
that govern the efficient production of ionizing photons. 
We will turn back to these puzzling trends later.

On the other hand, the middle panels of Figure \ref{fig:xiion_Muv_ewhb_Z}
illustrate that EW(\Hb) is well-correlated with \xiion\ in a positive manner 
with a relatively small uncertainty. 
Both the EMPG and the MPG samples fall on the same relationship,
with the following equation form:
\begin{equation}
\log \xi_{\rm ion}  = 23.53 + 0.92 \times \log {\rm EW} ({\rm H}\beta).
\label{eq:xiion_ewhb}
\end{equation}
Such a tight correlation has been suggested in earlier studies in the local universe 
\citep{chevallard2018}%
\footnote{
This paper originally presents a tight relationship between \xiion\ and 
EW(\OIII$\lambda\lambda 5007,4959$). 
We expect a similar positive correlation between \xiion\ and \Hb\ 
given the relatively narrow range \OIII$/$\Hb\ in their sample.
}, 
at $z\sim 1-2$ \citep{tang2019}, at $z\sim 3$ \citep{nakajima2020}, 
and even at $z\sim 4-5$ \citep{harikane2018,lam2019}.
This work confirms the same relationship for metal-poor galaxies,  
and also suggests it holds over the wider range of EW(\Hb) than
previously explored, 
down to $\sim 10$\,\AA\ and up to $\sim 600$\,\AA.
The stacked faint LAEs at $z\sim 4-5$ interestingly appears to fall above 
the relationship \citep{maseda2020}, although the large uncertainty remains 
both in \xiion\ and EW(\Ha).
Following the apparently universal trend as a function of 
EWs of Hydrogen Balmer lines, the production efficiency of ionizing photons 
would be mainly governed by the stellar age of the current star-formation.
This is reasonable as it probes the relative abundance of the youngest, 
most massive stars (age of $\lesssim 10$\,Myr) to that of less massive,
long-lived stars that can contribute to the non-ionizing UV radiation 
(age of $\lesssim 100$\,Myr).
Theoretically a metal-poorer stellar population would present a harder 
ionizing spectrum due to several effects such as metal blanketing, 
rotational hardening, and binary evolution
(e.g. \citealt{levesque2012,kewley2013_theory,eldridge2017}).
In the current sample, however, we do not see any significant secondary 
dependence of metallicity on the relationship, albeit with a caveat that 
the metallicity we refer to is the gas-phase oxygen abundance while 
the stellar metallicity, especially the iron abundance, controls the hardness
of ionizing spectrum (e.g., \citealt{steidel2016,cullen2019}).
We need such metal absorption studies for EMPGs to discuss the 
stellar metallicity dependences.
The lack of metallicity dependence on \xiion\ in our sample can be partly 
because we explore only the lowest metallicity range below the sub-solar value. 
Indeed, a weak-but-decreasing trend of \xiion\ with metallicity is seen 
in the $z=2-3$ sample on average in the high metallicity range of 
\Oabundance\ $>8.2$ \citep{shivaei2018}, 
although this may be a result of the correlation between EW(\Hb) and metallicity 
in the MOSDEF sample \citep{reddy2018_mosdef}.
Moreover, the assumption of zero escape fraction may not be realistic 
for extremely faint, metal-poor systems (see below).

We now revisit the puzzling trends found in the \Muv\ vs. \xiion\ plot
following the strong dependence of \xiion\ on EW(\Hb).
At the bright end of Figure \ref{fig:xiion_Muv_ewhb_Z}, 
the relatively large \xiion\ values reported on average in high-redshift galaxies
are primarily due to a large EW 
in galaxies typically found at high-redshift \citep{khostovan2016,faisst2016_ew,reddy2018_mosdef}.
Among these high-redshift galaxies,
less massive galaxies with a fainter \Muv\ tend to be younger and present
a larger EW, and hence have a harder \xiion\ to set up 
the weakly increasing trend. 
In particular, LAEs with a stronger EW(\Lya) show a larger EW(\Hb),
and are associated with a harder ionizing spectrum as marked with
orange symbols in Figure \ref{fig:xiion_Muv_ewhb_Z}.
\citep{nakajima2016,matthee2017,harikane2018,nakajima2018_z3laes,nakajima2020,maseda2020}.
At the fainter end of \Muv\ ($\gtrsim 15$) in the current plot, 
half of the EMPGs in the faint \Muv\ range have a low value of EW(\Hb) 
below $\sim 80$\,\AA\ (Figure \ref{fig:dist_Muv_ewhb_Z})
and thus lower the typical value of \xiion\ in the current compilation.
The MPGs, which are compiled in a less complete manner 
(\S \ref{sssec:samples_empgs_others}) show a biased distribution of EW(\Hb) 
toward larger values (Figure \ref{fig:dist_Mi_ewhb_Z_redshift}a)
and thus tend to have a larger \xiion\ than the EMPGs for a given \Muv.

Finally, we note again that the current \xiion\ measurements for the compiled 
galaxies assume no escaping ionizing photons.
In Figure \ref{fig:xiion_Muv_ewhb_Z}, we plot the intrinsic \xiion\ values 
only for some green pea galaxies at $z\simeq 0.3$ and LAEs at $z\simeq 3$
whose escape fraction of ionizing photons, \fesc, is directly measured
(\fesc\ $\sim 0.05$ to $\sim 0.5$; 
\citealt{izotov2016_nature,izotov2016_4more,schaerer2016,izotov2018_46per,schaerer2018,fletcher2019,nakajima2020}).
The intrinsic values are greater than the observed ones by a factor of 
$1/(1-f_{\rm esc})$, which ranges from $\sim 1.05$ to as large as $2$
in the above cases.
In this sense, the \xiion\ values currently estimated for the remaining (E)MPGs 
serve as lower-limits, to be precise.
This could be part of another reason why the \xiion\ of the (E)MPGs appear 
not so large at the extremely faint and low-metallicity regimes.
Indeed, galaxies with a bluer UV continuum slope, as blue as the intrinsic slope $\beta_0$,
tend to be more associated with a LyC leakage 
(e.g., \citealt{zackrisson2013,zackrisson2017}).
Following the trend between $\beta$ and metallicity (bottom panels in Figure \ref{fig:beta_Muv_ewhb_Z}),
EMPGs are thought to have a higher chance to present a non-zero \fesc. 
On the other hand, the \fesc\ uncertainty would not break the relationship
between \xiion\ and EW(\Hb) 
because EW(\Hb) is observed to become small by the same factor of 
$1/(1-f_{\rm esc})$ as \xiion\ 
(see also the slope of Equation (\ref{eq:xiion_ewhb}) is almost unity).
Although direct LyC observations for the local (E)MPGs are challenging because they are 
too near-by ($z\lesssim 0.05$) to be observed with HST/COS in the wavelength below 
the Lyman limit, indirect methods such as using the \Lya's spectral profile and the flux
at the systemic velocity (e.g., \citealt{verhamme2017, vanzella2019_ion2, naidu2021}) 
could help understand the local (E)MPGs' intrinsic nature of ionizing photon production/escape
and its dependence on the galaxy's properties.

\section{Summary} \label{sec:summary}

We investigate the optical-line metallicity indicators together with the 
fundamental ultra-violet (UV) properties of low-mass, extremely metal-poor galaxies (EMPGs)
to provide useful anchors for forthcoming spectroscopic studies in the early universe.
We make use of the EMPRESS sample \citep{kojima2020} and perform a follow-up
spectroscopic observation to enlarge the EMPG sample.
Compiling previously known metal-poor galaxies from the literature, 
we build a large sample of EMPGs ($N^o=103$) 
covering a large parameter space of 
magnitude (\Mi\ from $-19$ to $-7$ and \Muv\ from $-20$ to $-9$) and 
\Hb\ equivalent width (EW(\Hb) from $10$ to $600$\,\AA), 
i.e., wide ranges of stellar mass and star-formation rate.
Our main results regarding the metallicity indicators are summarized as follows.
\begin{enumerate}
\item Utilizing the largest EMPG sample as well as the stacked spectra of 
$120,000$ SDSS galaxies \citep{curti2017,curti2020}, 
we derive the relationships between strong optical line ratios and gas-phase metallicity
over the range of \Oabundance\ $=6.9$ to $8.9$ corresponding to $0.02$ to $2$ solar metallicity \Zsun\
fully based on the reliable metallicity measurements of the direct $T_e$ method.
\item We confirm that R23-index, (\OIII+\OII)/\Hb, 
shows the smallest scatter in the relation with the metallicity measurements
($\Delta$\,log\,(O/H) $=0.14$) over the full metallicity range, suggesting that 
R23-index is most reliable among various metallicity indicators 
over the wide range of metallicity.
Unlike R23-index, the other metallicity indicators do not use a sum of singly and doubly
ionized lines and cannot trace both low and high ionization gas.
A caveat is an R23-based metallicity becomes less accurate around 
\Oabundance\ $\sim 8.0\pm 0.2$ as compared to the lower and higher metallicity ranges
due to the two-branch nature. 
\item We find that the accuracy of the metallicity indicators, 
including the famous ones such as R3-index and N2-index, 
becomes significantly improved (by a factor of as large as 2),
if one uses \Hb\ equivalent width measurements 
that tightly correlate with ISM ionization states.
This application is supported by our \verb+CLOUDY+ photoionization modeling,
and suggested to work irrespective of redshift.
We argue that it is crucial to correct for the ISM ionization condition 
in estimating the metallicity if the strong line methods,  
especially when using only low- or high-ionization lines
are used.
Such a correction would be of particular importance
when discussing the mass-metallicity relation,
its dependence on star-formation activity, and its cosmic evolution.
\end{enumerate}
Moreover, the analysis of \textsl{GALEX} FUV and NUV band photometry for the EMPGs
allows us to characterize the UV properties of UV absolute magnitude \Muv,
UV continuum slope $\beta$, and ionizing photon production efficiency \xiion\
for the extreme population of EMPGs. The main findings are as follows.
\begin{enumerate}
\setcounter{enumi}{3}
\item We identify the UV slope $\beta$ is best-correlated with metallicity
below 12+log(O/H) $\lesssim 7.4$.
The most metal-deficient galaxies with \Oabundance\ $\sim 7$ on average 
show the lowest $\beta$ value almost close to the intrinsic UV continuum slope 
$\beta_0=-2.6$.
The negative correlation between $\beta$ and \Muv\ known in high-redshift UV-selected 
galaxies would be caused by a luminosity--metallicity relation.
\item We confirm the ionizing photon production efficiency \xiion\ is best-correlated 
with EWs of Hydrogen Balmer lines of \Ha\ and \Hb\ over a wide range from 
EW(\Hb) $=10$\,\AA\ to $600$\,\AA.
A large variation of \xiion\ is recognized even for galaxies with the faintest UV luminosity
(\Muv\ $\gtrsim -15$) and the lowest metallicity (\Oabundance\ $\lesssim 7.69$). 
The variations of \xiion\ as well as EW(\Hb) for a given metallicity worsen 
the accuracy of the metallicity diagnostics as discussed above.
\end{enumerate}
The metallicity-sensitive emission line ratios and the UV properties 
for the compiled 103 EMPGs are publicly available in the form of ASCII table 
as partly listed in Table \ref{tbl:compilation_example}.

\newpage

\begin{splitdeluxetable*}{lccccccccBccccccccBcccccccc}
\tablecaption{Properties of Compiled 103 EMPGs
\label{tbl:compilation_example}}
\tablewidth{0.99\columnwidth}
\tabletypesize{\scriptsize}
\tablehead{
\colhead{ID} &
\colhead{REDSHIFT} &
\colhead{12+LOG(O/H)} &
\colhead{e\_12+LOG(O/H)} &
\colhead{EW(H-beta)} &
\colhead{Mi} &
\colhead{Muv} &
\colhead{BETA} &
\colhead{LOG(XIION)} &
\colhead{R23-INDEX} &
\colhead{e\_R23-INDEX} &
\colhead{R2-INDEX} &
\colhead{e\_R2-INDEX} &
\colhead{R3-INDEX} &
\colhead{e\_R3-INDEX} &
\colhead{O32-INDEX} &
\colhead{e\_O32-INDEX} &
\colhead{N2-INDEX} &
\colhead{e\_N2-INDEX} &
\colhead{O3N2-INDEX} &
\colhead{e\_O3N2-INDEX} &
\colhead{S2-INDEX} &
\colhead{e\_S2-INDEX} &
\colhead{Ne3O2-INDEX} &
\colhead{e\_Ne3O2-INDEX} \\
\colhead{--} &
\colhead{--} &
\colhead{--} &
\colhead{--} &
\colhead{0.1nm} &
\colhead{mag} &
\colhead{mag} &
\colhead{--} &
\colhead{[Hz.J-1.10+7]} &
\colhead{--} &
\colhead{--} &
\colhead{--} &
\colhead{--} &
\colhead{--} &
\colhead{--} &
\colhead{--} &
\colhead{--} &
\colhead{--} &
\colhead{--} &
\colhead{--} &
\colhead{--} &
\colhead{--} &
\colhead{--} &
\colhead{--} &
\colhead{--} 
}
\startdata
J0036+0052 &  0.0282 &  7.390 &  0.080 &  79.7 &  -16.21 &  -15.82 &  -2.36 &  25.75 &  0.790 &  0.005 &  -0.124 &  0.011 &  0.605 &  0.006 &  0.729 &  0.011 &  -2.121 &  0.038 &  2.726 &  0.039 &  -1.390 &  0.015 &  -0.381 &  0.016 \\
J01074656+01035206 &  0.0020 &  7.679 &  0.118 &  40.0 &  -9.80 &  -12.12 &  -1.61 &  24.28 &  0.726 &  0.016 &  0.432 &  0.019 &  0.293 &  0.018 &  -0.139 &  0.017 &   &   &   &   &   &   &  -1.180 &  0.073 \\
J0113+0052No.1 &  0.0038 &  7.150 &  0.090 &  40.7 &  -10.22 &   &   &   &  0.520 &  0.022 &  -0.083 &  0.040 &  0.275 &  0.022 &  0.358 &  0.038 &  $<$-2.133 &   &   &   &  $<$-1.656 &   &  -0.541 &  0.067 \\
J0113+0052No.2 &  0.0038 &  7.320 &  0.040 &  20.8 &  -10.20 &   &   &   &  0.523 &  0.029 &  0.298 &  0.035 &  -0.002 &  0.030 &  -0.300 &  0.034 &  -1.539 &  0.058 &  1.537 &  0.064 &  -0.878 &  0.031 &  $<$-1.466 &   \\
J0113+0052No.3 &  0.0038 &  7.300 &  0.080 &  30.0 &  -10.23 &   &   &   &  0.633 &  0.024 &  0.184 &  0.035 &  0.312 &  0.026 &  0.127 &  0.032 &  $<$-2.128 &   &   &   &  -1.063 &  0.032 &  $<$-1.352 &   \\
\enddata
\tablecomments{%
Table \ref{tbl:compilation_example} is published in its entirety in the machine-readable format. A portion is shown here for guidance regarding its form and content.
}
\end{splitdeluxetable*}

\begin{acknowledgments}

We thank M. Curti for providing data of the stacked SDSS spectra shown in Figs.\ref{fig:Z_empirical_all1}--\ref{fig:Z_empirical_model2} at the high-metallicity regime, and the anonymous referee for helpful comments that improved our manuscript.
This paper includes data gathered with the 6.5 m Magellan Telescopes located at Las Campanas Observatory, Chile. We are grateful to the observatory personnel for help with the observations.

The Hyper Suprime-Cam (HSC) collaboration includes the astronomical communities of Japan and Taiwan, and Princeton University. The HSC instrumentation and software were developed by the National Astronomical Observatory of Japan (NAOJ), the Kavli Institute for the Physics and Mathematics of the Universe (Kavli IPMU), the University of Tokyo, the High Energy Accelerator Research Organization (KEK), the Academia Sinica Institute for Astronomy and Astrophysics in Taiwan (ASIAA), and Princeton University. Funding was contributed by the FIRST program from Japanese Cabinet Office, the Ministry of Education, Culture, Sports, Science and Technology (MEXT), the Japan Society for the Promotion of Science (JSPS), Japan Science and Technology Agency (JST), the Toray Science Foundation, NAOJ, Kavli IPMU, KEK, ASIAA, and Princeton University. 

This paper makes use of software developed for the Large Synoptic Survey Telescope. We thank the LSST Project for making their code available as free software at \url{http://dm.lsst.org}.

The Pan-STARRS1 Surveys (PS1) have been made possible through contributions of the Institute for Astronomy, the University of Hawaii, the Pan-STARRS Project Office, the Max-Planck Society and its participating institutes, the Max Planck Institute for Astronomy, Heidelberg and the Max Planck Institute for Extraterrestrial Physics, Garching, The Johns Hopkins University, Durham University, the University of Edinburgh, Queen’s University Belfast, the Harvard-Smithsonian Center for Astrophysics, the Las Cumbres Observatory Global Telescope Network Incorporated, the National Central University of Taiwan, the Space Telescope Science Institute, the National Aeronautics and Space Administration under Grant No. NNX08AR22G issued through the Planetary Science Division of the NASA Science Mission Directorate, the National Science Foundation under Grant No. AST-1238877, the University of Maryland, and Eotvos Lorand University (ELTE) and the Los Alamos National Laboratory.

Based in part on data collected at the Subaru Telescope and retrieved from the HSC data archive system, which is operated by Subaru Telescope and Astronomy Data Center at National Astronomical Observatory of Japan.

\end{acknowledgments}

\begin{acknowledgments}

This work is supported by the World Premier International
Research Center Initiative (WPI Initiative), MEXT, Japan, as
well as JSPS KAKENHI Grant Numbers
JP20K22373, JP20H00180, JP21H04467, JP21K13953, and JP21K03622.
JHK acknowledges the support from the National Research Foundation of Korea (NRF) grant, No. 2021M3F7A1084525, funded by the Korea government (MSIT).

\end{acknowledgments}

%

\vspace{5mm}
\facilities{Magellan (MagE), GALEX}



\bibliographystyle{aasjournal}{}
\bibliography{Refs_paper.bib}{}



\end{document}